\documentclass[reprint,amsmath,amssymb,aps]{revtex4-2}
\usepackage[colorlinks,citecolor=blue,linkcolor=blue,anchorcolor=blue,filecolor=blue,
urlcolor=blue]{hyperref}
\usepackage{graphicx}
\usepackage{ulem}
\usepackage{dcolumn}
\usepackage{bm}
\usepackage{todonotes}
\usepackage{bigints}
\usepackage{amsmath}
\usepackage{booktabs}

\newcommand{\vis}{\mathrm{vis}}
\newcommand{\dm}{\mathrm{dm}}
\newcommand{\resc}{\mathrm{resc}}
\newcommand{\sym}{\mathrm{sym}}
\newcommand{\kin}{\mathrm{kin}}
\newcommand{\pot}{\mathrm{pot}}
\newcommand{\had}{\mathrm{had}}
\newcommand{\rmf}{\mbox{\tiny RMF}}
\newcommand{\rmfsrc}{\mbox{\tiny RMF-SRC}}
\newcommand{\cru}{\mathrm{crust}}
\newcommand{\src}{\mbox{\tiny SRC}}
\newcommand{\sat}{\mathrm{sat}}
\newcommand{\ma}{\mathrm{max}}

\newcommand{\kfp}{k_{F_p}}
\newcommand{\kfn}{k_{F_n}}
\newcommand{\Ln}{\mathrm{ln}}

\begin{document}

\title{Effects of short-range correlations at high densities on neutron stars with and without DM content: role of the repulsive self-interaction}
\author{Odilon Louren\c{c}o$^1$, Everson H. Rodrigues$^2$, Carline Biesdorf$^1$, Mariana Dutra$^1$}
\affiliation{
$^1$Departamento de F\'isica e Laborat\'orio de Computa\c c\~ao Cient\'ifica Avan\c cada e Modelamento (Lab-CCAM), Instituto Tecnol\'ogico de Aeron\'autica, DCTA, 12228-900, S\~ao Jos\'e dos Campos, SP, Brazil
\\
$^2$Núcleo de Astrofísica e Cosmologia (Cosmo-Ufes) $\&$ Departamento de Física, Universidade Federal do Espírito Santo, 29075–910, Vitória, ES, Brazil
}

\date{\today}

\begin{abstract}
In this work, we investigate how short-range correlations affect relativistic hadronic models at high densities, with direct consequences for the structure of neutron stars, both with and without dark matter content. Two versions of the model are examined: one with vector self-interactions up to second order ($\omega_0^2$) and another including a fourth-order term ($\omega_0^4$). We show that SRC tend to soften the equation of state when only the quadratic term is present, but produce a noticeable stiffening once the $\omega_0^4$ term is included. The corresponding Tolman–Oppenheimer–Volkoff solutions for pure neutron stars indicate that short-range correlations reduce the maximum mass in the first case but increase it in the second. Extending the analysis to stars containing a fermionic dark matter component, within the two-fluid formalism, we verify that the same features appear in the respective mass–radius diagrams. In particular, the decrease of the maximum mass with increasing dark matter fraction is partly compensated by the SRC effects in the hadronic sector for the model with the fourth-order term. In all cases, the resulting parametrizations are consistent with recent astrophysical constraints, including the joint NICER–XMM-Newton analyses of the pulsars PSR~J0030+0451 and PSR~J0740+6620, as well as the gravitational-wave event GW190425.
\end{abstract}

\maketitle

\section{Introduction}
\label{sec:int}

The behavior of strongly interacting matter at supranuclear densities is a central topic in nuclear astrophysics, with neutron stars (NSs) serving as natural laboratories for probing matter under extreme conditions. Recent advances in multi-messenger astronomy - as the gravitational wave detections from binary mergers~\cite{Abbott_2017,Abbott_2020-2}, the X-ray timing from the NICER mission~\cite{Riley_2019,Miller_2019,Salmi_2022,Miller_2021,Vinciguerra_2024}, and the precision mass measurements from radio pulsars - have placed increasingly stringent constraints on the equation of state (EoS) of dense matter.

In conventional relativistic mean-field (RMF) models, the nucleon momentum distribution is typically described by a step function, corresponding to a free Fermi gas of independent particles. This approximation neglects the effects of nucleon-nucleon correlations at short distances, which have been experimentally observed and theoretically studied in recent years. Short-range correlations (SRC) modify this picture by introducing a high-momentum tail (HMT) in the distribution, reflecting the presence of correlated nucleon pairs with large relative momentum~\cite{sciencesrc1,naturesrc2,naturescr3,naturesrc4,hen2017,duer2019,cai,baoanli21,baoanli22,baoanli-aop,rios1,rios2,revsrc1,revsrc2} . These correlations are supported by electron-induced nucleon knockout experiments, and can be implemented in RMF models through phenomenological parametrizations that replace the step function with a distribution proportional to \(1/k^4\) above the Fermi surface. This modification has been shown to impact the thermodynamic properties of nuclear matter and the structure of neutron stars. Nonlinear RMF extensions, such as the inclusion of a quartic \(\omega^4\) term, have been shown to regulate the stiffness of the EoS at high densities~\cite{muller-serot96}. As demonstrated in Ref.~\cite{muller-serot96}, this term arises naturally in effective field theory and significantly affects neutron star mass predictions, even when models share identical saturation properties.

In parallel, the possibility that NSs may contain dark matter (DM) has gained increasing attention. Although this matter may not interact with the visible - also referred to as hadronic, ordinary, or normal matter - except through gravity, gravitational capture can lead to its accumulation in the stellar interior~\cite{kouvaris2008,leung2011,Brayeur_2012,das2021}. Fermionic DM candidates, such as the neutralino, can be modeled as relativistic Fermi gases with or without self-interactions~\cite{Nelson_2019,milva24}. Given the absence of confirmed non-gravitational couplings, a two-fluid formalism is often adopted, treating the visible and dark sectors as independent components coupled only through gravity~\cite{goldman1989,sandin2009,tolos2015dark,dengler2021erratum,dengler2022second,carline25}.

In this work we investigate how SRC affect the high-density behavior of relativistic hadronic models and the resulting implications for NS structure, both with and without DM content. We consider two versions of the RMF model: one with vector self-interactions up to second order, and another including a fourth-order term. By analyzing the asymptotic behavior of the energy and pressure, we show that SRC tend to soften the EoS when only the quadratic term is present, but produce a noticeable stiffening once the quartic term is included.

We extend our analysis to neutron stars containing a fermionic dark matter component, modeled within a two-fluid formalism in which the visible and dark sectors interact solely through gravity. In particular, we examine how SRC in the hadronic sector can partially compensate for the softening effects introduced by dark matter, especially in models with stronger vector self-interactions. All parametrizations are calibrated to reproduce key nuclear empirical parameters and are tested against recent observational constraints, including the NICER and XMM-Newton measurements of PSR J0030+0451 and PSR J0740+6620, as well as the gravitational-wave event GW190425.

The paper is organized as follows. In Sec.~\ref{sec:rmfsrc} we present the RMF model with SRC and derive the associated thermodynamic quantities. Sec.~\ref{sec:III} is devoted to the analysis of the high-density limit and its consequences for NS structure. In Sec.~\ref{sec:rmf-src-dm} we introduce the DM component using a two-fluid formalism and explore its impact on the mass-radius relation, highlighting the interplay between SRC and DM in satisfying current astrophysical constraints. Finally, our conclusions are summarized in Sec.~\ref{sec:summ}.

\section{Relativistic short-range correlated model}
\label{sec:rmfsrc}

We consider a typical Lagrangian density that incorporates meson self-interaction terms up to the fourth order in the meson fields. It takes the following form
\begin{align}
\mathcal{L}_{\had} &= \overline{\psi}(i\gamma^\mu\partial_\mu - M_{\mbox{\tiny nuc}})\psi 
+ g_\sigma\sigma\overline{\psi}\psi 
- g_\omega\overline{\psi}\gamma^\mu\omega_\mu\psi 
\nonumber \\ 
&- \frac{g_\rho}{2}\overline{\psi}\gamma^\mu\vec{b}_\mu\vec{\tau}\psi
+\frac{1}{2}(\partial^\mu \sigma \partial_\mu \sigma - m^2_\sigma\sigma^2)
-\frac{1}{4}F^{\mu\nu}F_{\mu\nu} 
\nonumber\\
&
+ \frac{1}{2}m^2_\omega\omega_\mu\omega^\mu 
-\frac{1}{4}\vec{B}^{\mu\nu}\vec{B}_{\mu\nu} 
+ \frac{1}{2}m^2_\rho\vec{b}_\mu\vec{b}^\mu
\nonumber \\
&- \frac{A}{3}\sigma^3 - \frac{B}{4}\sigma^4 + \frac{C}{4}(g_\omega^2\omega_\mu\omega^\mu)^2,
\label{eq:dlag}
\end{align}
where $\psi$ denotes the nucleon field (mass $M_{\mbox{\tiny nuc}}$), while $\sigma$, $\omega^\mu$, and $\vec{b}_\mu$ correspond to the scalar, vector, and isovector-vector meson fields, respectively (masses $m_\sigma$, $m_\omega$, and $m_\rho$). The second rank tensors are defined as $F_{\mu\nu} = \partial_\nu\omega_\mu - \partial_\mu\omega_\nu$ for the vector meson and $\vec{B}_{\mu\nu} = \partial_\mu\vec{b}_\nu - \partial_\nu\vec{b}_\mu - g_\rho(\vec{b}_\mu \times \vec{b}_\nu)$ for the isovector-vector meson. Generalizations of Eq.~\eqref{eq:dlag} with more sophisticated meson-meson interactions can be found in Refs.~\cite{baoanli08,dutra14,lattimer24}, for instance. The mean-field approximation consists in replacing all mediator fields by their respective ``classical'' values: $\sigma\rightarrow \left<\sigma\right>\equiv\sigma$, $\omega_\mu\rightarrow \left<\omega_\mu\right>\equiv\omega_0$, $\vec{b}_\mu\rightarrow \left<\vec{b}_\mu\right>\equiv b_{0(3)}$. The use of the Euler-Lagrangian equations, along with such an approximation, leads to the field equations given by
\begin{align}
\frac{\Phi}{G_\sigma^2} + a\Phi^2 + b\Phi^3 &= \rho_s,
\label{eq:phi}
\\
W\left(\frac{1}{G_\omega^2} + CW^2\right) &= \rho,
\label{eq:w}
\\
\frac{2R}{G_\rho^2} &= (2y -1)\rho, 
\label{eq:r}
\\
[\gamma^\mu (i\partial_\mu - V) - M^*]\psi &= 0,
\end{align}
with the effective nucleon mass written as
\begin{align}
M^* &= M_{\mbox{\tiny nuc}} - \Phi,
\label{eq:mnuc}
\end{align}
where $\Phi=g_\sigma\sigma$, $W=g_\omega\omega_0$, $R=g_\rho b_{0(3)}$, $G_i^2=(g_i/m_i)^2$ ($i=\sigma,\omega,\rho$), $a=A/g_\sigma^3$, $b=B/g_\sigma^4$, and $V=W+\tau_3R/2$ with $\tau_3=1$ for protons and $-1$ for neutrons. The proton fraction of the system is defined as $y=\rho_p/\rho$, and the total density is $\rho=\rho_p+\rho_n$, with $\rho_p$~($\rho_n$) being the proton~(neutron) density. The total scalar density is $\rho_s =\left<\overline{\psi}\psi\right>={\rho_s}_p+{\rho_s}_n$.

Using the Lagrangian density in Eq.~(\ref{eq:dlag}) together with the associated energy-momentum tensor $T^{\mu\nu}$, one can derive the energy density and pressure of the system. The respective expressions are given by
\begin{align}
\mathcal{E}_{\had} &= \mathcal{E}_{\kin,p} + \mathcal{E}_{\kin,n} + \mathcal{E}_{\pot}
\label{eq:eden}
\end{align}
with the potential part written as
\begin{align}
\mathcal{E}_{\pot} &= -\frac{W^2}{2G_\omega^2} - \frac{C}{4}W^4 + W\rho 
- \frac{R^2}{2G_\rho^2} + \frac{R}{2}(2y-1)\rho
\nonumber\\
&+ \frac{1}{2G_\sigma^2}\Phi^2 + \frac{a}{3}\Phi^3 + \frac{b}{4}\Phi^4,
\end{align}
or, equivalently as
\begin{align}
\mathcal{E}_{\pot} &= \frac{1}{2G_\omega^2}W^2 + \frac{3C}{4}W^4 + \frac{1}{2G_\rho^2}R^2 
+ \frac{1}{2G_\sigma^2}\Phi^2
\nonumber\\
&+ \frac{a}{3}\Phi^3 + \frac{b}{4}\Phi^4,
\end{align}
and
\begin{align}
P_{\had} &= P_{\kin,p} + P_{\kin,n} + \frac{1}{2G_\omega^2}W^2 + \frac{C}{4}W^4 + \frac{1}{2G_\rho^2} R^2 
\nonumber\\
&-\frac{1}{2G_\sigma^2}\Phi^2 - \frac{a}{3}\Phi^3 - \frac{b}{4}\Phi^4.
\label{eq:press}
\end{align}
At the mean-field level, the inclusion of SRC in the hadronic model impacts the kinetic contributions of the respective thermodynamical quantities, due to the modification of the nucleon momentum distribution function in this case written as~\cite{cai}
\begin{eqnarray}
n_{n,p}(k) = \left\{ 
\begin{array}{ll}
\Delta_{n,p}, & 0<k<k_{F\,{n,p}}
\\ \\
\dfrac{C_{n,p}\,k_{F\,{n,p}}^4}{k^4}, & k_{F\,{n,p}}<k<\phi_{n,p} k_{F\,{n,p}}.
\end{array} 
\right.
\label{eqhtm}
\end{eqnarray}
Note the replacement of the usual step function with one that includes the HMT. The quantities $\Delta_{n,p}$, $C_{n,p}$, and $\phi_{n,p}$ could, in principle, be treated as independent parameters. However, they are constrained by the following normalization condition,
\begin{equation}
\frac{1}{\pi^2} \int_0^{\infty}dk\,k^2\,n_{n,p}(k) = \frac{({k_{F\,{n,p}}})^3}{3\pi^2} = \rho_{p,n}.
\label{norm}
\end{equation}
This relation allows one to eliminate one of the parameters in terms of the other two. In this case, we write $\Delta_{n,p}$ as a function of $C_{n,p}$ and $\phi_{n,p}$. Moreover, theoretical results based on self-consistent Green's function approaches~\cite{green} and Brueckner-Hartree-Fock models~\cite{bhf} reveal that the depletion factor $\Delta_{n,p}$ behaves nearly linearly with the isospin asymmetry, $\delta=1-2y$. This empirical trend, together with the normalization condition, leads to the following expressions
\begin{align}
\Delta_{n,p}(y) &= 1 - 3C_{n,p}(y)[1-1/\phi_{n,p}(y)],
\\
C_{n,p}(y)&= C_0[1 \pm C_1(1-2y)],
\\
\phi_{n,p}(y)&= \phi_0[1 \pm \phi_1(1-2y)],
\end{align}
where the plus sign applies to neutrons and the minus sign to protons in the last two equations. The constants adopted in these parametrizations are $C_0 = 0.161$, $C_1 = -0.25$, $\phi_0 = 2.38$, and $\phi_1 = -0.56$~\cite{cai}. These values reproduce a HMT contribution of about 28\% in symmetric nuclear matter~(SNM) and approximately 1.5\% in the case of pure neutron matter~(PNM)~\cite{cai}.

The inclusion of SRC lead to the following generalized equations for the protons and neutrons kinetic contributions to the energy density,
\begin{align} 
&\mathcal{E}_{\kin,n,p} = \frac{\Delta_{n,p}(y)}{\pi^2} \int_0^{{k_{F\,{n,p}}}} 
k^2dk({k^{2}+M^{* 2}})^{1/2}
\nonumber\\
&+ \frac{C_{n,p}(y)}{\pi^2}\int_{k_{F\,{n,p}}}^{\phi_{n,p}(y){k_{F\,{n,p}}}} 
\frac{{k_F}_{n,p}^4}{k^2}\, dk({k^{2}+M^{* 2}})^{1/2},
\label{eq:dekin}
\end{align}
and pressure,
\begin{eqnarray} 
P_{\kin,n,p} &=&  
\frac{\Delta_{n,p}(y)}{3\pi^2} \int_0^{k_{F\,{n,p}}}  
\frac{k^4dk}{\left({k^{2}+M^{*2}}\right)^{1/2}} 
\nonumber\\
&+& \frac{C_{n,p}(y)}{3\pi^2} \int_{k_{F\,{n,p}}}^{\phi_{n,p}(y){k_{F\,{n,p}}}} 
\frac{{k_F}_{n,p}^4dk}{\left({k^{2}+M^{*2}}\right)^{1/2}}.
\label{eq:pkin}
\end{eqnarray}
The scalar densities are also modified to
\begin{align}
&{\rho_s}_{n,p} = 
\frac{M^*\Delta_{n,p}(y)}{\pi^2} \int_0^{k_{F\,{n,p}}}  
\frac{k^2dk}{\left({k^{2}+M^{*2}}\right)^{1/2}} 
\nonumber\\
&+ \frac{M^*C_{n,p}(y)}{\pi^2} \int_{k_{F\,{n,p}}}^{\phi_{n,p}(y){k_{F\,{n,p}}}} 
\frac{{k_F}_{n,p}^4}{k^2}  \frac{dk}{\left({k^{2}+M^{*2}}\right)^{1/2}}.
\label{eq:rhos}
\end{align}

Note that in the limit $\phi_0=1$ and $\phi_1=0$, one obtains $\Delta_{n,p}=\phi
_p=\phi_n=1$, and the momentum distribution reduces to the standard step function, recovering the conventional RMF expressions without a high-momentum tail.

A specific quantity that can be derived from the energy density is the chemical potential. For the case of RMF models, its general definition is given by~\cite{jerome24}
\begin{align}
\mu_q &= \frac{\partial\mathcal{E}_{\had}}{\partial\rho_q}\Bigg|_{\rho_{\bar{q}},W,R,\Phi} 
\nonumber\\
&= \frac{\partial(\mathcal{E}_{\kin,p}+\mathcal{E}_{\kin,n})}{\partial \rho_q}\Bigg|_{\rho_{\bar{q}},W,R,\Phi} + \frac{\partial\mathcal{E}_{\pot}}{\partial \rho_q}\Bigg|_{\rho_{\bar{q}},W,R,\Phi},
\end{align}
in which $q$ represents a particle of a given isospin-index, and ${\bar q}$ describes the other one. The full expressions of $\mu_p$ and $\mu_n$ are 
\begin{align}
\mu_{p,n} &= \mu^{p,n}_\kin + W \pm \frac{R}{2},
\end{align}
where 
\begin{align}
\mu^{p,n}_\kin = \Delta_{p,n}(y)E^*_{F_{p,n}} + \mu^{p,n}_{\kin(\src)},
\end{align}
with $E^*_{F_{p,n}} = (k^2_{F_{p,n}} + M^{*2})^{1/2}$,
\begin{align}
&\mu^p_{\kin(\src)} = 3C_p(y)\left[E^*_{F_p} - \frac{\left(\phi_p^2(y)\kfp^2 + {M^{*2}}\right)^{1/2}}{\phi_p(y)} \right]
\nonumber\\
&+ 4C_p(y)\kfp\Ln\left[\frac{\phi_p(y)\kfp + \left(\phi_p^2(y)\kfp^2 + {M^{*2}}\right)^{1/2}}{\kfp +E^*_{F_p}}\right]  
\nonumber\\
&+ \frac{2}{\pi^2}\frac{\rho_n}{\rho^2}\eta_p,
\label{eq:mupkinsrc}
\end{align}
and
\begin{align}
&\mu^n_{\kin(\src)} = 3C_n(y)\left[E^*_{F_n} - \frac{\left(\phi_n^2(y)\kfn^2 + {M^*}^2\right)^{1/2}}{\phi_n(y)} \right] 
\nonumber\\
&+ 4C_n(y)\kfn\Ln\left[\frac{\phi_n(y)\kfn + \left(\phi_n^2(y)\kfn^2 + {M^*}^2\right)^{1/2}}{\kfn + E^*_{F_n}}\right]  
\nonumber\\
&+ \frac{2}{\pi^2}\frac{\rho_p}{\rho^2}\eta_n. 
\label{eq:munkinsrc}
\end{align}
In Refs.~\cite{souza2020,dutra_mnras22,lourenco_prd2022_1,lourenco_prd2022_2,pelicer2023}, an approximate version of Eqs.~\eqref{eq:mupkinsrc}--\eqref{eq:munkinsrc} was used, in which $\Delta_{p,n}$, $C_{p,n}$, and $\phi_{p,n}$ were considered independent of $\rho_p$ and $\rho_n$. In that case, the quantities $\eta_{p,n}$ vanish. Here, we take into account the full dependence on $\rho_p$ and $\rho_n$, thus obtaining the complete expressions presented earlier. The details of these derivations, as well as the expressions for $\eta_{p,n}$, are given in the Appendix.

\section{High density limit and related implications for neutron stars}\label{sec:III}

As in Ref.~\cite{muller-serot96}, we begin our analysis in the high-density regime by first noting that the effective mass in typical RMF models is positive and lies within the range $0 \le M^* \le M$, with the limit $M^* \sim 0$ being attained at high densities. This leads to a finite value for $\Phi = M_{\mbox{\tiny nuc}} - M^*$ and, consequently, for the last three terms in Eqs.\eqref{eq:eden}–\eqref{eq:press} in this regime. Physically, this feature is interpreted as the capability of the models to reproduce, at least partially, the chiral-symmetry restoration. In this case, the nucleon mass is reduced as a consequence of its interaction with the medium. In particular, at $\rho\rightarrow\infty$ the kinetic terms of energy density and pressure are modified to 
\begin{align}
&\mathcal{E}_{\kin} \sim \sum_{i=p,n}\left[\frac{\Delta_i}{\pi^2} \int_0^{k_{F_i}}k^3 dk 
+ \frac{C_ik_{F_i}^4}{\pi^2} \int_{k_{F_i}}^{\phi_ik_{F_i}}k^{-1}dk\right]
\nonumber\\
&=\sum_{i=p,n}\left[\frac{\Delta_i}{4\pi^2}k_{F_i}^4 + \frac{C_i}{\pi^2} k_{F_p}^4 \ln(\phi_i)\right]
\nonumber\\
&= \frac{3(3\pi^2)^{1/3}}{4}\rho^{4/3}[y^{4/3}\mathcal{A}_{\src,p}(y) + (1 - y)^{4/3}\mathcal{A}_{\src,n}(y)],
\label{eq:ekinh}
\end{align}
and
\begin{align}
&P_{\kin} \sim \frac{1}{3}\sum_{i=p,n}\left[\frac{\Delta_i}{\pi^2} \int_0^{k_{F_i}}k^3 dk 
+ \frac{C_ik_{F_i}^4}{\pi^2} \int_{k_{F_i}}^{\phi_ik_{F_i}}k^{-1}dk\right]
\nonumber\\
&= \frac{(3\pi^2)^{1/3}}{4}\rho^{4/3}[y^{4/3}\mathcal{A}_{\src,p}(y) + (1 - y)^{4/3}\mathcal{A}_{\src,n}(y)]
\nonumber\\
&= \frac{1}{3}\lim_{\rho\to\infty} \mathcal{E}_{\kin},
\label{eq:pkinh}
\end{align}
with
\begin{align}
\mathcal{A}_{\src,i}(y) = \Delta_i(y) + 4C_i(y)\ln[\phi_i(y)]
\end{align}
for $i=p,n$. For the case of the model without SRC included, this expression is reduced to $\mathcal{A}_{\src,n,p}=1$.

\subsection{Model up to second order in the vector field}
\label{sub1}

We start with the simplest case, in which the Lagrangian density contains terms only up to second order in the vector field, i.e. $C=0$. In this situation, we obtain $W=G_\omega^2\rho$ and $R=(2y-1)G_\rho^2\rho/2$. Hence, we have
\begin{align}
&\lim_{\rho \to \infty} \mathcal{E}_{\had} \sim \frac{1}{8}\left[4G_\omega^2 + (2y - 1)^2G_\rho^2 \right]\rho^2
\nonumber\\
&+\frac{3(3\pi^2)^{1/3}}{4}\rho^{4/3}[y^{4/3}\mathcal{A}_{\src,p}(y) + (1 - y)^{4/3}\mathcal{A}_{\src,n}(y)]
\label{eq:ehadh}
\end{align}
Note that, since the leading term in the energy density scales as $\rho^2$, the following relationship holds:
\begin{equation}
\rho^{4/3} \simeq
\frac{8^{2/3}}{\left[4G_\omega^2 + (2y-1)^2G_\rho^2\right]^{2/3}}\mathcal{E}_{\had}^{2/3}.
\label{eq:rhoh}
\end{equation}
For the pressure, one obtains
\begin{align}
&\lim_{\rho \to \infty} P_{\had} \sim \frac{1}{8}\left[4G_\omega^2 + (2y - 1)^2G_\rho^2 \right]\rho^2
\nonumber\\
&+\frac{(3\pi^2)^{1/3}}{4}\rho^{4/3}[y^{4/3}\mathcal{A}_{\src,p}(y) + (1 - y)^{4/3}\mathcal{A}_{\src,n}(y)]
\nonumber\\
&= \frac{3(3\pi^2)^{1/3}}{4}\rho^{4/3}[y^{4/3}\mathcal{A}_{\src,p}(y) + (1 - y)^{4/3}\mathcal{A}_{\src,n}(y)] 
\nonumber\\
&- \frac{(3\pi^2)^{1/3}}{2}\rho^{4/3}[y^{4/3}\mathcal{A}_{\src,p}(y) + (1 - y)^{4/3}\mathcal{A}_{\src,n}(y)]
\nonumber\\
&+ \frac{1}{8}\left[4G_\omega^2 + (2y - 1)^2G_\rho^2 \right]\rho^2,
\end{align}
which, using Eqs.~\eqref{eq:ehadh} and \eqref{eq:rhoh}, can be rewritten as
\begin{align}
P_{\had} &\sim \mathcal{E}_{\had} - \alpha'(y,\mathcal{A}_{\src,p}(y),\mathcal{A}_{\src,n}(y))\mathcal{E}_{\had}^{2/3} + \dots
\label{eq:pressaprox1}
\end{align}
with
\begin{align}
&\alpha'(y,\mathcal{A}_{\src,p}(y),\mathcal{A}_{\src,n}(y)) \equiv \alpha_{\src}(y) =
\nonumber\\
&=\frac{(72\pi^4)^{2/3}}{6\pi^2}\frac{[y^{4/3}\mathcal{A}_{\src,p}(y) + (1 - y)^{4/3}\mathcal{A}_{\src,n}(y)]}{[4G_\omega^2 + (2y-1)^2G_\rho^2]^{2/3}}.
\label{eq:alphap}
\end{align}

\begin{figure}[!htb]
\centering
\includegraphics[scale=0.5]{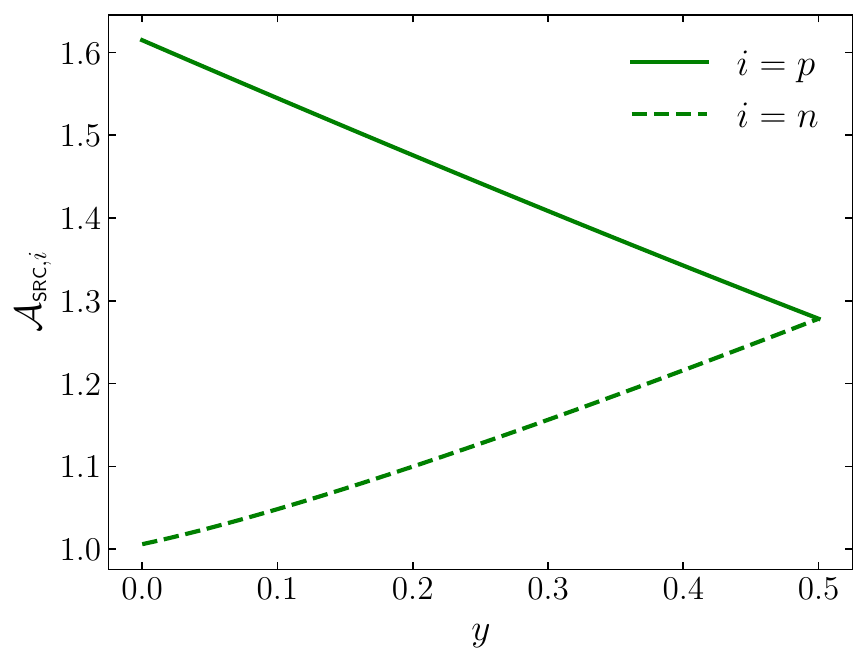}
\caption{SRC-induced enhancement factors \(A_{\text{SRC},i}\) for neutrons and protons as functions of the proton fraction $y$, illustrating the impact of short-range correlations on high-momentum tails.}
\label{fig:asrc}
\end{figure}

It is important to highlight the modification induced by SRC in the coefficient of the nonleading term in the high-density expansion of the pressure. If the \mbox{RMF-SRC} model adopts the same values of $G_\omega^2$ and $G_\rho^2$ as the original RMF model, then the impact of SRC can be assessed solely by analyzing the possible values of $\mathcal{A}_{\src,i}$, as shown in Fig.~\ref{fig:asrc}.

From the figure, we note that $\mathcal{A}_{\src,i}$ is greater than~$1$. Consequently, the coefficient $\alpha_{\src}(y)$ is always larger than its counterpart without SRC, namely, $\alpha(y) \equiv \alpha'(y, \mathcal{A}_{\src,p} = 1, \mathcal{A}_{\src,n} = 1)$. As a consequence, the inclusion of SRC softens the model at the high-density regime. However, this may no longer be true if one considers a refitting of the model’s free coupling constants when SRC are taken into account. In this scenario, both models, RMF and \mbox{RMF-SRC}, are constrained to reproduce the same set of nuclear empirical parameters (NEPs): saturation density ($\rho_{\sat}$), binding energy ($E_{\sat}$), effective mass ratio ($m^*_{\sat}= M^*_{\sat}/M{\mbox{\tiny nuc}}$), incompressibility ($K_{\sat}$), and symmetry energy ($E_{\sym,2}$), all quantities at $\rho=\rho_{\sat}$. As a result, the inclusion of SRC necessarily generates different values for $G^2_\sigma$, $G^2_\omega$, $G^2_\rho$, $a$, and~$b$, compared to the model without this phenomenology implemented in the system. In order to carry out our investigation within a realistic and well-established theoretical framework, we make use in Secs.~\ref{sub1} and~\ref{sub2}, of the NEPs related to the NL3* parametrization~\cite{nl3s}, systematically tested against experimental data for a large set of spherical nuclei, including $^{16}\rm O$, $^{34}\rm Si$, $^{40}\rm Ca$, $^{48}\rm Ca$, $^{52}\rm Ca$, $^{54}\rm Ca$, $^{48}\rm Ni$, $^{56}\rm Ni$, $^{78}\rm Ni$, $^{90}\rm Zr$, $^{100}\rm Sn$, $^{132}\rm Sn$, and $^{208}\rm Pb$. Furthermore, it extends its applicability beyond finite nuclei, providing a good description of some macroscopic properties of NSs, such as their mass-radius relationship (a detailed analysis involving more than $400$ RMF parametrizations is available in Ref.~\cite{brett-jerome}). 

\begin{figure}[!htb]
\centering
\includegraphics[scale=0.5]{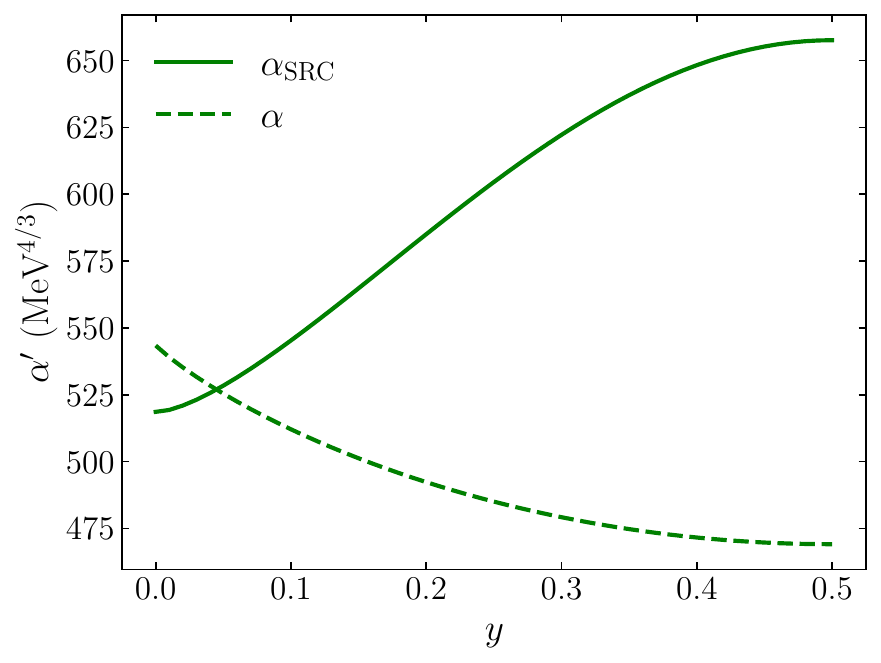}
\caption{Coefficient of the high-density pressure expansion as a function of proton fraction $y$, comparing RMF models with and without SRC for the NL3* parametrization.}
\label{fig:coef}
\end{figure}

In Fig.~\ref{fig:coef}, the coefficient~$\alpha$ with and without SRC included is displayed for the NL3* model.
As we can see from the figure, the SRC affect the coefficient in Eq.~\eqref{eq:pressaprox1} in the following way: $\alpha_{\src} < \alpha$ from $y = 0$ (pure neutron matter) up to $y \sim 0.04$. From this point on, an inversion in the behavior is observed. Hence, we identify a critical proton fraction above which the hardening of the EoS changes to a softening. In the case of stellar matter, the first region ($0 \le y \lesssim 0.04$) corresponds, in general, to densities below the saturation density. 

\begin{figure*}[!htb]
\centering
\includegraphics[scale=0.7]{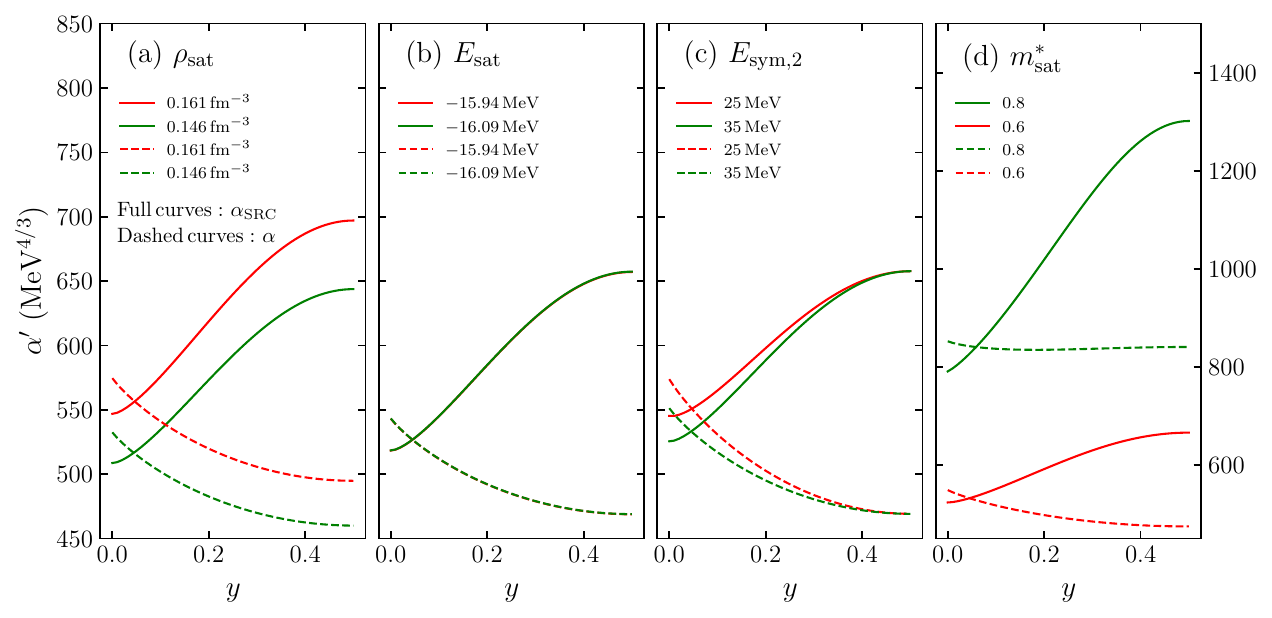}
\caption{Variation of the high-density pressure expansion coefficient with proton fraction $y$, shown for different nuclear empirical parameters: $\rho_{\text{sat}}$, $E_{\text{sat}}$, $E_{\text{sym},2}$, and $m^*_{\text{sat}}$.}
\label{fig:coefnep}
\end{figure*}

For the sake of completeness, we also investigate the effect of the NEPs on this analysis. The results are depicted in Fig.~\ref{fig:coefnep}.
In particular, we choose the variation of $\rho_{\text{sat}}$ and $E_{\text{sat}}$ to be consistent with the maximum and minimum values associated with the models in the $D_{4\sym}$ group described in Ref.~\cite{brett-jerome}, to which the NL3* model belongs. This group is a more selective subset of energy density functionals that satisfy stringent constraints from low-energy nuclear data as well as boundaries for the symmetry energy. For the other NEPs, we extend the ranges to $25~\mbox{MeV}\leq E_{\sym,2}\leq 35~\mbox{MeV}$ and $0.6\leq m^*_{\sat}\leq 0.8$, in order to encompass a wider range of possible values for these quantities. As shown in Fig.~\ref{fig:coef}, the qualitative behavior remains unchanged even for these broader NEP ranges, indicating that the predominant effect of the SRC in the RMF model with $C=0$ is to soften the corresponding EoS at the high-density regime.

\subsection{Fourth order in the vector field}
\label{sub2}

At this point we proceed by analyzing the situation in which $C\ne 0$. In this case $R$ remains the same, but the field equation for $W$ becomes $W\left(1 + CG_\omega^2 W^2\right)\sim CG_\omega^2 W^3 = G_\omega^2\rho$, leading to $W\sim C^{-1/3}\rho^{1/3}$ and 
\begin{widetext}
\begin{align}
\lim_{\rho \to \infty}&\mathcal{E}_{\had} \sim 
\frac{3(3\pi^2)^{1/3}}{4}[y^{4/3}\mathcal{A}_{\src,p}(y) + (1 - y)^{4/3}\mathcal{A}_{\src,n}(y)]\rho^{4/3}
+ \frac{\rho^{2/3}}{2G_\omega^2C^{2/3}} + \frac{3\rho^{4/3}}{4C^{1/3}}
+ \frac{1}{8}(2y - 1)^2G_\rho^2\rho^2 
\nonumber\\
&= \frac{3(3\pi^2)^{1/3}}{4}\left[y^{4/3}\mathcal{A}_{\src,p}(y)+(1-y)^{4/3}\mathcal{A}_{\src,n}(y)
+ \frac{1}{(3\pi^2C)^{1/3}}\right]\rho^{4/3}
+ \frac{\rho^{2/3}}{2G_\omega^2C^{2/3}}
+ \frac{1}{8}(2y - 1)^2G_\rho^2\rho^2.
\label{eq:ehadhc}
\end{align}
For the pressure, the high-density limit yields
\begin{align}
\lim_{\rho \to \infty}P_{\had} &\sim 
\frac{(3\pi^2)^{1/3}}{4}[y^{4/3}\mathcal{A}_{\src,p}(y) + (1 - y)^{4/3}\mathcal{A}_{\src,n}(y)]\rho^{4/3}
+ \frac{\rho^{2/3}}{2G_\omega^2C^{2/3}} + \frac{\rho^{4/3}}{4C^{1/3}}
+ \frac{1}{8}(2y - 1)^2G_\rho^2\rho^2 
\nonumber\\
&= \frac{(3\pi^2)^{1/3}}{4}\left[y^{4/3}\mathcal{A}_{\src,p}(y)+(1-y)^{4/3}\mathcal{A}_{\src,n}(y)
+ \frac{1}{(3\pi^2C)^{1/3}}\right]\rho^{4/3}
+ \frac{\rho^{2/3}}{2G_\omega^2C^{2/3}}
+ \frac{1}{8}(2y - 1)^2G_\rho^2\rho^2
\nonumber\\
&= \mathcal{E}_{\had} - \frac{(3\pi^2)^{1/3}}{2}\left[y^{4/3}\mathcal{A}_{\src,p}(y)+(1-y)^{4/3}\mathcal{A}_{\src,n}(y) + \frac{1}{(3\pi^2C)^{1/3}}\right]\rho^{4/3}.
\label{eq:presshadhc1}
\end{align}
From Eq.~\eqref{eq:ehadhc}, we find $\rho^{4/3}=\{8\mathcal{E}_{\had}/[(2y-1)^2G_\rho^2]\}^{2/3}$. This expression is used in Eq.~\eqref{eq:presshadhc1} to finally obtain
\begin{align}
P_{\had} &\sim \mathcal{E}_{\had} - \beta'(y,\mathcal{A}_{\src,p}(y),\mathcal{A}_{\src,n}(y))\mathcal{E}_{\had}^{2/3} + \dots
\label{eq:pressaprox2}
\end{align}
with
\begin{align}
\beta'(y,\mathcal{A}_{\src,p}&(y),\mathcal{A}_{\src,n}(y)) \equiv \beta_{\src}(y) =
\nonumber\\
&=\frac{1}{6\pi^2}\left[\frac{72\pi^4}{(2y-1)^2G_\rho^2}\right]^{2/3}\left[y^{4/3}\mathcal{A}_{\src,p}(y)+(1-y)^{4/3}\mathcal{A}_{\src,n}(y) + \frac{1}{(3\pi^2C)^{1/3}}\right].
\label{eq:betap}
\end{align}
\end{widetext}
As in the previous case ($C=0$), the corresponding coefficient for a system without SRC is obtained by setting $\beta(y) \equiv \beta'(y, \mathcal{A}_{\src,p} = 1, \mathcal{A}_{\src,n} = 1)$.

A direct inspection of Eq.~\eqref{eq:betap} reveals that the inclusion of SRC leads to $\beta_{\src} > \beta$, provided the strength of the $\rho$-meson field is kept fixed when comparing the RMF and \mbox{RMF-SRC} models. This result arises from the fact that $\beta'$ does not depend on $G_\omega^2$, as in the $C = 0$ case, and we assume, in principle, that the constant $C$ is the same for models with and without SRC. The increase in the coefficient reflects a stronger suppression of the pressure at high densities, resulting in a softer EoS. Thus, SRC introduces a direct softening effect on the high-density behavior of the hadronic EoS. However, this conclusion does not hold when the refitting procedure is taken into account. Notice that the expression for the symmetry energy at $\rho=\rho_\sat$ for the \mbox{RMF-SRC} model reads~(quadratic contribution)
\begin{align}
E_{\sym,2}&\equiv \mathcal{S}_2(\rho_\sat,1/2) = \frac{1}{8}\frac{\partial^2(\mathcal{E}_{\had}/\rho)}{\partial y^2}\Bigg|_{\rho=\rho_\sat,y=1/2} 
\nonumber\\
&= E^{\src}_{\kin,2}(\rho_\sat,M^*_\sat) +  \frac{G_\rho^2}{8}\rho_\sat,
\label{eq:esym2}
\end{align}
with the kinetic term given by~\cite{cai}
\begin{widetext}
\begin{align}
E^{\src}_{\kin,2}(&\rho_\sat,M^*_\sat) = \frac{{k_F^\sat}^2}{6 E_F^*} \left[ 1 - 3 C_0 \left( 1 - \frac{1}{\phi_0} \right) \right] 
- 3 E_F^* C_0 \left[ C_1 \left( 1 - \frac{1}{\phi_0} \right) + \frac{\phi_1}{\phi_0} \right] \nonumber \\
&- \frac{9 M_\sat^{*4}}{8 {k_F^\sat}^3} \frac{C_0 \phi_1 (C_1 - \phi_1)}{\phi_0} 
\left[ \frac{2 {k_F^\sat}}{M_\sat^*} \left( \left( \frac{{k_F^\sat}}{M_\sat^*} \right)^2 + 1 \right)^{3/2} 
- \frac{{k_F^\sat}}{M_\sat^*} \left( \left( \frac{{k_F^\sat}}{M_\sat^*} \right)^2 + 1 \right)^{1/2} 
- \operatorname{arcsinh} \left( \frac{{k_F^\sat}}{M_\sat^*} \right) \right] \nonumber \\
&+ \frac{2 {k_F^\sat} C_0 (6 C_1 + 1)}{3} \left[ \operatorname{arcsinh} \left( \frac{\phi_0 {k_F^\sat}}{M_\sat^*} \right) 
- \sqrt{1 + \left( \frac{M_\sat^*}{\phi_0 {k_F^\sat}} \right)^2} 
- \operatorname{arcsinh} \left( \frac{{k_F^\sat}}{M_\sat^*} \right) 
+ \sqrt{1 + \left( \frac{M_\sat^*}{{k_F^\sat}} \right)^2} \right] \nonumber \\
&+ \frac{3 {k_F^\sat} C_0}{2} \left[ \frac{(1 + 3 \phi_1)^2}{9} 
\left( \frac{\phi_0 {k_F^\sat}}{F_F^*} - \frac{2 F_F^* (3 \phi_1 - 1)}{9 \phi_0 {k_F^\sat}} \right) 
- \frac{1}{9} \frac{{k_F^\sat}}{E_F^*} + \frac{4 E_F^*}{9 {k_F^\sat}} \right] \nonumber \\
&+ \frac{C_0 (4 + 3 C_1)}{3} \left[ \frac{F_F^* (1 + 3 \phi_1)}{\phi_0} - E_F^* \right],
\label{eq:esym2kinsrc}
\end{align}
\end{widetext}
where $E_F^* = (M^{*2}_\sat + k_F^{\sat\,2})^{1/2}$, $F_F^* = [M^{*2}_\sat + (\phi_0 {k_F^\sat})^2]^{1/2}$, and $k_F^\sat=(3\pi^2\rho_\sat/2)^{1/3}$. Then, it is clear that the constant $G_\rho^2$ is determined solely by $M^*_\sat$, $E_{\sym,2}$, and $\rho_\sat$, according to Eq.~\eqref{eq:esym2}, namely, $G_\rho^2 = 8[E_{\sym,2} - E^{\src}_{\kin,2}(\rho_\sat, M^*_\sat)]/\rho_\sat$. This formulation highlights the key role of the kinetic contribution to the symmetry energy in determining how SRC affect the hadronic EoS. In particular, if the relation $E^{\src}_{\kin,2} < E_{\kin,2}$ holds, then the value of $G_\rho^2$ increases in the \mbox{RMF-SRC} model. This outcome arises because, during the refitting process, both $E_{\sym,2}$ and $\rho_\sat$ are kept fixed, regardless of whether SRC effects are included. In Ref.~\cite{cai}, the authors show that this condition is satisfied across a broad range of $M^*_\sat$. The same behavior is observed for the range of $\rho_\sat$ and $m^*_\sat$ used in Fig.~\ref{fig:coefnep}, as shown in Table~\ref{tab:ekin2}.

\begin{table}[!htb]
\centering
\caption{Kinetic contribution to the symmetry energy $E_{\text{kin},2}$ with and without SRC, for different values of saturation density $\rho_{\text{sat}}$ and effective mass $m^*_{\text{sat}}$.}
\begin{tabular*}{\linewidth}{@{\extracolsep{\fill}}lccc}
\toprule
\midrule 
$\rho_\sat$ (fm$^{-3}$) & $m^*_\sat$ &  $E_{\kin,2}$ (MeV) & $E^{\src}_{\kin,2}$ (MeV) \\
\midrule
$0.146$  & $0.6$ & $17.54$ & $-16.03$ \\
$0.146$  & $0.8$ & $13.68$ & $-13.45$ \\
$0.161$  & $0.6$ & $18.62$ & $-16.85$ \\
$0.161$  & $0.8$ & $14.55$ & $-14.20$ \\
\bottomrule
\end{tabular*}
\label{tab:ekin2}
\end{table}

\begin{figure*}[!htb]
\centering
\includegraphics[scale=0.66]{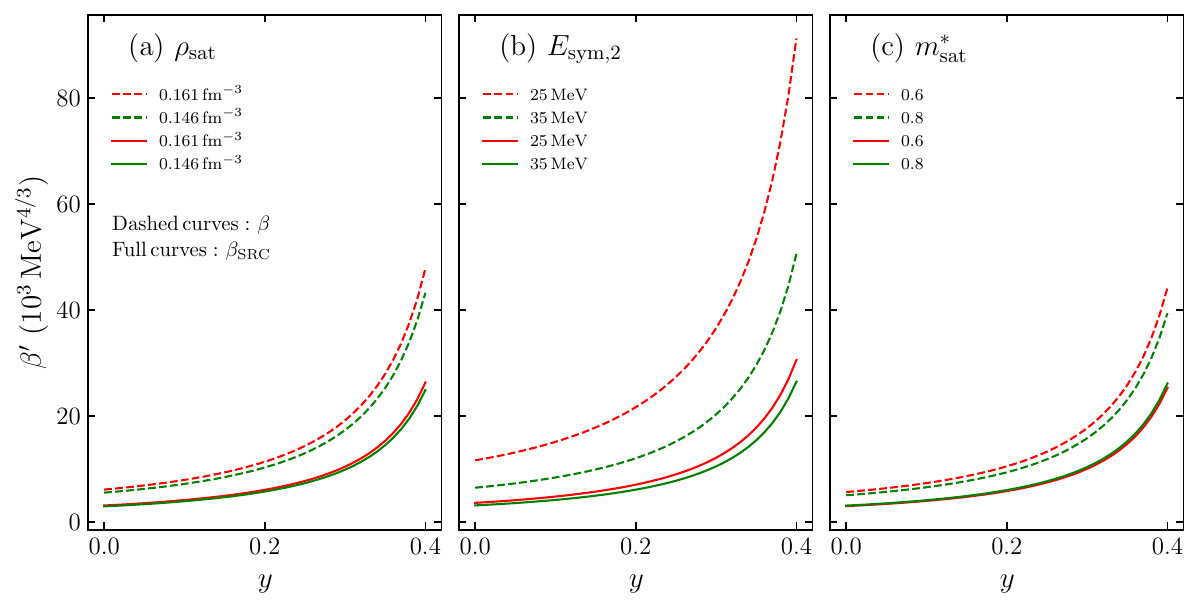}
\caption{High-density pressure expansion coefficient $B_{\text{SRC}}(y)$ from Eq.~\eqref{eq:pressaprox2}, plotted against proton fraction $y$ for $C = 0.01$ and varying nuclear empirical parameters.}
\label{fig:coefcnep}
\end{figure*}

As shown in Ref.~\cite{cai}, we also find negative values for the kinetic term of the symmetry energy in the \mbox{RMF-SRC} model. As a result, $G_\rho^2$ is consistently larger than its corresponding value in the RMF model. This increase may outweigh the rise in the numerator of $\beta_{\src}$, leading to an effective reduction in the coefficient of Eq.~\eqref{eq:pressaprox2}. This behavior is indeed observed, as the reader can verify in Fig.~\ref{fig:coefcnep}.

Therefore, it is clear that the hadronic EoS becomes stiffer when SRC effects are incorporated into the model with $C \neq 0$. Moreover, this behavior is observed across all proton fractions analyzed; that is, no inversion occurs as seen in $\alpha'$, see Fig.~\ref{fig:coef}.

\subsection{Impact for neutron stars}

\begin{figure}[!htb]
\centering
\includegraphics[scale=0.5]{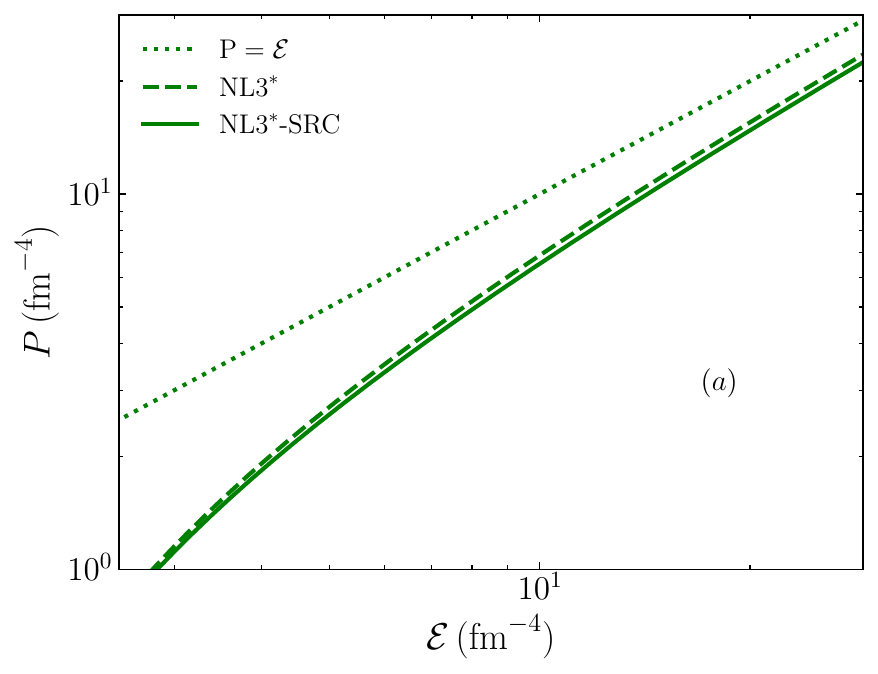}
\includegraphics[scale=0.5]{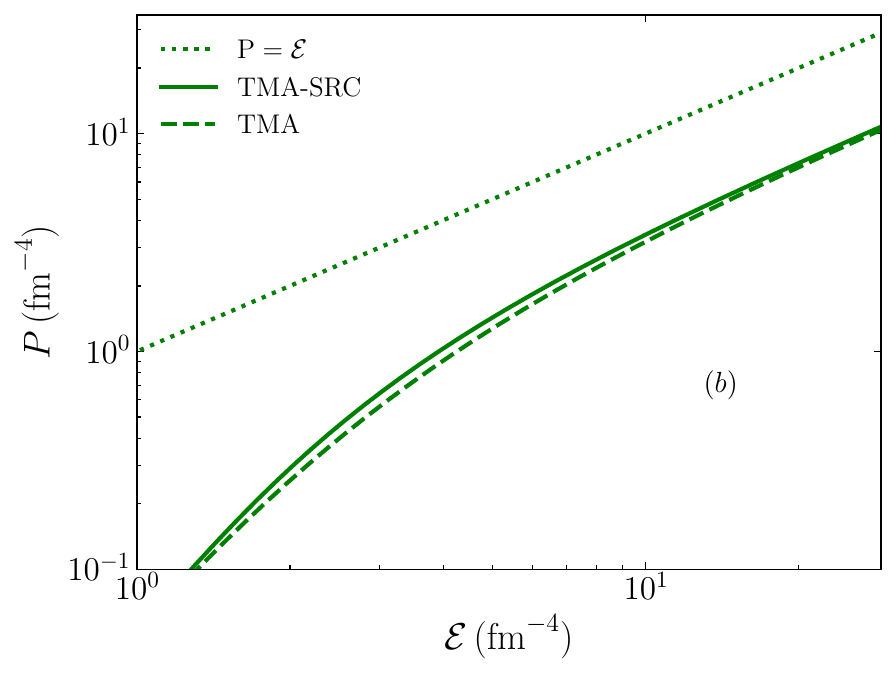}
\caption{Pressure–energy density relations for RMF models with and without SRC in the stellar matter regime. Panel (a): models with $C = 0$; panel (b): models with $C \ne 0$.}
\label{fig:eos}
\end{figure}

We now analyze how the results obtained in the previous sections affect the description of NSs. For this purpose, we consider stellar matter in beta equilibrium by imposing charge neutrality, $\rho_p - \rho_e = \rho_\mu$, and chemical equilibrium, $\mu_n - \mu_p = \mu_e$ with $\mu_\mu = \mu_e$, in a system composed of protons, neutrons, electrons, and muons. The total energy density and pressure are given by $\mathcal{E} = \mathcal{E}_{\text{had}} + \mathcal{E}_e + \mathcal{E}_\mu$ and $P = P_{\text{had}} + P_e + P_\mu$, respectively, where $\mathcal{E}_i$ and $P_i$ ($i = e, \mu$) denote the kinetic contributions from electrons (assumed massless) and muons with mass $m_\mu = 105.7~\text{MeV}$. For simplicity, our analysis focuses exclusively on the NS core. In particular, Fig.~\ref{fig:eos}{\color{blue}a} shows the EoS corresponding to the NL3*/\mbox{NL3*-SRC} models with $C = 0$.

\begin{figure}[!htb]
\centering
\includegraphics[scale=0.5]{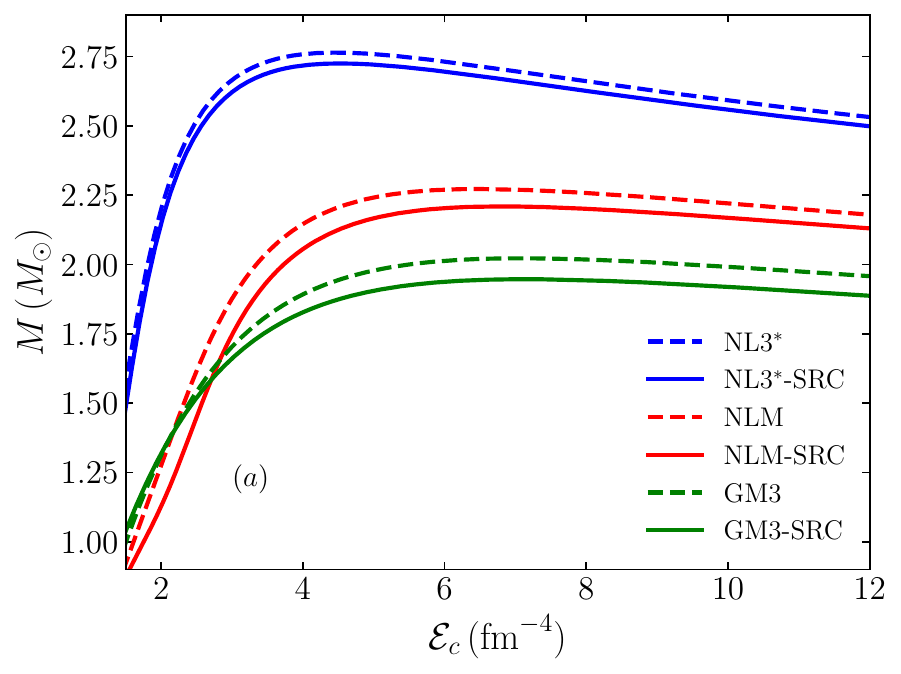}
\includegraphics[scale=0.5]{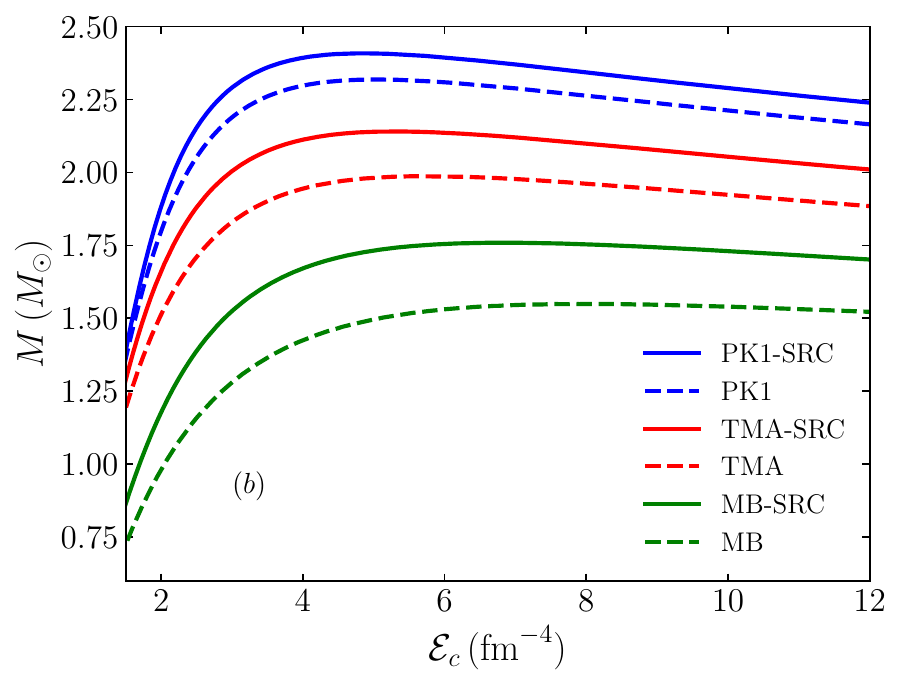}
\caption{Neutron star mass versus central energy density for RMF models with and without SRC. Panel (a): models with~$C=0$~(NL3*~\cite{nl3s}, NLM~\cite{nlm}, GM3~\cite{gm3}); panel (b): models with~$C\ne 0$ (PK1~\cite{pk1}, TMA~\cite{tma}, MB~\cite{mb}).}
\label{fig:star}
\end{figure}

Note that the SRC-induced behavior observed in the asymptotic regime also emerges within the energy density range relevant to NS modeling. Consequently, the softening of the EoS caused by SRC is evident in the mass-radius diagrams of high-mass stars, which correspond to higher central densities—or, equivalently, higher central energy densities—resulting in less massive NSs. This trend is confirmed in Fig.~\ref{fig:star}{\color{blue}a}. 
In this figure, the NS mass, calculated from the solution of the TOV equations~\cite{tov39,tov39a}, is plotted as a function of the central energy density for different parameterizations of the model with and without SRC included, all of which incorporate self-interactions of the vector field up to second order ($C = 0$). For all models that include SRC, less massive stars are observed, at least in the range $\mathcal{E}_c \gtrsim 2$~fm$^{-4}$. 

In contrast, Fig.~\ref{fig:eos}{\color{blue}b} displays the EoS for models with $C \ne 0$, where the self-interaction of the vector meson field includes a fourth-order term. Once again, the influence of SRC at $\rho \to \infty$ extends across the density range relevant to NS modeling, despite a slower approach to the asymptotic regime compared to the $C = 0$ case, due to the relation $\beta' > \alpha'$. These findings indicate that incorporating SRC into the RMF model enhances its ability to support more massive NSs. 

This behavior is particularly relevant in light of recent astrophysical observations reporting compact stars with masses around or above two solar masses, which impose stringent constraints on viable nuclear models. To illustrate this case, Fig.~\ref{fig:star}{\color{blue}b} shows parameterizations where the fourth-order term in $W$ is present in the vector channel ($C \ne 0$). A clear increase in the stellar mass is observed across the entire range of central energy densities analyzed. Notably, this increase is more pronounced than the decrease seen in the $C = 0$ scenario. In summary, the influence of SRC on NS properties is highly sensitive to the structure of the vector self-interaction in the model. While SRC tend to soften the EoS when $C = 0$, they can induce significant stiffening for $C \ne 0$, thereby enabling consistency with observations of massive NSs. For the sake of completeness, in Table~\ref{tab:mmax} we show the maximum mass values obtained for each parametrization shown in Fig.~\ref{fig:star}, in order to quantify the effects induced by SRC in each type of model.
\begin{table}[!htb]
\centering
\caption{Maximum stellar mass, in units of $M_\odot$, for the mass-radius diagrams presented in Fig.~\ref{fig:star}.}
\begin{tabular*}{\linewidth}{@{\extracolsep{\fill}}lcc}
\toprule
\midrule 
models & $M_\ma$  &  $M^{\src}_\ma$  \\
\midrule
$C=0$\\
NL3*/NL3*-SRC  & $2.76$ & $2.73$  \\
NLM/NLM-SRC    & $2.27$ & $2.21$ \\
GM3/GM3-SRC    & $2.02$ & $1.95$ \\
$C\ne0$\\
PK1/PK1-SRC    & $2.32$ & $2.41$ \\
TMA/TMA-SRC    & $1.99$ & $2.14$ \\
MB/MB-SRC      & $1.55$ & $1.76$ \\
\bottomrule
\end{tabular*}
\label{tab:mmax}
\end{table}

It should be noted that Fig.~\ref{fig:star} includes dynamically unstable stars ($\partial M / \partial \mathcal{E}_c < 0$) to emphasize that the asymptotic high-density behavior persists across the density range relevant for NSs. Omitting these segments would obscure this continuity and potentially mislead interpretation, so we retain them to clearly illustrate the impact of SRC throughout the entire domain.

\section{Consequences in neutron stars with DM content}
\label{sec:rmf-src-dm}

Within the mean-field framework, strongly interacting matter containing a DM component can be modeled through distinct approaches. A common strategy involves Higgs-portal scenarios~\cite{higgsportal1,higgsportal2}, in which the dark sector interacts with standard matter via Higgs boson exchange~\cite{rmfdm13,rmfdm2,rmfdm3,abdul,rmfdm6,rmfdm11,rmfdm10,rmfdm8,rmfdm7,rmfdm12,nitr}. In this case, both sectors are described by a unified Lagrangian density. However, due to the smallness of the Higgs field values as a function of density in the mean-field approximation, the system behaves effectively as two independent subsystems with no practical interaction between them.

Experimental findings from direct detection searches~\cite{LUX:2017ree,XENON:2018voc,PandaX-II:2017hlx,Billard:2021uyg,Schumann:2019eaa} support this picture, as they provide only upper bounds on the elastic scattering cross-section. In view of the apparent absence of non-gravitational couplings—or, at most, the presence of a very weak interaction—it is physically well motivated to describe the two sectors inside compact stars as two independent fluids interacting exclusively through gravity. Such a two-fluid formalism has been extensively studied; see, e.g., Refs.~\cite{qian14,Collier:2022cpr,Shakeri:2022dwg,Miao:2022rqj,Emma:2022xjs,Hong:2024sey,Karkevandi:2021ygv,Liu:2023ecz,Ivanytskyi:2019wxd,Buras-Stubbs:2024don,Rutherford:2022xeb,Thakur:2023aqm,Mahapatra:2024ywx,Thakur:2024mxs,carline25,liu25}. In this work, we adopt a specific implementation in which the total pressure is given by \( P(r) = P_{\vis}(r) + P_{\dm}(r) \), where \( r \) denotes the radial coordinate from the center of the star. The subscript ``vis'' refers to the visible sector of the system, specifically the hadronic component.

The generalized and dimensionless Tolman–Oppenheimer–Volkoff (TOV) equations for this formalism, expressed in natural units (\( \hbar = G = c = 1 \)), are given by~\cite{laura24,jurgen06}:


\begin{align}
\frac{dP'_{\vis}(r')}{dr'} &= \frac{\left[\mathcal{E}'_{\vis}(r') + P'_{\vis}(r')\right]\left[m'(r') + 4\pi r'^3 P'(r')\right]}{r'\left[2m'(r')-r'\right]},
\label{eq:pvisdimless}
\\
\frac{dP'_{\dm}(r')}{dr'} &= \frac{\left[\mathcal{E}'_{\dm}(r') + P'_{\dm}(r')\right]\left[m'(r') + 4\pi r'^3 P'(r')\right]}{r'\left[2m'(r')-r'\right]},
\label{eq:pdmdimless}
\\
\dfrac{dm'_{\vis}(r')}{dr'} &= 4\pi r'^2\mathcal{E}'_{\vis}(r'),
\quad
\dfrac{dm'_{\dm}(r')}{dr'} = 4\pi r'^2\mathcal{E}'_{\dm}(r'),
\label{eq:massdimless}
\end{align}
where \( m'(r') = m'_{\vis}(r') + m'_{\dm}(r') \) is the total dimensionless mass enclosed within the dimensionless radius~\( r' \). The quantities \( m'_{\vis} \) and \( m'_{\dm} \) represent the visible and DM contributions, respectively. Further details on the derivation of these coupled equations can be found in Ref.~\cite{qian14}. All primed variables are dimensionless, which facilitates numerical integration of the TOV equations. Physical units are recovered using the following scaling relations: $P_{\vis,\dm}=M^4_{\resc}P'_{\vis,\dm}$, $\mathcal{E}_{\vis,\dm}=M^4_{\resc}\mathcal{E}'_{\vis,\dm}$, $m_{\vis,\dm}=(M^3_p/M^2_{\resc})m'_{\vis,\dm}$, and $r=(M_p/M^2_{\resc})r'$, where \( M_{\resc} \) is a common rescaling mass and \( M_p = 1.22 \times 10^{19} \)~GeV is the Planck mass~\cite{pdg24}. Note that \( m_{\vis,\dm} \) and \( r \) must still be converted to solar masses (\( M_\odot \)) and kilometers (km), respectively.

For the dark sector, we adopt a system composed of fermionic particles. It is modeled as a relativistic Fermi gas of non-interacting particles with mass $m_\chi$, since dark fermions obey the Pauli exclusion principle and, as a result, a degeneracy pressure naturally arises, enabling the system to counterbalance gravitational collapse. However, as in the visible sector, we also include self-repulsion mediated by a vector field, which allows for the formation of more massive compact stars. The proposed Lagrangian density is given by~\cite{rmfdm1,qian14,Thakur:2023aqm,milva24}
\begin{align}
\mathcal{L}_{\dm} &= \bar{\chi}\left[ \gamma_{\mu} (i\partial^{\mu} - g_VV^{\mu}) - m_{\chi} \right]\chi + \frac{1}{2}m^{2}_VV_{\mu}V^{\mu}
\nonumber\\
&- \frac{1}{4}(\partial_\mu V_\nu - \partial_\nu V_\mu)(\partial^\mu V^\nu - \partial^\nu V^\mu),
\label{eq:lagdendm}
\end{align}
where the dark vector field and its associated mass are denoted by $V_\mu$ and $m_V$, respectively. The spinor representing the dark fermion is denoted by $\chi$.

\begin{figure}[!htb]
\centering
\includegraphics[scale=0.41]{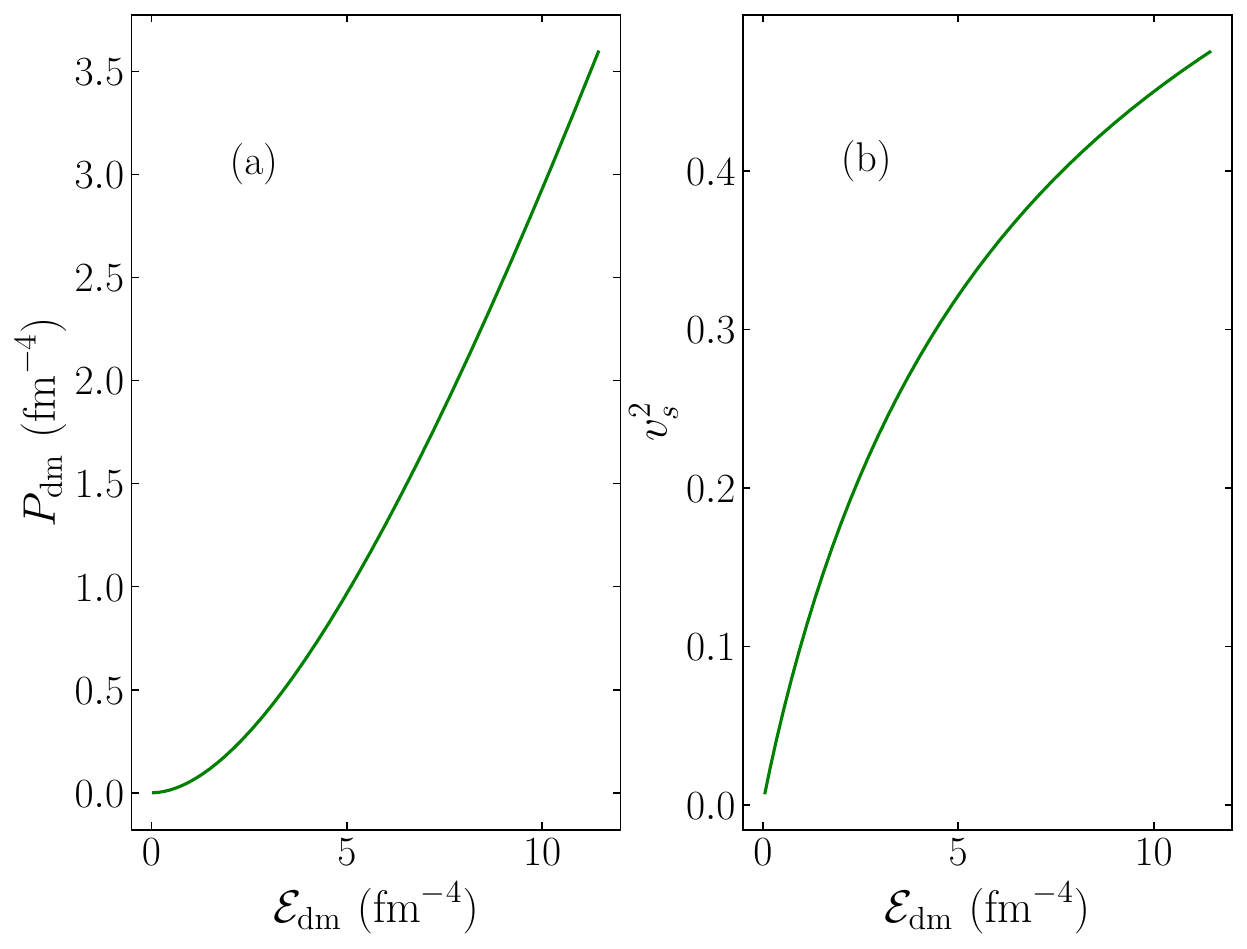}
\caption{(a) Pressure of fermionic DM and (b) squared sound speed, shown as functions of the dark energy density.}
\label{fig:eosvs2fdm}
\end{figure}

Analogously to calculations performed in relativistic hadronic models—where the Lagrangian density shares the same mathematical structure—we apply the mean-field approximation to evaluate the pressure and energy density associated with the dark sector. The resulting expressions are
\begin{align}
P_{\dm} &= \frac{1}{3\pi^2}\int_0^{k_{F\chi}}\hspace{-0.2cm}dk\,\frac{k^4}{(k^2 + m_\chi^2)^{1/2}}
+\frac{1}{2}C_V^2\rho_\chi^2,
\label{eq:pfer}
\end{align}
and
\begin{align}
\mathcal{E}_{\dm} &= \frac{1}{\pi^2}\int_0^{k_{F\chi}}\hspace{-0.2cm}dk\,k^2(k^2 + m_\chi^2)^{1/2}
+\frac{1}{2}C_V^2\rho_\chi^2,
\label{eq:defer}
\end{align}
where $C_V=g_V/m_V$, and $\rho_\chi=k_{F\chi}^3/(3\pi^2)$. Here, $k_{F\chi}$ corresponds to the Fermi momentum of the dark particle. From this point onward, we adopt the values $C_V = 3.26$~fm and $m_\chi = 1.9$~GeV for the free parameters of the model~\cite{Thakur:2023aqm}. These values are consistent with the ranges $0.5~\mbox{GeV} \leqslant m_\chi \leqslant 4.5~\mbox{GeV}$~\cite{Thakur:2023aqm,calmet21} and $0.1~\mbox{fm} \leqslant C_V \leqslant 5~\mbox{fm}$~\cite{Thakur:2023aqm,qian14}. In Fig.~\ref{fig:eosvs2fdm}, we present the relation between $P_{\dm}$ and $\mathcal{E}_{\dm}$, along with the corresponding squared sound speed, defined as $v^2_s = \partial P_{\dm} / \partial \mathcal{E}_{\dm}$. The figure shows that the DM model remains fully causal within the range of energy densities considered.

In the hadronic sector, we employ a reference parametrization in which the NEPs are identical to those of the NL3* model, as discussed in Secs.~\ref{sub1} and~\ref{sub2}. For the NL3* extension that includes a fourth-order term in the vector field, we adopt the choice $C = 0.01$. Accordingly, the EoSs for the ``visible'' sector, under charge neutrality and chemical equilibrium, are given by $\mathcal{E}_\vis = \mathcal{E}_{\had} + \mathcal{E}_{\cru} + \mathcal{E}_e + \mathcal{E}_\mu$ and $P_\vis = P_{\had} + P_{\cru} + P_e + P_\mu$. The contribution of the stellar crust, represented by $\mathcal{E}_{\cru}$ and $P_{\cru}$, is divided into outer and inner regions. The outer crust is described by the Baym-Pethick-Sutherland (BPS) EoS in the density range $0.6295 \times 10^{-11}\,\mathrm{fm}^{-3} \leq \rho \leq 0.199 \times 10^{-3}\,\mathrm{fm}^{-3}$ (Table~5 of~\cite{bps}). The inner crust is modeled using the Skyrme nuclear interaction SLy4~\cite{sly4} within a compressible liquid-drop approach, covering the density range $0.20905 \times 10^{-3}\,\mathrm{fm}^{-3} \leq \rho \leq 0.73174 \times 10^{-1}\,\mathrm{fm}^{-3}$ (Table~3 of~\cite{douchin}). The EoS of the core, described by the hadronic model, is considered starting from $\rho \simeq 0.5\rho_\sat$.

The numerical solution of Eqs.~\eqref{eq:pvisdimless}–\eqref{eq:massdimless} is carried out by prescribing the central conditions $m'_{\vis}(0) = m'_{\dm}(0) = 0$, $P'_{\vis}(0) = P'^c_{\vis}$, and $P'_{\dm}(0) = P'^c_{\dm}$. The dimensionless central pressures $P'^c_{\vis}$ and $P'^c_{\dm}$ are determined by the respective equations of state (EoSs) of the visible and dark sectors. The system is then integrated outward using a fourth-order Runge–Kutta algorithm, yielding the radial profiles of dimensionless pressure and enclosed mass. The dimensionless stellar radii of the two components, $R'_{\vis}$ and $R'_{\dm}$, are defined by the vanishing of their respective pressures, i.e., $P'_{\vis}(R'_{\vis}) = 0$ and $P'_{\dm}(R'_{\dm}) = 0$, within a prescribed numerical tolerance. The corresponding dimensionless masses are given by $M'_{\vis} = m'_{\vis}(R'_{\vis})$ and $M'_{\dm} = m'_{\dm}(R'_{\dm})$, such that the total dimensionless mass of the configuration is $M' = M'_{\vis} + M'_{\dm}$. The overall dimensionless radius of the star is taken as the larger of the two, i.e., $R' = \max(R'_{\vis}, R'_{\dm})$. Configurations for which $R' = R'_{\dm} > R'_{\vis}$ are interpreted as possessing a dark-matter halo. By repeating the integration for all central pressure values consistent with the input EoSs, the mass–radius relation is obtained. Final results are expressed in physical units, with stellar mass $M$ in units of $M_\odot$ and radius $R$ in kilometers.

In this approach, each stellar configuration is assumed to contain a prescribed fraction of DM, expressed by the fixed ratio $F_{\dm} = M_{\dm} / M$. To implement this constraint, the required input for the TOV system must be specified - namely, the central pressures and energy densities of both components: $P^c_{\vis}$, $\mathcal{E}^c_{\vis}$, $P^c_{\dm}$, and $\mathcal{E}^c_{\dm}$. 
For the dark sector, we establish its connection with the hadronic part by assuming a proportionality between the two energy densities, $\mathcal{E}_{\dm} = f\,\mathcal{E}_{\vis}$, with $f$ taken as a constant parameter~\cite{qian14}. The two-fluid TOV equations are then solved over a broad range of values of $f$. For each integration associated with a specific $f$, the corresponding DM fraction $F_{\dm}$ is calculated, and only the $(M, R)$ pairs fulfilling the imposed value of $F_{\dm}$ are retained. In this way, the resulting set of solutions defines the mass–radius relation of NSs admixed with a fixed dark-matter content.

Within the two-fluid framework, stability requires that, at the onset of instability, the particle numbers of the visible ($N_{\text{vis}}$) and dark ($N_{\text{dm}}$) sectors remain stationary under variations in the central energy densities, $\mathcal{E}^c_{\text{vis}}$ and $\mathcal{E}^c_{\text{dm}}$. The quantities $N_{\vis}$ and $N_{\dm}$ are obtained from the relations
\begin{align}
\frac{dN'_{\vis}}{dr'} = \frac{4\pi r'^2\rho'_{\vis}}{\sqrt{1 - 2m'(r')/r'}}, \
\frac{dN'_{\dm}}{dr'} = \frac{4\pi r'^2\rho'_{\dm}}{\sqrt{1 - 2m'(r')/r'}},
\end{align}
where $N'_{\vis,\dm} = (M_\resc^3 / M_p^3) N_{\vis,\dm}$ and $\rho'_{\vis,\dm} = \rho_{\vis,\dm} / M_\resc^3$. Both equations must be solved consistently with the respective two-fluid TOV system. As discussed in~\cite{mauricio23,laura24,pitz25,carline25}, the stability criterion is formulated in terms of the following matrix equation:
\begin{align}
&
\begin{bmatrix}
\delta N_{\vis} \\ 
\delta N_{\dm}
\end{bmatrix}
=
\nonumber\\
&=
\begin{bmatrix}
(\partial N_{\vis}/\partial \mathcal{E}^c_{\vis})_{\mathcal{E}^c_{\dm}} & 
(\partial N_{\vis}/\partial \mathcal{E}^c_{\dm})_{\mathcal{E}^c_{\vis}} \\ 
(\partial N_{\dm}/\partial \mathcal{E}^c_{\vis})_{\mathcal{E}^c_{\dm}} & 
(\partial N_{\dm}/\partial \mathcal{E}^c_{\dm})_{\mathcal{E}^c_{\vis}}
\end{bmatrix}
\begin{bmatrix}
\delta \mathcal{E}^c_{\vis} \\ 
\delta \mathcal{E}^c_{\dm}
\end{bmatrix}
= 0.
\end{align}
It should be noted that, when evaluating the derivatives of $N_{\vis,\dm}$ with respect to $\mathcal{E}^c$, the central energy density of the other sector is held fixed. The eigenvalues of this system, denoted here by $\kappa_1$ and $\kappa_2$, define the stability condition, which requires $\kappa_1 > 0$ and $\kappa_2 > 0$ for equilibrium configurations to be considered stable.

\begin{figure}[!htb]
\centering
\includegraphics[scale=0.57]{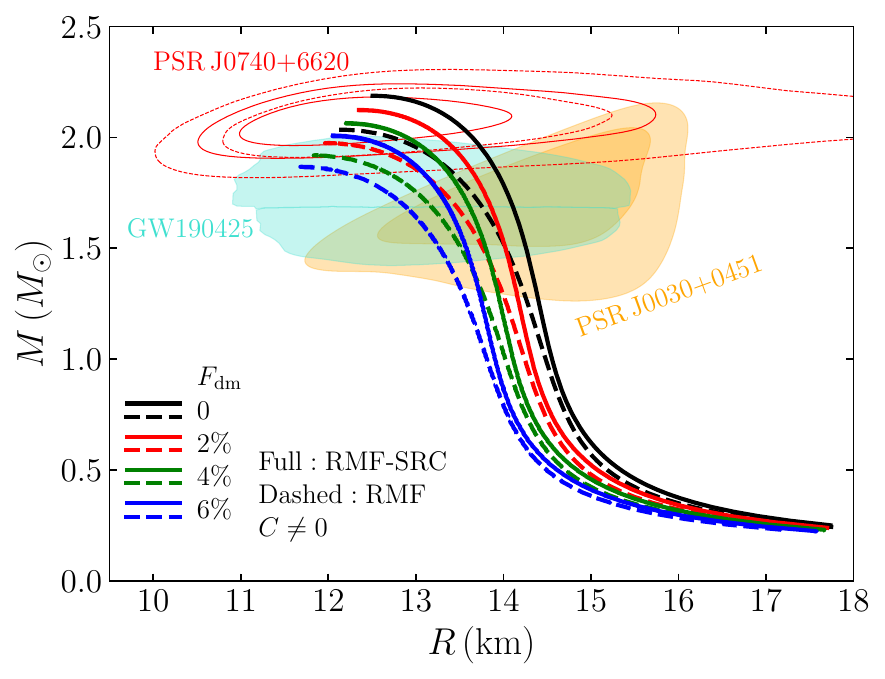}
\caption{Mass-radius diagrams for different fractions of DM in each star ($F_\dm$). Results for the hadronic model in which $C=0.005$, with~(\mbox{RMF-SRC}: full curves) and without~(RMF: dashed curves) short-range correlations. It is also plotted jointly NICER$-$XMM-Newton data concerning the pulsars PSR J0030+0451 (orange contours)~\cite{Vinciguerra_2024}, and PSR J0740+6620 (solid~\cite{Salmi_2024} and dashed~\cite{Dittmann_2024} red contours). We also compared our results with data from the GW190425 event (turquoise contours)~\cite{Abbott_2020-2}.}
\label{fig:mr-DM_C}
\end{figure}

In Fig.~\ref{fig:mr-DM_C}, we present the mass–radius diagrams obtained using the hadronic model with a fourth-order vector field term ($C \ne 0$), both with and without SRC, for different DM contents, defined here by the value of $F_\dm$. As in Secs.~\ref{sub1} and~\ref{sub2}, we impose that the NEPs of both the RMF and \mbox{RMF-SRC} parametrizations are identical to those of the NL3* model, except for the symmetry energy slope at saturation density, given by $L_{\sym,2} = 3\rho_\sat(\partial \mathcal{S}_2 / \partial \rho)_{\\rho = \rho_\sat}$.

As evident from the results, the general effect previously observed in ordinary NSs without DM also appears in admixed DM stars — namely, an increase in stellar mass due to the inclusion of SRC. This increase is observed across all values of $F_\dm$ analyzed, as shown in Table~\ref{tab:mr-DM_C}, which reports the maximum stable stellar masses ($M_\ma$) and corresponding radii ($R_\ma$).
\begin{table}[!htb]
\centering
\caption{Maximum stable stellar mass, in units of $M_\odot$, and respective radius, in units of km, for the different mass-radius diagrams presented in Fig.~\ref{fig:mr-DM_C}.}
\begin{tabular*}{\linewidth}{@{\extracolsep{\fill}}lcccc}
\toprule
\midrule 
$F_\dm$ ($\%$) & $M^{\rmf}_\ma$  &  $R^{\rmf}_\ma$  & $M^{\rmfsrc}_\ma$ & $R^{\rmfsrc}_\ma$ \\
\midrule
$0$  & $2.03$ & $12.12$ & $2.19$ & $12.51$ \\
$2$  & $1.97$ & $11.99$ & $2.12$ & $12.35$ \\
$4$  & $1.92$ & $11.82$ & $2.06$ & $12.21$ \\
$6$  & $1.87$ & $11.68$ & $2.01$ & $12.06$ \\
\bottomrule
\end{tabular*}
\label{tab:mr-DM_C}
\end{table}

Another feature observed in Fig.~\ref{fig:mr-DM_C} is that the radii predicted by the pure hadronic \mbox{RMF-SRC model} ($F_\dm = 0$) are larger than those obtained with the corresponding RMF model (black curves). In other words, the softer the EoS — RMF in this case — the smaller the stellar radius, implying a higher compactness for a given mass $M$. Consequently, a higher central pressure — or, equivalently, a greater central energy density — is required to support the star against gravitational collapse. This effect, namely the increase of $\mathcal{E}_c$ for a softer EoS at fixed $M$, is also confirmed in Fig.~\ref{fig:star}. Furthermore, the same pattern is observed when a finite amount of DM is included in the star, i.e., when $F_\dm \ne 0$. This behavior arises from the dominance of the hadronic component over the DM EoS, which ultimately governs the global stellar structure. In particular, the radius of the final admixed NS is always determined by the visible sector, $R_\vis$. For all \mbox{$M$–$R$} curves analyzed, we consistently find $R_\vis > R_\dm$, indicating that the DM distribution never extends beyond the hadronic surface. As a result, no stellar configuration with a DM halo is obtained within the present set of parameters.

At this point, it is instructive to highlight the impact of incorporating DM through different formalisms. A direct comparison between Fig.~\ref{fig:mr-DM_C} of the present work and Fig.~5 of Ref.~\cite{lourenco_prd2022_1} underscores the distinct modeling strategies adopted in the two studies. In Ref.~\cite{lourenco_prd2022_1}, DM is treated within a single-fluid formalism and coupled to hadrons via Higgs boson exchange, whereas in the present work, we employ a two-fluid approach in which the dark and hadronic sectors interact exclusively through gravity. Although both studies include SRC and a quartic vector self-interaction term, the numerical difference in the constant $C$ is small and does not lead to significant modifications in the neutron star structure. Consequently, the main differences observed in the mass–radius relations arise from the distinct DM EoSs, the corresponding TOV formulations (single-fluid versus two-fluid), and the choice of RMF parametrization. In summary, these comparisons indicate that the effects of SRC remain robust across both frameworks, i.e., the increasing of the NS mass, while the detailed impact of DM is strongly dependent on the adopted EoS and fluid formalism. 

\begin{figure}[!htb]
\centering
\includegraphics[scale=0.57]{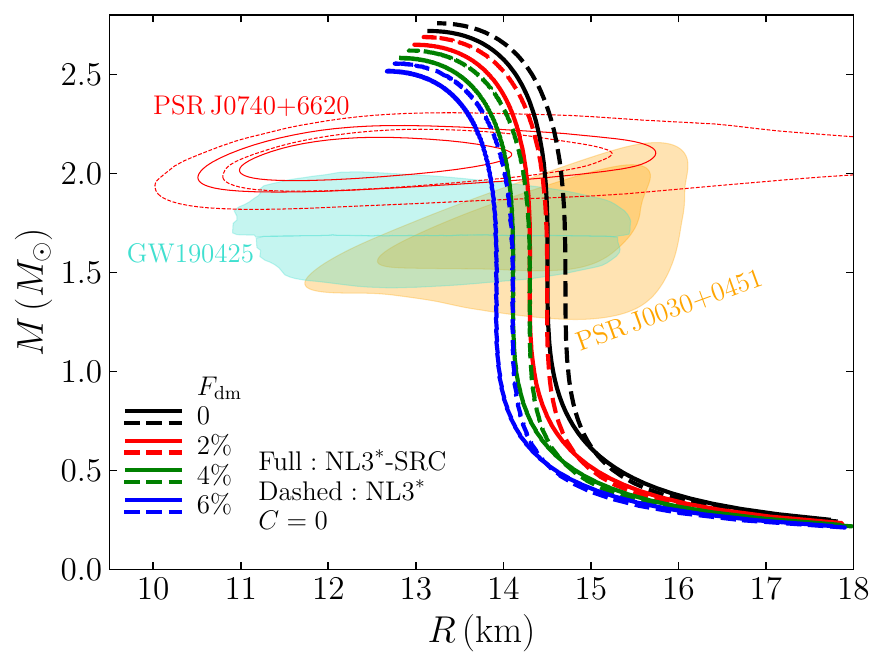}
\caption{Mass-radius diagrams for different fractions of DM in each star ($F_\dm$). Results for the NL3* model~($C=0$), with~(full curves) and without~(dashed curves) short-range correlations. It is also plotted jointly NICER$-$XMM-Newton data concerning the pulsars PSR J0030+0451 (orange contours)~\cite{Vinciguerra_2024}, and PSR J0740+6620 (solid~\cite{Salmi_2024} and dashed~\cite{Dittmann_2024} red contours). We also compared our results with data from the GW190425 event (turquoise contours)~\cite{Abbott_2020-2}.}
\label{fig:mr-DM}
\end{figure}

The mass-radius diagrams for the model with vector field only up to second order ($C=0$) are shown in Fig.~\ref{fig:mr-DM}. 

As demonstrated in the previous sections, in the specific case of pure NSs ($F_{\text{dm}} = 0$), the absence of higher-order self-interaction terms in the vector meson field leads to a systematic stiffening of the hadronic EoS. Consequently, both the maximum mass and the corresponding stellar radius increase, as illustrated by the solid black curves in Figs.~\ref{fig:mr-DM} and~\ref{fig:mr-DM_C}. This trend is consistent with earlier findings reported in Ref.~\cite{muller-serot96}, which highlighted the role of nonlinear vector meson contributions in softening the EoS at high densities. Upon incorporating SRC, the EoS becomes effectively softer, resulting in less massive and more compact NSs within this class of models. The same qualitative behavior is observed for NSs admixed with DM ($F_\dm \ne 0$), indicating that the hadronic sector continues to govern the global structure of the star even in the presence of a dark component. Once again, no evidence of DM halo configurations is found for the cases shown in Fig.~\ref{fig:mr-DM}, suggesting that DM remains gravitationally bound within the stellar core. Finally, for the sake of completeness, we provide data regarding the last stable star in Table~\ref{tab:mr-DM}.
\begin{table}[!htb]
\centering
\caption{Maximum stable stellar mass, in units of $M_\odot$, and respective radius, in units of km, for the different mass-radius diagrams presented in Fig.~\ref{fig:mr-DM}.}
\begin{tabular*}{\linewidth}{@{\extracolsep{\fill}}lcccc}
\toprule
\midrule 
$F_\dm$ ($\%$) & $M^{\mbox{\tiny NL3*}}_\ma$  &  $R^{\mbox{\tiny NL3*}}_\ma$  & $M^{\mbox{\tiny NL3*-SRC}}_\ma$ & $R^{\mbox{\tiny NL3*-SRC}}_\ma$ \\
\midrule
$0$  & $2.76$ & $13.24$ & $2.72$ & $13.17$ \\
$2$  & $2.69$ & $13.09$ & $2.65$ & $13.00$ \\
$4$  & $2.62$ & $12.92$ & $2.58$ & $12.83$ \\
$6$  & $2.56$ & $12.78$ & $2.52$ & $12.69$ \\
\bottomrule
\end{tabular*}
\label{tab:mr-DM}
\end{table}

\begin{figure}[!htb]
\centering
\includegraphics[scale=0.53]{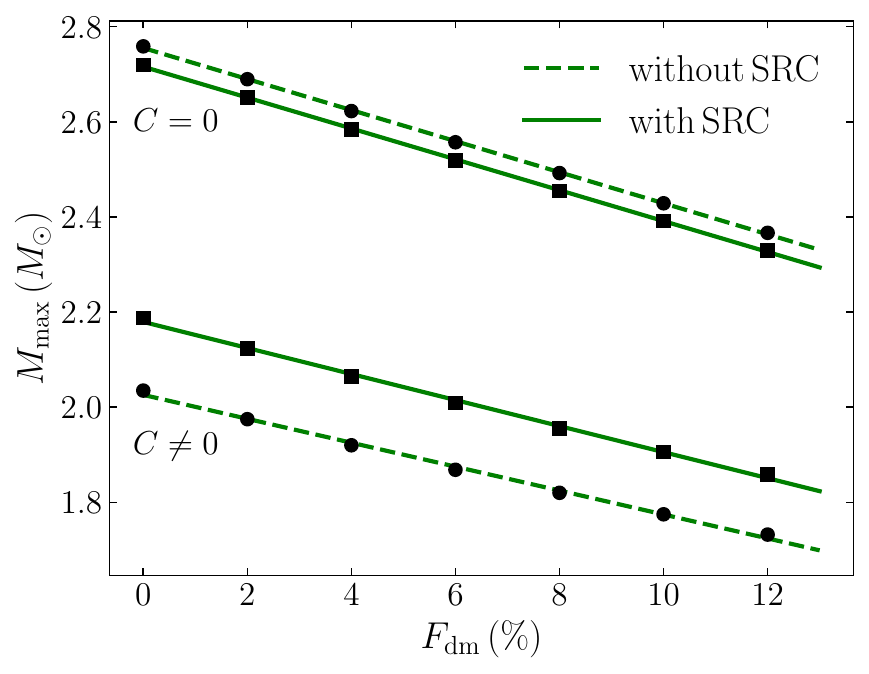}
\caption{Maximum mass as a function of the mass fraction. Results extracted from the diagrams displayed in Figs.~\ref{fig:mr-DM_C} and~\ref{fig:mr-DM} for the models with (squares and fitted full curves) and without (circles and dashed fitted curves) SRC included.}
\label{fig:mmax}
\end{figure}

Fig.~\ref{fig:mmax} illustrates how the mass of the last stable star, as determined from Figs.~\ref{fig:mr-DM_C} and~\ref{fig:mr-DM}, varies as a function of the DM mass fraction $F_{\text{dm}}$.
A clear linear dependence is observed for both models analyzed, at least within the range of $F_{\text{dm}}$ values considered here. Notably, the inclusion of SRC leads to nearly parallel fitting curves, a feature confirmed by the respective slope values: $-0.0326$ (NL3*), $-0.0324$ (\mbox{NL3*-SRC}) for the model with $C = 0$, and $-0.0251$ (RMF), $-0.0274$ (\mbox{RMF-SRC}) for the model with $C \neq 0$, all in units of $M_\odot$. The decrease in stellar mass with increasing DM fraction is a behavior also observed in other scenarios; see, for instance, Refs.~\cite{Karkevandi:2021ygv,jurgen16,Ivanytskyi:2019wxd,ellis18,nitr}. Qualitatively, this reduction—when compared with pure compact stars—is attributed to the formation of a DM core, a structure typically found for heavy DM particles, either bosonic or fermionic. In the fermionic model adopted here, we set $m_\chi = 1.9$~GeV, which leads to the formation of DM cores in all models investigated. Conversely, diffuse halo structures enveloping the final admixed star are observed for models with lighter DM particles and, in general, lead to an increase in the maximum stable mass~\cite{laura24,carline25}. Furthermore, interactions among dark particles can also produce such an increase~\cite{carline25}. Possible microscopic mechanisms responsible for the formation of DM cores or halos in NSs—such as neutron conversion into dark particles and DM production via bremsstrahlung in neutron–neutron scatterings—are discussed in the Appendix of Ref.~\cite{ellis18}.

From Fig.~\ref{fig:mmax}, we compute $\Delta M$, defined as the difference between the maximum stellar mass obtained from the model with and without SRC. For the case $C \ne 0$, where $\Delta M > 0$, we find $0.13 < \Delta M/M_\odot < 0.15$ across the range of $F_\dm$ considered. This interval is larger than that reported in Ref.~\cite{lourenco_prd2022_1}. In that study, a one-fluid formalism was employed to describe DM-admixed NSs, with the DM content controlled by the Fermi momentum of the dark particle ($k_F^\dm$). The hadronic sector was modeled by a relativistic framework including a quartic vector self-interaction term, and an increase in the maximum stellar mass was also observed when SRC effects were taken into account. From Fig.~8 of Ref.~\cite{lourenco_prd2022_1}, we obtain the corresponding range in $\Delta M$ given by $0.059 < \Delta M/M_\odot < 0.066$ for $0 < k_F^\dm < 0.06~\mathrm{GeV}$.

Finally, we emphasize that all parametrizations presented in Figs.~\ref{fig:mr-DM_C} and~\ref{fig:mr-DM} are consistent with the most recent astrophysical observations, including the joint NICER–XMM-Newton analyses of the pulsars PSR~J0030+0451~\cite{Vinciguerra_2024} and PSR~J0740+6620~\cite{Salmi_2024,Dittmann_2024}, as well as the gravitational-wave event GW190425~\cite{Abbott_2020-2}. In the case of the model with $C \ne 0$, it is worth noting that the inclusion of SRC enables the final admixed NSs to support a larger amount of DM, since the reduction in the maximum stable mass with increasing $F_\dm$ is partially compensated by the corresponding increase induced by SRC.

\section{Summary and concluding remarks}
\label{sec:summ}

In this work, we have analyzed the high-density regime of relativistic hadronic mean-field models, expanded up to second ($\omega_0^2$) and fourth order ($\omega_0^4$) in the repulsive vector field. In the latter case, the self-interaction strength is governed by the constant $C$. We have also investigated the impact of incorporating SRC effects on the structure of these models. This phenomenology alters the kinetic contributions to the energy density, with direct consequences for the thermodynamical EoSs derived from it—such as the pressure and the chemical potentials of protons and neutrons—for which exact analytical expressions are provided.

We have shown that SRC affect the coefficient of the nonleading term in the pressure–energy density expansion in distinct ways, depending on the form of the vector self-interaction. In the model with $C = 0$, the inclusion of SRC leads to an increase in this coefficient across a wide range of proton fractions, resulting in a softened EoS. This trend is consistently reproduced not only for a single RMF parametrization but also across various sets of nuclear empirical parameters, including the saturation density, binding energy, symmetry energy, and effective mass. In contrast, for models with $C \ne 0$, SRC play an opposite role: the modified EoS becomes stiffer across different parametrizations, characterized by distinct values of $\rho_{\sat}$, $E_{\sym,2}$, and $m^*_{\sat}$, due to the decrease of the respective coefficient.

The impact of these features was investigated in astrophysical compact systems, focusing on two scenarios: NSs in beta equilibrium—subject to charge neutrality and chemical equilibrium—both with and without DM content. In the first case, we found that the inclusion of SRC leads to less massive stars for models with $C = 0$. In contrast, when SRC are incorporated into models featuring a fourth-order vector self-interaction ($C \ne 0$), the resulting stars become more massive compared to those without SRC. This trend is consistently observed across the different parametrizations analyzed. Moreover, the effect identified in the asymptotic high-density regime also manifests within the energy density range relevant for the structure of pure NSs.

To model NSs admixed with DM, we employed the two-fluid formalism~\cite{qian14,Collier:2022cpr,Shakeri:2022dwg,Miao:2022rqj,Emma:2022xjs,Hong:2024sey,Karkevandi:2021ygv,Liu:2023ecz,Ivanytskyi:2019wxd,Buras-Stubbs:2024don,Rutherford:2022xeb,Thakur:2023aqm,Mahapatra:2024ywx,Thakur:2024mxs,carline25}, in which the DM component was described by a fermionic model featuring a repulsive interaction mediated by a dark vector field. The resulting nonsuperluminal EoS, derived within the mean-field approximation, depended on two free parameters: the dark fermion mass, $m_\chi$, and the ratio between the coupling strength of the vector interaction and the vector field mass, $C_V$. We solved the generalized TOV equations, coupled to differential equations for the particle numbers—used to establish stellar stability conditions as detailed in Refs.~\cite{mauricio23,laura24,pitz25,carline25}—to obtain mass–radius diagrams for various values of the DM mass fraction, $F_{\text{dm}}$.

We showed that the general effect observed in pure NSs ($F_\dm = 0$) regarding the inclusion of SRC is reproduced for configurations with $F_\dm > 0$ in both classes of models, i.e., those constructed with $C = 0$ and $C \ne 0$. This outcome reflects the dominance of the hadronic sector, which controls the overall stellar structure, with no dark halo formation in this case. Typically, such a structure is found for light DM particles (here we use $m_\chi = 1.9$~GeV). Furthermore, our results indicated a systematic linear decrease in the maximum stable stellar mass, $M_{\text{max}}$, as a function of $F_{\text{dm}}$ for models containing the $\omega_0^2$ and $\omega_0^4$ terms. The inclusion of SRC did not alter this linear trend; instead, it produced nearly parallel fitting curves for this relationship. In particular, for the model including the $\omega_0^4$ term, we found that the reduction in $M_{\text{max}}$ due to the presence of DM is partially compensated by the increase resulting from SRC effects in the hadronic sector.

In addition to the analysis presented in this work, we highlighted the compatibility of our results with recent astrophysical data from the joint NICER–XMM analysis of the pulsars PSR~J0030+0451~\cite{Vinciguerra_2024} and PSR~J0740+6620~\cite{Salmi_2024,Dittmann_2024}, where modeling of the respective pulse profiles was used to infer macroscopic properties such as mass and radius. We also compared our findings with the gravitational-wave event GW190425~\cite{Abbott_2020-2}, which reported the coalescence of a binary system with a total mass of approximately $3.4\,M_\odot$, with component masses consistent with NSs in the range of $1.1\,M_\odot$ to $2.5\,M_\odot$.

Beyond these current astrophysical analyses, upcoming observational programs such as STROBE-X~\cite{strobex}, eXTP~\cite{zand2019}, Athena~\cite{athena}, and NewAthena~\cite{newathena2024} are expected to play a crucial role in deepening our understanding of compact stars and their potential DM content. With its high effective area and microsecond timing precision~\cite{Santangelo2024}, eXTP will enable detailed measurements of pulsations and quasiperiodic oscillations, providing essential constraints on the stellar mass–radius relation. In addition, its remarkable sensitivity to surface thermal emission will allow stringent observational tests of theoretical scenarios predicting DM capture and accumulation in compact stars. The synergy between these forthcoming X-ray missions and gravitational-wave observations, such as those expected from the proposed Einstein Telescope~\cite{Koehn2024}, will help establish tighter limits on the dense-matter EoS and shed light on the microscopic properties of DM inside NSs.

\section{Appendix}
\label{sec:app}

Here we present the derivation of the kinetic contribution of the chemical potentials, with some important consequences. It follows:
\begin{widetext}

\begin{align}
\mu^{p,n}_{\kin} &\equiv \frac{\partial (\mathcal{E}_{\kin,p} + \mathcal{E}_{\kin,n})}{\partial \rho_{p,n}} \Bigg|_{\rho_n, M^*}
= \frac{1}{\pi^2} \frac{\partial \Delta_p(y)}{\partial \rho_{p,n}}\int_0^{k_{Fp}} dk\, k^2 \left( k^2 + M^{*2} \right)^{1/2}
+ \frac{\Delta_p(y)}{\pi^2} \frac{\partial}{\partial \rho_{p,n}} \int_0^{k_{Fp}} dk\, k^2 \left( k^2 + M^{*2} \right)^{1/2}
\nonumber \\
&+ \frac{k_{Fp}^4}{\pi^2} \frac{\partial C_p}{\partial \rho_{p,n}} \int_{k_{Fp}}^{\phi_p(y) k_{Fp}}dk\frac{(k^2 + M^{*2})^{1/2}}{k^2}
+ \frac{4\, C_p(y)\, k_{Fp}^3}{\pi^2} \frac{\partial k_{Fp}}{\partial \rho_{p,n}} \int_{k_{Fp}}^{\phi_p(y) k_{Fp}}dk\frac{(k^2 + M^{*2})^{1/2}}{k^2}
\nonumber \\
&+ \frac{C_p(y)\, k_{Fp}^4}{\pi^2} \frac{\partial}{\partial \rho_{p,n}} \int_{k_{Fp}}^{\phi_p(y) k_{Fp}}dk\frac{(k^2 + M^{*2})^{1/2}}{k^2}
+ \frac{1}{\pi^2} \frac{\partial \Delta_n(y)}{\partial \rho_{p,n}}\int_0^{k_{Fn}} dk\, k^2 \left( k^2 + M^{*2} \right)^{1/2} 
\nonumber\\
&+ \frac{\Delta_n(y)}{\pi^2} \frac{\partial}{\partial \rho_{p,n}} \int_0^{k_{Fn}} dk\, k^2 \left( k^2 + M^{*2} \right)^{1/2}
+ \frac{k_{Fn}^4}{\pi^2} \frac{\partial C_n}{\partial \rho_{p,n}} \int_{k_{Fn}}^{\phi_n(y) k_{Fn}}dk\frac{(k^2 + M^{*2})^{1/2}}{k^2}
\nonumber\\
&+ \frac{4\, C_n(y)\, k_{Fn}^3}{\pi^2} \frac{\partial k_{Fn}}{\partial \rho_{p,n}} \int_{k_{Fn}}^{\phi_n(y) k_{Fn}}dk\frac{(k^2 + M^{*2})^{1/2}}{k^2}
+ \frac{C_n(y)\, k_{Fn}^4}{\pi^2} \frac{\partial}{\partial \rho_{p,n}} \int_{k_{Fn}}^{\phi_n(y) k_{Fn}}dk\frac{(k^2 + M^{*2})^{1/2}}{k^2}.
\label{eq:mukinp}
\end{align}

\subsection{Derivation of $\texorpdfstring{\mu^p_\kin}{mu^p_kin}$}

We start with the proton chemical potential, $\mu^p_\kin$. Due to $\partial k_{Fn}/\partial \rho_p = 0$ and, since there is no dependence on $\rho_p$ in integrals whose limits depend only on $k_{Fn}$, the seventh and ninth terms in Eq.~\eqref{eq:mukinp} vanish (we are taking the derivative with respect to $\rho_p$). Furthermore, we also have that
\begin{align}
\frac{\Delta_p(y)}{\pi^2} \frac{\partial}{\partial \rho_p} \int_0^{k_{Fp}} dk\, k^2 ( k^2 + M^{*2} )^{1/2} = \Delta_p(y)( \kfp^2 + M^{*2})^{1/2},
\end{align}
\begin{align}
\frac{4\, C_p(y)\, k_{Fp}^3}{\pi^2}\frac{\partial k_{Fp}}{\partial \rho_p}\int_{k_{Fp}}^{\phi_p(y) k_{Fp}}dk\frac{(k^2 + M^{*2})^{1/2}}{k^2}
&= 4\, C_p(y)\, k_{Fp} \left[\ln \left(\frac{k_{Fp} \phi_p(y) + \sqrt{k_{Fp}^2 \phi_p^2(y) + M^{*2}}}{k_{Fp} + \sqrt{k_{Fp}^2 + M^{*2}}}\right)
+ \frac{\sqrt{k_{Fp}^2 + M^{*2}}}{k_{Fp}}\right.
\nonumber\\
&- \left.\frac{\sqrt{k_{Fp}^2 \phi_p^2(y) + M^{*2}}}{k_{Fp} \phi_p(y)}\right],
\end{align}
\begin{align}
\frac{C_p(y)\, k_{Fp}^4}{\pi^2}\frac{\partial}{\partial \rho_p}\int_{k_{Fp}}^{\phi_p(y)\, k_{Fp}}dk\frac{(k^2 + M^{*2})^{1/2}}{k^2}
&= C_p(y)\frac{\left( \phi_p^2(y)\, k_{Fp}^2 + M^{*2} \right)^{1/2}}{\phi_p(y)}
+ C_p(y)\, k_{Fp}\, \frac{\left( \phi_p^2(y)\, k_{Fp}^2 + M^{*2} \right)^{1/2}}{\phi_p^2(y)}\, \frac{\partial \phi_p(y)}{\partial k_{Fp}}
\nonumber\\
&- C_p(y)\left( k_{Fp}^2 + M^{*2} \right)^{1/2},
\end{align}
and
\begin{align}
\frac{C_n(y)\, k_{Fn}^4}{\pi^2} \frac{\partial}{\partial \rho_p} \int_{k_{Fn}}^{\phi_n(y) k_{Fn}}dk\frac{(k^2 + M^{*2})^{1/2}}{k^2} &= 
\frac{C_n(y)}{k_{Fp}^2} \, k_{Fn}^3 \, \frac{\left( \phi_n^2(y)\, k_{Fn}^2 + M^{*2} \right)^{1/2}}{\phi_n^2(y)}\frac{\partial \phi_n(y)}{\partial k_{Fp}}.
\end{align}
If we consider as a first approximation that $\Delta_{p,n}$, $C_{p,n}$, and $\phi_{p,n}$ are constants, as in Refs.~\cite{souza2020,dutra_mnras22,lourenco_prd2022_1,lourenco_prd2022_2,pelicer2023}, then $\mu_\kin^p$ reads
\begin{align}
\mu^p_\kin &= \Delta_p(y)(k^2_{F_{p}} + M^{*2})^{1/2} + \tilde{\mu}^{p}_{\kin(\src)},
\end{align}
with
\begin{align}
&\tilde{\mu}^p_{\kin(\src)} = 3C_p(y)\left[(k^2_{F_{p}} + M^{*2})^{1/2} - \frac{\left(\phi_p^2(y)\kfp^2 + {M^{*2}}\right)^{1/2}}{\phi_p(y)} \right]
+ 4C_p(y)\kfp\Ln\left[\frac{\phi_p(y)\kfp + \left(\phi_p^2(y)\kfp^2 + {M^{*2}}\right)^{1/2}}{\kfp + (\kfp^2 + {M^{*2}})^{1/2}}\right].
\end{align}
On the other hand, we use the following results
\begin{align}
\frac{\partial \phi_p}{\partial k_{Fp}} 
= \frac{k_{Fp}^2}{\pi^2} \frac{\partial \phi_p}{\partial \rho_p} 
= \frac{k_{Fp}^2}{\pi^2} \frac{\rho_n}{\rho^2} \frac{\partial \phi_p}{\partial y}
= \frac{k_{Fp}^2}{\pi^2} \frac{\rho_n}{\rho^2}\,2 \phi_0 \phi_1,\quad
\frac{\partial \phi_n}{\partial k_{Fp}} 
= \frac{k_{Fp}^2}{\pi^2} \, \frac{\rho_n}{\rho^2} \frac{\partial \phi_n}{\partial y} 
= -\frac{k_{Fp}^2}{\pi^2} \frac{\rho_n}{\rho^2} \, 2 \phi_0 \phi_1,
\end{align}
\begin{align}
\frac{\partial C_p}{\partial \rho_p} = \frac{\rho_n}{\rho^2} \, \frac{\partial C_p}{\partial y} 
= \frac{\rho_n}{\rho^2} \, 2 C_0 C_1,\quad
\frac{\partial C_n}{\partial \rho_p} = \frac{\rho_n}{\rho^2} \, \frac{\partial C_n}{\partial y} 
= -\frac{\rho_n}{\rho^2} \, 2 C_0 C_1,
\end{align}
\begin{align}
\frac{\partial \Delta_p}{\partial \rho_p} = -6 \left[ C_0 C_1 \left( 1 - \frac{1}{\phi_p(y)} \right) + \frac{C_p(y)}{\phi_p^2(y)} \phi_0 \phi_1 \right] \, \frac{\rho_n}{\rho^2},\quad
\frac{\partial \Delta_n}{\partial \rho_p} = 6 \left[ C_0 C_1 \left( 1 - \frac{1}{\phi_n(y)} \right) + \frac{C_n(y)}{\phi_n^2(y)} \phi_0 \phi_1 \right] \, \frac{\rho_n}{\rho^2},
\end{align}
in order to obtain the full expression, given by
\begin{align}
\mu^p_\kin &= \Delta_p(y)(k^2_{F_{p}} + M^{*2})^{1/2} + \tilde{\mu}^{p}_{\kin(\src)} + \frac{2}{\pi^2}\frac{\rho_n}{\rho^2}\eta_p,
\end{align}
with
\begin{align}
 \eta_p  &= \phi_0\phi_1\left[\frac{C_p(y)}{\phi_p^2(y)}\kfp^3\left(\phi_p^2(y)\kfp^2 + {M^*}^2\right)^{1/2} 
 - \frac{C_n(y)}{\phi_n^2(y)}\kfn^3\left(\phi_n^2(y)\kfn^2 + {M^*}^2\right)^{1/2}\right]
 \nonumber \\ 
 &+C_0C_1\left[\kfp^4\int_{\kfp}^{\phi_p(y)\kfp} dk \frac{(k^2 + {M^*}^2)^{1/2}}{k^2} 
 -\kfn^4\int_{\kfn}^{\phi_n(y)\kfn} dk \frac{(k^2 + {M^*}^2)^{1/2}}{k^2}\right] 
 \nonumber\\ 
 &-3\left[C_0C_1\left(1 - \frac{1}{\phi_p(y)}\right) + \phi_0\phi_1\frac{C_p(y)}{\phi_p^2(y)}\right]\int_{0}^{\kfp} dk\, k^2(k^2 + {M^*}^2)^{1/2}
 \nonumber\\
 &+3\left[C_0C_1\left(1 - \frac{1}{\phi_n(y)}\right) + \phi_0\phi_1\frac{C_n(y)}{\phi_n^2(y)}\right]\int_{0}^{\kfn} dk\, k^2 (k^2 + {M^*}^2)^{1/2}.
\end{align}

\subsection{Derivation of $\texorpdfstring{\mu^n_\kin}{mu^n_kin}$}

Notice that $\partial k_{Fp}/\partial \rho_n = 0$ and there is no dependence on $\rho_n$ in integrals whose limits depend only on $k_{Fp}$. Therefore, the second and fourth terms in Eq.~\eqref{eq:mukinp} disappear (derivative with respect to $\rho_n$ in the calculation of $\mu^n_\kin$). In this case one has,
\begin{align}
\frac{C_p(y)\, k_{Fp}^4}{\pi^2}\frac{\partial}{\partial \rho_n}\int_{k_{Fp}}^{\phi_p(y)\, k_{Fp}}dk\frac{(k^2 + M^{*2})^{1/2}}{k^2}
&= \frac{C_p(y)}{k_{Fn}^2} \, k_{Fp}^3 \, \frac{\left( \phi_p^2(y)\, k_{Fp}^2 + M^{*2} \right)^{1/2}}{\phi_p^2(y)}\frac{\partial \phi_p(y)}{\partial k_{Fn}},
\end{align}
\begin{align}
\frac{\Delta_n(y)}{\pi^2} \frac{\partial}{\partial \rho_{n}} \int_0^{k_{Fn}} dk\, k^2 \left( k^2 + M^{*2} \right)^{1/2} = \Delta_n(y)( \kfn^2 + M^{*2})^{1/2},
\end{align}
\begin{align}
\frac{4\, C_n(y)\, k_{Fn}^3}{\pi^2} \frac{\partial k_{Fn}}{\partial \rho_{n}} \int_{k_{Fn}}^{\phi_n(y) k_{Fn}}\hspace{-0.3cm}dk\frac{(k^2 + M^{*2})^{1/2}}{k^2} 
&= 4\, C_n(y)\, k_{Fn} \left[\ln \left(\frac{k_{Fn} \phi_n(y) + \sqrt{k_{Fn}^2 \phi_n^2(y) + M^{*2}}}{k_{Fn} + \sqrt{k_{Fn}^2 + M^{*2}}}\right)
+ \frac{\sqrt{k_{Fn}^2 + M^{*2}}}{k_{Fn}}\right.
\nonumber\\
&- \left.\frac{\sqrt{k_{Fn}^2 \phi_n^2(y) + M^{*2}}}{k_{Fn} \phi_n(y)}\right],
\end{align}
\begin{align}
\frac{C_n(y)\, k_{Fn}^4}{\pi^2} \frac{\partial}{\partial \rho_{n}} \int_{k_{Fn}}^{\phi_n(y) k_{Fn}}\hspace{-0.3cm}dk\frac{(k^2 + M^{*2})^{1/2}}{k^2}
&= C_n(y)\frac{\left( \phi_n^2(y)\, k_{Fn}^2 + M^{*2} \right)^{1/2}}{\phi_n(y)}
+ C_n(y)\, k_{Fn}\, \frac{\left( \phi_n^2(y)\, k_{Fn}^2 + M^{*2} \right)^{1/2}}{\phi_n^2(y)}\, \frac{\partial \phi_n(y)}{\partial k_{Fn}}
\nonumber\\
&- C_n(y)\left( k_{Fn}^2 + M^{*2} \right)^{1/2}.
\end{align}
Again, the use of the approximation taken in last section generates
\begin{align}
\mu^n_\kin &= \Delta_n(y)(k^2_{F_{n}} + M^{*2})^{1/2} + \tilde{\mu}^{n}_{\kin(\src)},
\end{align}
with
\begin{align}
&\tilde{\mu}^n_{\kin(\src)} = 3C_n(y)\left[(k^2_{F_{n}} + M^{*2})^{1/2} - \frac{\left(\phi_n^2(y)\kfn^2 + {M^{*2}}\right)^{1/2}}{\phi_n(y)} \right]
+ 4C_n(y)\kfn\Ln\left[\frac{\phi_n(y)\kfn + \left(\phi_n^2(y)\kfn^2 + {M^{*2}}\right)^{1/2}}{\kfn + \left(\kfn^2 + {M^{*2}}\right)^{1/2}}\right].
\end{align}
Finally, the results presented below, namely,
\begin{align}
\frac{\partial \phi_p}{\partial k_{Fn}} = -\frac{k_{Fn}^2}{\pi^2} \frac{\rho_p}{\rho^2}\,2 \phi_0 \phi_1,\quad
\frac{\partial \phi_n}{\partial k_{Fn}} = \frac{k_{Fn}^2}{\pi^2} \frac{\rho_p}{\rho^2} \, 2 \phi_0 \phi_1,\quad
\frac{\partial C_p}{\partial \rho_n} = -\frac{\rho_p}{\rho^2} \, 2 C_0 C_1,\quad
\frac{\partial C_n}{\partial \rho_n} = \frac{\rho_p}{\rho^2} \, 2 C_0 C_1,
\end{align}
\begin{align}
\frac{\partial \Delta_p}{\partial \rho_n} = 6 \left[ C_0 C_1 \left( 1 - \frac{1}{\phi_p(y)} \right) + \frac{C_p(y)}{\phi_p^2(y)} \phi_0 \phi_1 \right] \, \frac{\rho_p}{\rho^2},\quad
\frac{\partial \Delta_n}{\partial \rho_n} = -6 \left[ C_0 C_1 \left( 1 - \frac{1}{\phi_n(y)} \right) + \frac{C_n(y)}{\phi_n^2(y)} \phi_0 \phi_1 \right] \, \frac{\rho_p}{\rho^2},
\end{align}
are used to find the complete expression given by
\begin{align}
\mu^n_\kin &= \Delta_n(y)(k^2_{F_{n}} + M^{*2})^{1/2} + \tilde{\mu}^{n}_{\kin(\src)} + \frac{2}{\pi^2}\frac{\rho_p}{\rho^2}\eta_n,
\end{align}
with
\begin{align}
 \eta_n  &= \phi_0\phi_1\left[-\frac{C_p(y)}{\phi_p^2(y)}\kfp^3\left(\phi_p^2(y)\kfp^2 + {M^*}^2\right)^{1/2} 
 + \frac{C_n(y)}{\phi_n^2(y)}\kfn^3\left(\phi_n^2(y)\kfn^2 + {M^*}^2\right)^{1/2}\right]
 \nonumber \\ 
 &+C_0C_1\left[-\kfp^4\int_{\kfp}^{\phi_p(y)\kfp} dk \frac{(k^2 + {M^*}^2)^{1/2}}{k^2} 
 +\kfn^4\int_{\kfn}^{\phi_n(y)\kfn} dk \frac{(k^2 + {M^*}^2)^{1/2}}{k^2}\right] 
 \nonumber\\ 
 &+3\left[C_0C_1\left(1 - \frac{1}{\phi_p(y)}\right) + \phi_0\phi_1\frac{C_p(y)}{\phi_p^2(y)}\right]\int_{0}^{\kfp} dk\, k^2(k^2 + {M^*}^2)^{1/2}
 \nonumber\\
 &-3\left[C_0C_1\left(1 - \frac{1}{\phi_n(y)}\right) + \phi_0\phi_1\frac{C_n(y)}{\phi_n^2(y)}\right]\int_{0}^{\kfn} dk\, k^2 (k^2 + {M^*}^2)^{1/2} = -\eta_p.
\end{align}

\subsection{General consequences}

By considering the kinetic part of the model, we have
\begin{align}
\rho_p\mu^p_\kin  &+ \rho_n\mu^n_\kin = \rho_p\Delta_p(y)(k^2_{F_{p}} + M^{*2})^{1/2} + \rho_p\tilde{\mu}^{p}_{\kin(\src)} + \frac{2}{\pi^2}\frac{\rho_p\rho_n}{\rho^2}\eta_p
\nonumber\\
&+ \rho_n\Delta_n(y)(k^2_{F_{n}} + M^{*2})^{1/2} + \rho_n\tilde{\mu}^{n}_{\kin(\src)} + \frac{2}{\pi^2}\frac{\rho_n\rho_p}{\rho^2}\eta_n
\nonumber\\
&= \rho_p\Delta_p(y)(k^2_{F_{p}} + M^{*2})^{1/2} + \rho_p\tilde{\mu}^{p}_{\kin(\src)} + \rho_n\Delta_n(y)(k^2_{F_{n}} + M^{*2})^{1/2} + \rho_n\tilde{\mu}^{n}_{\kin(\src)} + \frac{2\rho_p\rho_n}{\pi^2\rho^2}(\eta_p+\eta_n).
\end{align}
\end{widetext}
Note that due to $\eta_n = -\eta_p$, the last term in the expression above vanishes, and since the remaining expression is equal to $\mathcal{E}_{\kin,p} + \mathcal{E}_{\kin,n} + P_{\kin,p} + P_{\kin,n}$, the Euler relation is satisfied for both the exact expression of $\mu^{p,n}_\kin$ and its approximate version, in which $\Delta_{p,n}$, $C_{p,n}$, and $\phi_{p,n}$ are taken as constants.

Another interesting result concerns the symmetric nuclear matter regime. In this particular case, where $y = 1/2$ and $k_{Fp} = k_{Fn}$, we have $C_p = C_n = C_0$, $\phi_p = \phi_n = \phi_0$, and $\Delta_p = \Delta_n = 1 - 3 C_0(1 - 1/\phi_0)$. Consequently, $\eta_p = \eta_n = 0$, and the expression for $\mu^{p,n}_\kin$ coincides with that obtained using the approximate form. This is a significant feature, as it shows that the approximate version of $\mu^{p,n}_\kin$ can be reliably used to determine the coupling constants of the model with SRC included, calculated at saturation density, and for $y = 1/2$.

\begin{figure}[!htb]
\centering
\includegraphics[scale=0.5]{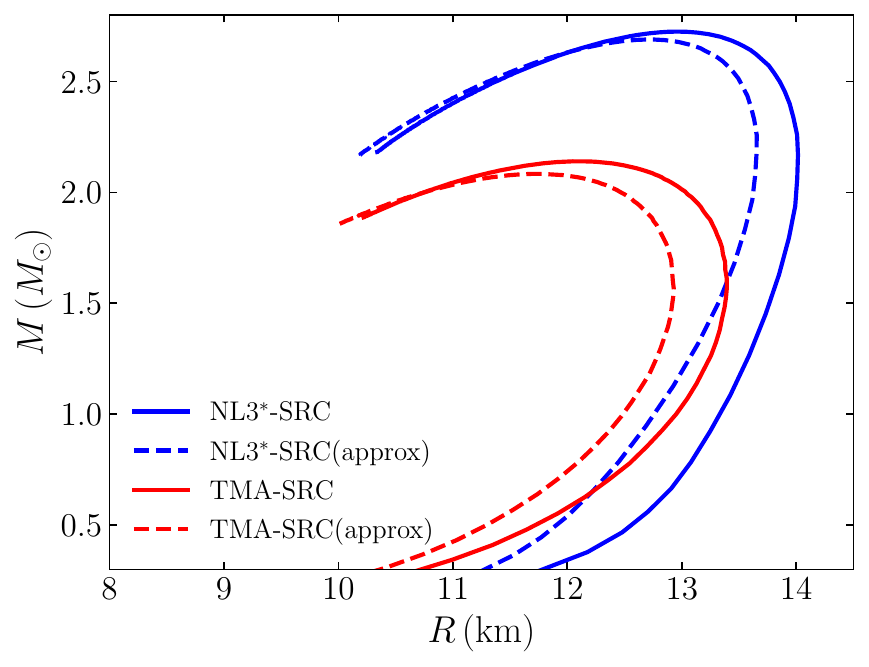}
\caption{Neutron star mass-radius diagrams for two different parametrizations of the SRC model with $\eta_p\ne 0$ and $\eta_n\ne 0$ (full curves) and $\eta_p=\eta_n=0$. Crust EoS is not included.}
\label{fig:approx}
\end{figure}
At this point, we also emphasize the impact on the NS structure arising from the additional terms in the full expression of the chemical potentials when SRC are included. Since in the beta-equilibrated matter the condition $\mu_n - \mu_p = \mu_e = \mu_\mu$ must be satisfied, together with charge neutrality, $\rho_p = \rho_e + \rho_\mu$, the terms $2\rho_n\eta_p/(\pi^2\rho^2)$ and $2\rho_p\eta_n/(\pi^2\rho^2)$ modify the quantities determined from these equations, namely the proton fraction and the electron density. As a consequence, the total EoS used as input for solving the TOV equations (in either the one- or two-fluid formalism) differs from that obtained when the approximation is employed, as it depends on $y$ and $\rho_e$. In Fig.~\ref{fig:approx}, we present the mass-radius relations for some parametrizations considered in this work, explicitly comparing the two cases.

\begin{acknowledgements}
This work has been done as a part of the Project INCT-F\'isica Nuclear e Aplica\c{c}\~oes, under No. 408419/2024-5. It is also supported by Conselho Nacional de Desenvolvimento Cient\'ifico e Tecnol\'ogico (CNPq) under Grants No. 444797/2024-6 (O.L., M.D.), No. 307255/2023-9 (O.L.), No. 301779/2025-2 (M.D.), No. 401565/2023-8~(Universal - O.L., M.D.), No. 409736/2025-2~(Universal - O.L., M.D.), and Funda\c{c}\~ao de Amparo \`a Pesquisa do Estado de S\~ao Paulo (FAPESP) under Thematic Project No. 2024/17816-8 (O.L., M.D.), and Project No 2025/06216-2 (C.~B.).
\end{acknowledgements}

\bibliographystyle{apsrev4-2}
\bibliography{references}

\begin{thebibliography}{105}%
\makeatletter
\providecommand \@ifxundefined [1]{%
 \@ifx{#1\undefined}
}%
\providecommand \@ifnum [1]{%
 \ifnum #1\expandafter \@firstoftwo
 \else \expandafter \@secondoftwo
 \fi
}%
\providecommand \@ifx [1]{%
 \ifx #1\expandafter \@firstoftwo
 \else \expandafter \@secondoftwo
 \fi
}%
\providecommand \natexlab [1]{#1}%
\providecommand \enquote  [1]{``#1''}%
\providecommand \bibnamefont  [1]{#1}%
\providecommand \bibfnamefont [1]{#1}%
\providecommand \citenamefont [1]{#1}%
\providecommand \href@noop [0]{\@secondoftwo}%
\providecommand \href [0]{\begingroup \@sanitize@url \@href}%
\providecommand \@href[1]{\@@startlink{#1}\@@href}%
\providecommand \@@href[1]{\endgroup#1\@@endlink}%
\providecommand \@sanitize@url [0]{\catcode `\\12\catcode `\$12\catcode `\&12\catcode `\#12\catcode `\^12\catcode `\_12\catcode `\%12\relax}%
\providecommand \@@startlink[1]{}%
\providecommand \@@endlink[0]{}%
\providecommand \url  [0]{\begingroup\@sanitize@url \@url }%
\providecommand \@url [1]{\endgroup\@href {#1}{\urlprefix }}%
\providecommand \urlprefix  [0]{URL }%
\providecommand \Eprint [0]{\href }%
\providecommand \doibase [0]{https://doi.org/}%
\providecommand \selectlanguage [0]{\@gobble}%
\providecommand \bibinfo  [0]{\@secondoftwo}%
\providecommand \bibfield  [0]{\@secondoftwo}%
\providecommand \translation [1]{[#1]}%
\providecommand \BibitemOpen [0]{}%
\providecommand \bibitemStop [0]{}%
\providecommand \bibitemNoStop [0]{.\EOS\space}%
\providecommand \EOS [0]{\spacefactor3000\relax}%
\providecommand \BibitemShut  [1]{\csname bibitem#1\endcsname}%
\let\auto@bib@innerbib\@empty
\bibitem [{\citenamefont {et~al.}(2017)}]{Abbott_2017}%
  \BibitemOpen
  \bibfield  {author} {\bibinfo {author} {\bibfnamefont {A.}~\bibnamefont {et~al.}},\ }\href {https://doi.org/10.3847/2041-8213/aa920c} {\bibfield  {journal} {\bibinfo  {journal} {The Astrophysical Journal Letters}\ }\textbf {\bibinfo {volume} {848}},\ \bibinfo {pages} {L13} (\bibinfo {year} {2017})}\BibitemShut {NoStop}%
\bibitem [{\citenamefont {Abbott}\ \emph {et~al.}(2020)\citenamefont {Abbott}, \citenamefont {Abbott}, \citenamefont {Abbott}, \citenamefont {Abraham}, \citenamefont {Acernese} \emph {et~al.}}]{Abbott_2020-2}%
  \BibitemOpen
  \bibfield  {author} {\bibinfo {author} {\bibfnamefont {B.~P.}\ \bibnamefont {Abbott}}, \bibinfo {author} {\bibfnamefont {R.}~\bibnamefont {Abbott}}, \bibinfo {author} {\bibfnamefont {T.~D.}\ \bibnamefont {Abbott}}, \bibinfo {author} {\bibfnamefont {S.}~\bibnamefont {Abraham}}, \bibinfo {author} {\bibfnamefont {F.}~\bibnamefont {Acernese}}, \emph {et~al.},\ }\href {https://doi.org/10.3847/2041-8213/ab75f5} {\bibfield  {journal} {\bibinfo  {journal} {The Astrophysical Journal Letters}\ }\textbf {\bibinfo {volume} {892}},\ \bibinfo {pages} {L3} (\bibinfo {year} {2020})}\BibitemShut {NoStop}%
\bibitem [{\citenamefont {Riley}\ \emph {et~al.}(2019)\citenamefont {Riley}, \citenamefont {Watts}, \citenamefont {Bogdanov}, \citenamefont {Ray}, \citenamefont {Ludlam}, \citenamefont {Guillot}, \citenamefont {Arzoumanian}, \citenamefont {Baker}, \citenamefont {Bilous}, \citenamefont {Chakrabarty}, \citenamefont {Gendreau}, \citenamefont {Harding}, \citenamefont {Ho}, \citenamefont {Lattimer}, \citenamefont {Morsink},\ and\ \citenamefont {Strohmayer}}]{Riley_2019}%
  \BibitemOpen
  \bibfield  {author} {\bibinfo {author} {\bibfnamefont {T.~E.}\ \bibnamefont {Riley}}, \bibinfo {author} {\bibfnamefont {A.~L.}\ \bibnamefont {Watts}}, \bibinfo {author} {\bibfnamefont {S.}~\bibnamefont {Bogdanov}}, \bibinfo {author} {\bibfnamefont {P.~S.}\ \bibnamefont {Ray}}, \bibinfo {author} {\bibfnamefont {R.~M.}\ \bibnamefont {Ludlam}}, \bibinfo {author} {\bibfnamefont {S.}~\bibnamefont {Guillot}}, \bibinfo {author} {\bibfnamefont {Z.}~\bibnamefont {Arzoumanian}}, \bibinfo {author} {\bibfnamefont {C.~L.}\ \bibnamefont {Baker}}, \bibinfo {author} {\bibfnamefont {A.~V.}\ \bibnamefont {Bilous}}, \bibinfo {author} {\bibfnamefont {D.}~\bibnamefont {Chakrabarty}}, \bibinfo {author} {\bibfnamefont {K.~C.}\ \bibnamefont {Gendreau}}, \bibinfo {author} {\bibfnamefont {A.~K.}\ \bibnamefont {Harding}}, \bibinfo {author} {\bibfnamefont {W.~C.~G.}\ \bibnamefont {Ho}}, \bibinfo {author} {\bibfnamefont {J.~M.}\ \bibnamefont {Lattimer}}, \bibinfo {author} {\bibfnamefont {S.~M.}\ \bibnamefont {Morsink}},\ and\ \bibinfo
  {author} {\bibfnamefont {T.~E.}\ \bibnamefont {Strohmayer}},\ }\href {https://doi.org/10.3847/2041-8213/ab481c} {\bibfield  {journal} {\bibinfo  {journal} {The Astrophysical Journal Letters}\ }\textbf {\bibinfo {volume} {887}},\ \bibinfo {pages} {L21} (\bibinfo {year} {2019})}\BibitemShut {NoStop}%
\bibitem [{\citenamefont {Miller}\ \emph {et~al.}(2019)\citenamefont {Miller}, \citenamefont {Lamb}, \citenamefont {Dittmann}, \citenamefont {Bogdanov}, \citenamefont {Arzoumanian}, \citenamefont {Gendreau}, \citenamefont {Guillot}, \citenamefont {Harding}, \citenamefont {Ho}, \citenamefont {Lattimer}, \citenamefont {Ludlam}, \citenamefont {Mahmoodifar}, \citenamefont {Morsink}, \citenamefont {Ray}, \citenamefont {Strohmayer}, \citenamefont {Wood}, \citenamefont {Enoto}, \citenamefont {Foster}, \citenamefont {Okajima}, \citenamefont {Prigozhin},\ and\ \citenamefont {Soong}}]{Miller_2019}%
  \BibitemOpen
  \bibfield  {author} {\bibinfo {author} {\bibfnamefont {M.~C.}\ \bibnamefont {Miller}}, \bibinfo {author} {\bibfnamefont {F.~K.}\ \bibnamefont {Lamb}}, \bibinfo {author} {\bibfnamefont {A.~J.}\ \bibnamefont {Dittmann}}, \bibinfo {author} {\bibfnamefont {S.}~\bibnamefont {Bogdanov}}, \bibinfo {author} {\bibfnamefont {Z.}~\bibnamefont {Arzoumanian}}, \bibinfo {author} {\bibfnamefont {K.~C.}\ \bibnamefont {Gendreau}}, \bibinfo {author} {\bibfnamefont {S.}~\bibnamefont {Guillot}}, \bibinfo {author} {\bibfnamefont {A.~K.}\ \bibnamefont {Harding}}, \bibinfo {author} {\bibfnamefont {W.~C.~G.}\ \bibnamefont {Ho}}, \bibinfo {author} {\bibfnamefont {J.~M.}\ \bibnamefont {Lattimer}}, \bibinfo {author} {\bibfnamefont {R.~M.}\ \bibnamefont {Ludlam}}, \bibinfo {author} {\bibfnamefont {S.}~\bibnamefont {Mahmoodifar}}, \bibinfo {author} {\bibfnamefont {S.~M.}\ \bibnamefont {Morsink}}, \bibinfo {author} {\bibfnamefont {P.~S.}\ \bibnamefont {Ray}}, \bibinfo {author} {\bibfnamefont {T.~E.}\ \bibnamefont {Strohmayer}}, \bibinfo
  {author} {\bibfnamefont {K.~S.}\ \bibnamefont {Wood}}, \bibinfo {author} {\bibfnamefont {T.}~\bibnamefont {Enoto}}, \bibinfo {author} {\bibfnamefont {R.}~\bibnamefont {Foster}}, \bibinfo {author} {\bibfnamefont {T.}~\bibnamefont {Okajima}}, \bibinfo {author} {\bibfnamefont {G.}~\bibnamefont {Prigozhin}},\ and\ \bibinfo {author} {\bibfnamefont {Y.}~\bibnamefont {Soong}},\ }\href {https://doi.org/10.3847/2041-8213/ab50c5} {\bibfield  {journal} {\bibinfo  {journal} {The Astrophysical Journal Letters}\ }\textbf {\bibinfo {volume} {887}},\ \bibinfo {pages} {L24} (\bibinfo {year} {2019})}\BibitemShut {NoStop}%
\bibitem [{\citenamefont {Salmi}\ \emph {et~al.}(2022)\citenamefont {Salmi}, \citenamefont {Vinciguerra}, \citenamefont {Choudhury}, \citenamefont {Riley}, \citenamefont {Watts}, \citenamefont {Remillard}, \citenamefont {Ray}, \citenamefont {Bogdanov}, \citenamefont {Guillot}, \citenamefont {Arzoumanian}, \citenamefont {Chirenti}, \citenamefont {Dittmann}, \citenamefont {Gendreau}, \citenamefont {Ho}, \citenamefont {Miller}, \citenamefont {Morsink}, \citenamefont {Wadiasingh},\ and\ \citenamefont {Wolff}}]{Salmi_2022}%
  \BibitemOpen
  \bibfield  {author} {\bibinfo {author} {\bibfnamefont {T.}~\bibnamefont {Salmi}}, \bibinfo {author} {\bibfnamefont {S.}~\bibnamefont {Vinciguerra}}, \bibinfo {author} {\bibfnamefont {D.}~\bibnamefont {Choudhury}}, \bibinfo {author} {\bibfnamefont {T.~E.}\ \bibnamefont {Riley}}, \bibinfo {author} {\bibfnamefont {A.~L.}\ \bibnamefont {Watts}}, \bibinfo {author} {\bibfnamefont {R.~A.}\ \bibnamefont {Remillard}}, \bibinfo {author} {\bibfnamefont {P.~S.}\ \bibnamefont {Ray}}, \bibinfo {author} {\bibfnamefont {S.}~\bibnamefont {Bogdanov}}, \bibinfo {author} {\bibfnamefont {S.}~\bibnamefont {Guillot}}, \bibinfo {author} {\bibfnamefont {Z.}~\bibnamefont {Arzoumanian}}, \bibinfo {author} {\bibfnamefont {C.}~\bibnamefont {Chirenti}}, \bibinfo {author} {\bibfnamefont {A.~J.}\ \bibnamefont {Dittmann}}, \bibinfo {author} {\bibfnamefont {K.~C.}\ \bibnamefont {Gendreau}}, \bibinfo {author} {\bibfnamefont {W.~C.~G.}\ \bibnamefont {Ho}}, \bibinfo {author} {\bibfnamefont {M.~C.}\ \bibnamefont {Miller}}, \bibinfo {author}
  {\bibfnamefont {S.~M.}\ \bibnamefont {Morsink}}, \bibinfo {author} {\bibfnamefont {Z.}~\bibnamefont {Wadiasingh}},\ and\ \bibinfo {author} {\bibfnamefont {M.~T.}\ \bibnamefont {Wolff}},\ }\href {https://doi.org/10.3847/1538-4357/ac983d} {\bibfield  {journal} {\bibinfo  {journal} {The Astrophysical Journal}\ }\textbf {\bibinfo {volume} {941}},\ \bibinfo {pages} {150} (\bibinfo {year} {2022})}\BibitemShut {NoStop}%
\bibitem [{\citenamefont {Miller}\ \emph {et~al.}(2021)\citenamefont {Miller}, \citenamefont {Lamb}, \citenamefont {Dittmann}, \citenamefont {Bogdanov}, \citenamefont {Arzoumanian}, \citenamefont {Gendreau}, \citenamefont {Guillot}, \citenamefont {Ho}, \citenamefont {Lattimer}, \citenamefont {Loewenstein}, \citenamefont {Morsink}, \citenamefont {Ray}, \citenamefont {Wolff}, \citenamefont {Baker}, \citenamefont {Cazeau}, \citenamefont {Manthripragada}, \citenamefont {Markwardt}, \citenamefont {Okajima}, \citenamefont {Pollard}, \citenamefont {Cognard}, \citenamefont {Cromartie}, \citenamefont {Fonseca}, \citenamefont {Guillemot}, \citenamefont {Kerr}, \citenamefont {Parthasarathy}, \citenamefont {Pennucci}, \citenamefont {Ransom},\ and\ \citenamefont {Stairs}}]{Miller_2021}%
  \BibitemOpen
  \bibfield  {author} {\bibinfo {author} {\bibfnamefont {M.~C.}\ \bibnamefont {Miller}}, \bibinfo {author} {\bibfnamefont {F.~K.}\ \bibnamefont {Lamb}}, \bibinfo {author} {\bibfnamefont {A.~J.}\ \bibnamefont {Dittmann}}, \bibinfo {author} {\bibfnamefont {S.}~\bibnamefont {Bogdanov}}, \bibinfo {author} {\bibfnamefont {Z.}~\bibnamefont {Arzoumanian}}, \bibinfo {author} {\bibfnamefont {K.~C.}\ \bibnamefont {Gendreau}}, \bibinfo {author} {\bibfnamefont {S.}~\bibnamefont {Guillot}}, \bibinfo {author} {\bibfnamefont {W.~C.~G.}\ \bibnamefont {Ho}}, \bibinfo {author} {\bibfnamefont {J.~M.}\ \bibnamefont {Lattimer}}, \bibinfo {author} {\bibfnamefont {M.}~\bibnamefont {Loewenstein}}, \bibinfo {author} {\bibfnamefont {S.~M.}\ \bibnamefont {Morsink}}, \bibinfo {author} {\bibfnamefont {P.~S.}\ \bibnamefont {Ray}}, \bibinfo {author} {\bibfnamefont {M.~T.}\ \bibnamefont {Wolff}}, \bibinfo {author} {\bibfnamefont {C.~L.}\ \bibnamefont {Baker}}, \bibinfo {author} {\bibfnamefont {T.}~\bibnamefont {Cazeau}}, \bibinfo {author}
  {\bibfnamefont {S.}~\bibnamefont {Manthripragada}}, \bibinfo {author} {\bibfnamefont {C.~B.}\ \bibnamefont {Markwardt}}, \bibinfo {author} {\bibfnamefont {T.}~\bibnamefont {Okajima}}, \bibinfo {author} {\bibfnamefont {S.}~\bibnamefont {Pollard}}, \bibinfo {author} {\bibfnamefont {I.}~\bibnamefont {Cognard}}, \bibinfo {author} {\bibfnamefont {H.~T.}\ \bibnamefont {Cromartie}}, \bibinfo {author} {\bibfnamefont {E.}~\bibnamefont {Fonseca}}, \bibinfo {author} {\bibfnamefont {L.}~\bibnamefont {Guillemot}}, \bibinfo {author} {\bibfnamefont {M.}~\bibnamefont {Kerr}}, \bibinfo {author} {\bibfnamefont {A.}~\bibnamefont {Parthasarathy}}, \bibinfo {author} {\bibfnamefont {T.~T.}\ \bibnamefont {Pennucci}}, \bibinfo {author} {\bibfnamefont {S.}~\bibnamefont {Ransom}},\ and\ \bibinfo {author} {\bibfnamefont {I.}~\bibnamefont {Stairs}},\ }\href {https://doi.org/10.3847/2041-8213/ac089b} {\bibfield  {journal} {\bibinfo  {journal} {The Astrophysical Journal Letters}\ }\textbf {\bibinfo {volume} {918}},\ \bibinfo {pages}
  {L28} (\bibinfo {year} {2021})}\BibitemShut {NoStop}%
\bibitem [{\citenamefont {Vinciguerra}\ \emph {et~al.}(2024)\citenamefont {Vinciguerra}, \citenamefont {Salmi}, \citenamefont {Watts}, \citenamefont {Choudhury}, \citenamefont {Riley}, \citenamefont {Ray}, \citenamefont {Bogdanov}, \citenamefont {Kini}, \citenamefont {Guillot}, \citenamefont {Chakrabarty}, \citenamefont {Ho}, \citenamefont {Huppenkothen}, \citenamefont {Morsink}, \citenamefont {Wadiasingh},\ and\ \citenamefont {Wolff}}]{Vinciguerra_2024}%
  \BibitemOpen
  \bibfield  {author} {\bibinfo {author} {\bibfnamefont {S.}~\bibnamefont {Vinciguerra}}, \bibinfo {author} {\bibfnamefont {T.}~\bibnamefont {Salmi}}, \bibinfo {author} {\bibfnamefont {A.~L.}\ \bibnamefont {Watts}}, \bibinfo {author} {\bibfnamefont {D.}~\bibnamefont {Choudhury}}, \bibinfo {author} {\bibfnamefont {T.~E.}\ \bibnamefont {Riley}}, \bibinfo {author} {\bibfnamefont {P.~S.}\ \bibnamefont {Ray}}, \bibinfo {author} {\bibfnamefont {S.}~\bibnamefont {Bogdanov}}, \bibinfo {author} {\bibfnamefont {Y.}~\bibnamefont {Kini}}, \bibinfo {author} {\bibfnamefont {S.}~\bibnamefont {Guillot}}, \bibinfo {author} {\bibfnamefont {D.}~\bibnamefont {Chakrabarty}}, \bibinfo {author} {\bibfnamefont {W.~C.~G.}\ \bibnamefont {Ho}}, \bibinfo {author} {\bibfnamefont {D.}~\bibnamefont {Huppenkothen}}, \bibinfo {author} {\bibfnamefont {S.~M.}\ \bibnamefont {Morsink}}, \bibinfo {author} {\bibfnamefont {Z.}~\bibnamefont {Wadiasingh}},\ and\ \bibinfo {author} {\bibfnamefont {M.~T.}\ \bibnamefont {Wolff}},\ }\href
  {https://doi.org/10.3847/1538-4357/acfb83} {\bibfield  {journal} {\bibinfo  {journal} {The Astrophysical Journal}\ }\textbf {\bibinfo {volume} {961}},\ \bibinfo {pages} {62} (\bibinfo {year} {2024})}\BibitemShut {NoStop}%
\bibitem [{\citenamefont {Hen}\ \emph {et~al.}(2014)\citenamefont {Hen} \emph {et~al.}}]{sciencesrc1}%
  \BibitemOpen
  \bibfield  {author} {\bibinfo {author} {\bibfnamefont {O.}~\bibnamefont {Hen}} \emph {et~al.},\ }\href {https://doi.org/10.1126/science.1256785} {\bibfield  {journal} {\bibinfo  {journal} {Science}\ }\textbf {\bibinfo {volume} {346}},\ \bibinfo {pages} {614} (\bibinfo {year} {2014})}\BibitemShut {NoStop}%
\bibitem [{\citenamefont {{CLAS Collaboration}}(2018)}]{naturesrc2}%
  \BibitemOpen
  \bibfield  {author} {\bibinfo {author} {\bibnamefont {{CLAS Collaboration}}},\ }\href {https://doi.org/10.1038/s41586-018-0400-z} {\bibfield  {journal} {\bibinfo  {journal} {Nature}\ }\textbf {\bibinfo {volume} {560}},\ \bibinfo {pages} {617} (\bibinfo {year} {2018})}\BibitemShut {NoStop}%
\bibitem [{\citenamefont {{CLAS Collaboration}}(2019)}]{naturescr3}%
  \BibitemOpen
  \bibfield  {author} {\bibinfo {author} {\bibnamefont {{CLAS Collaboration}}},\ }\href {https://doi.org/10.1038/s41586-019-0925-9} {\bibfield  {journal} {\bibinfo  {journal} {Nature}\ }\textbf {\bibinfo {volume} {566}},\ \bibinfo {pages} {354} (\bibinfo {year} {2019})}\BibitemShut {NoStop}%
\bibitem [{\citenamefont {{Schmidt}}\ \emph {et~al.}(2020)\citenamefont {{Schmidt}} \emph {et~al.}}]{naturesrc4}%
  \BibitemOpen
  \bibfield  {author} {\bibinfo {author} {\bibfnamefont {A.}~\bibnamefont {{Schmidt}}} \emph {et~al.},\ }\href {https://doi.org/10.1038/s41586-020-2021-6} {\bibfield  {journal} {\bibinfo  {journal} {Nature}\ }\textbf {\bibinfo {volume} {578}},\ \bibinfo {pages} {540} (\bibinfo {year} {2020})}\BibitemShut {NoStop}%
\bibitem [{\citenamefont {Hen}\ \emph {et~al.}(2017)\citenamefont {Hen}, \citenamefont {Miller}, \citenamefont {Piasetzky},\ and\ \citenamefont {Weinstein}}]{hen2017}%
  \BibitemOpen
  \bibfield  {author} {\bibinfo {author} {\bibfnamefont {O.}~\bibnamefont {Hen}}, \bibinfo {author} {\bibfnamefont {G.~A.}\ \bibnamefont {Miller}}, \bibinfo {author} {\bibfnamefont {E.}~\bibnamefont {Piasetzky}},\ and\ \bibinfo {author} {\bibfnamefont {L.~B.}\ \bibnamefont {Weinstein}},\ }\href {https://doi.org/10.1103/RevModPhys.89.045002} {\bibfield  {journal} {\bibinfo  {journal} {Rev. Mod. Phys.}\ }\textbf {\bibinfo {volume} {89}},\ \bibinfo {pages} {045002} (\bibinfo {year} {2017})}\BibitemShut {NoStop}%
\bibitem [{\citenamefont {{Duer}}\ \emph {et~al.}(2019)\citenamefont {{Duer}} \emph {et~al.}}]{duer2019}%
  \BibitemOpen
  \bibfield  {author} {\bibinfo {author} {\bibfnamefont {M.}~\bibnamefont {{Duer}}} \emph {et~al.},\ }\href {https://doi.org/10.1016/j.physletb.2019.07.039} {\bibfield  {journal} {\bibinfo  {journal} {Physics Letters B}\ }\textbf {\bibinfo {volume} {797}},\ \bibinfo {eid} {134792} (\bibinfo {year} {2019})}\BibitemShut {NoStop}%
\bibitem [{\citenamefont {Cai}\ and\ \citenamefont {Li}(2016)}]{cai}%
  \BibitemOpen
  \bibfield  {author} {\bibinfo {author} {\bibfnamefont {B.-J.}\ \bibnamefont {Cai}}\ and\ \bibinfo {author} {\bibfnamefont {B.-A.}\ \bibnamefont {Li}},\ }\href {https://doi.org/10.1103/PhysRevC.93.014619} {\bibfield  {journal} {\bibinfo  {journal} {Phys. Rev. C}\ }\textbf {\bibinfo {volume} {93}},\ \bibinfo {pages} {014619} (\bibinfo {year} {2016})}\BibitemShut {NoStop}%
\bibitem [{\citenamefont {Guo}\ \emph {et~al.}(2021)\citenamefont {Guo}, \citenamefont {Li},\ and\ \citenamefont {Yong}}]{baoanli21}%
  \BibitemOpen
  \bibfield  {author} {\bibinfo {author} {\bibfnamefont {W.-M.}\ \bibnamefont {Guo}}, \bibinfo {author} {\bibfnamefont {B.-A.}\ \bibnamefont {Li}},\ and\ \bibinfo {author} {\bibfnamefont {G.-C.}\ \bibnamefont {Yong}},\ }\href {https://doi.org/10.1103/PhysRevC.104.034603} {\bibfield  {journal} {\bibinfo  {journal} {Phys. Rev. C}\ }\textbf {\bibinfo {volume} {104}},\ \bibinfo {pages} {034603} (\bibinfo {year} {2021})}\BibitemShut {NoStop}%
\bibitem [{\citenamefont {Cai}\ and\ \citenamefont {Li}(2022{\natexlab{a}})}]{baoanli22}%
  \BibitemOpen
  \bibfield  {author} {\bibinfo {author} {\bibfnamefont {B.-J.}\ \bibnamefont {Cai}}\ and\ \bibinfo {author} {\bibfnamefont {B.-A.}\ \bibnamefont {Li}},\ }\href {https://doi.org/10.1103/PhysRevC.105.064607} {\bibfield  {journal} {\bibinfo  {journal} {Phys. Rev. C}\ }\textbf {\bibinfo {volume} {105}},\ \bibinfo {pages} {064607} (\bibinfo {year} {2022}{\natexlab{a}})}\BibitemShut {NoStop}%
\bibitem [{\citenamefont {Cai}\ and\ \citenamefont {Li}(2022{\natexlab{b}})}]{baoanli-aop}%
  \BibitemOpen
  \bibfield  {author} {\bibinfo {author} {\bibfnamefont {B.-J.}\ \bibnamefont {Cai}}\ and\ \bibinfo {author} {\bibfnamefont {B.-A.}\ \bibnamefont {Li}},\ }\href {https://doi.org/https://doi.org/10.1016/j.aop.2022.169062} {\bibfield  {journal} {\bibinfo  {journal} {Annals of Physics}\ }\textbf {\bibinfo {volume} {444}},\ \bibinfo {pages} {169062} (\bibinfo {year} {2022}{\natexlab{b}})}\BibitemShut {NoStop}%
\bibitem [{\citenamefont {Carbone}\ \emph {et~al.}(2012)\citenamefont {Carbone}, \citenamefont {Polls},\ and\ \citenamefont {Rios}}]{rios1}%
  \BibitemOpen
  \bibfield  {author} {\bibinfo {author} {\bibfnamefont {A.}~\bibnamefont {Carbone}}, \bibinfo {author} {\bibfnamefont {A.}~\bibnamefont {Polls}},\ and\ \bibinfo {author} {\bibfnamefont {A.}~\bibnamefont {Rios}},\ }\href {https://doi.org/10.1209/0295-5075/97/22001} {\bibfield  {journal} {\bibinfo  {journal} {Europhysics Letters}\ }\textbf {\bibinfo {volume} {97}},\ \bibinfo {pages} {22001} (\bibinfo {year} {2012})}\BibitemShut {NoStop}%
\bibitem [{\citenamefont {Rios}\ \emph {et~al.}(2014)\citenamefont {Rios}, \citenamefont {Polls},\ and\ \citenamefont {Dickhoff}}]{rios2}%
  \BibitemOpen
  \bibfield  {author} {\bibinfo {author} {\bibfnamefont {A.}~\bibnamefont {Rios}}, \bibinfo {author} {\bibfnamefont {A.}~\bibnamefont {Polls}},\ and\ \bibinfo {author} {\bibfnamefont {W.~H.}\ \bibnamefont {Dickhoff}},\ }\href {https://doi.org/10.1103/PhysRevC.89.044303} {\bibfield  {journal} {\bibinfo  {journal} {Phys. Rev. C}\ }\textbf {\bibinfo {volume} {89}},\ \bibinfo {pages} {044303} (\bibinfo {year} {2014})}\BibitemShut {NoStop}%
\bibitem [{\citenamefont {{M{\"u}ller}}\ and\ \citenamefont {{Serot}}(1996)}]{muller-serot96}%
  \BibitemOpen
  \bibfield  {author} {\bibinfo {author} {\bibfnamefont {H.}~\bibnamefont {{M{\"u}ller}}}\ and\ \bibinfo {author} {\bibfnamefont {B.~D.}\ \bibnamefont {{Serot}}},\ }\href {https://doi.org/10.1016/0375-9474(96)00187-X} {\bibfield  {journal} {\bibinfo  {journal} {Nucl. Phys. A}\ }\textbf {\bibinfo {volume} {606}},\ \bibinfo {pages} {508} (\bibinfo {year} {1996})}\BibitemShut {NoStop}%
\bibitem [{\citenamefont {Kouvaris}(2008)}]{kouvaris2008}%
  \BibitemOpen
  \bibfield  {author} {\bibinfo {author} {\bibfnamefont {C.}~\bibnamefont {Kouvaris}},\ }\href {https://doi.org/10.1103/PhysRevD.77.023006} {\bibfield  {journal} {\bibinfo  {journal} {Phys. Rev. D}\ }\textbf {\bibinfo {volume} {77}},\ \bibinfo {pages} {023006} (\bibinfo {year} {2008})}\BibitemShut {NoStop}%
\bibitem [{\citenamefont {Leung}\ \emph {et~al.}(2011)\citenamefont {Leung}, \citenamefont {Chu},\ and\ \citenamefont {Lin}}]{leung2011}%
  \BibitemOpen
  \bibfield  {author} {\bibinfo {author} {\bibfnamefont {S.-C.}\ \bibnamefont {Leung}}, \bibinfo {author} {\bibfnamefont {M.-C.}\ \bibnamefont {Chu}},\ and\ \bibinfo {author} {\bibfnamefont {L.-M.}\ \bibnamefont {Lin}},\ }\href {https://doi.org/10.1103/PhysRevD.84.107301} {\bibfield  {journal} {\bibinfo  {journal} {Phys. Rev. D}\ }\textbf {\bibinfo {volume} {84}},\ \bibinfo {pages} {107301} (\bibinfo {year} {2011})}\BibitemShut {NoStop}%
\bibitem [{\citenamefont {Brayeur}\ and\ \citenamefont {Tinyakov}(2012)}]{Brayeur_2012}%
  \BibitemOpen
  \bibfield  {author} {\bibinfo {author} {\bibfnamefont {L.}~\bibnamefont {Brayeur}}\ and\ \bibinfo {author} {\bibfnamefont {P.}~\bibnamefont {Tinyakov}},\ }\href {https://doi.org/10.1103/PhysRevLett.109.061301} {\bibfield  {journal} {\bibinfo  {journal} {Phys. Rev. Lett.}\ }\textbf {\bibinfo {volume} {109}},\ \bibinfo {pages} {061301} (\bibinfo {year} {2012})}\BibitemShut {NoStop}%
\bibitem [{\citenamefont {Kain}(2021)}]{das2021}%
  \BibitemOpen
  \bibfield  {author} {\bibinfo {author} {\bibfnamefont {B.}~\bibnamefont {Kain}},\ }\href {https://doi.org/10.1103/PhysRevD.103.043009} {\bibfield  {journal} {\bibinfo  {journal} {Phys. Rev. D}\ }\textbf {\bibinfo {volume} {103}},\ \bibinfo {pages} {043009} (\bibinfo {year} {2021})}\BibitemShut {NoStop}%
\bibitem [{\citenamefont {Nelson}\ \emph {et~al.}(2019)\citenamefont {Nelson}, \citenamefont {Reddy},\ and\ \citenamefont {Zhou}}]{Nelson_2019}%
  \BibitemOpen
  \bibfield  {author} {\bibinfo {author} {\bibfnamefont {A.~E.}\ \bibnamefont {Nelson}}, \bibinfo {author} {\bibfnamefont {S.}~\bibnamefont {Reddy}},\ and\ \bibinfo {author} {\bibfnamefont {D.}~\bibnamefont {Zhou}},\ }\href {https://doi.org/10.1088/1475-7516/2019/07/012} {\bibfield  {journal} {\bibinfo  {journal} {Journal of Cosmology and Astroparticle Physics}\ }\textbf {\bibinfo {volume} {2019}}\bibinfo  {number} { (07)},\ \bibinfo {pages} {012}}\BibitemShut {NoStop}%
\bibitem [{\citenamefont {Mariani}\ \emph {et~al.}(2024)\citenamefont {Mariani}, \citenamefont {Albertus}, \citenamefont {Alessandroni}, \citenamefont {Orsaria}, \citenamefont {P\'erez-Garc\'ia},\ and\ \citenamefont {Ranea-Sandoval}}]{milva24}%
  \BibitemOpen
\bibfield  {number} {  }\bibfield  {author} {\bibinfo {author} {\bibfnamefont {M.}~\bibnamefont {Mariani}}, \bibinfo {author} {\bibfnamefont {C.}~\bibnamefont {Albertus}}, \bibinfo {author} {\bibfnamefont {M.~d.~R.}\ \bibnamefont {Alessandroni}}, \bibinfo {author} {\bibfnamefont {M.~G.}\ \bibnamefont {Orsaria}}, \bibinfo {author} {\bibfnamefont {M.~A.}\ \bibnamefont {P\'erez-Garc\'ia}},\ and\ \bibinfo {author} {\bibfnamefont {I.~F.}\ \bibnamefont {Ranea-Sandoval}},\ }\href {https://doi.org/10.1093/mnras/stad3658} {\bibfield  {journal} {\bibinfo  {journal} {Monthly Notices of the Royal Astronomical Society}\ }\textbf {\bibinfo {volume} {527}},\ \bibinfo {pages} {6795} (\bibinfo {year} {2024})}\BibitemShut {NoStop}%
\bibitem [{\citenamefont {Goldman}\ and\ \citenamefont {Nussinov}(1989)}]{goldman1989}%
  \BibitemOpen
  \bibfield  {author} {\bibinfo {author} {\bibfnamefont {I.}~\bibnamefont {Goldman}}\ and\ \bibinfo {author} {\bibfnamefont {S.}~\bibnamefont {Nussinov}},\ }\href {https://doi.org/10.1103/PhysRevD.40.3221} {\bibfield  {journal} {\bibinfo  {journal} {Phys. Rev. D}\ }\textbf {\bibinfo {volume} {40}},\ \bibinfo {pages} {3221} (\bibinfo {year} {1989})}\BibitemShut {NoStop}%
\bibitem [{\citenamefont {Sandin}\ and\ \citenamefont {Ciarcelluti}(2009)}]{sandin2009}%
  \BibitemOpen
  \bibfield  {author} {\bibinfo {author} {\bibfnamefont {F.}~\bibnamefont {Sandin}}\ and\ \bibinfo {author} {\bibfnamefont {P.}~\bibnamefont {Ciarcelluti}},\ }\href {https://doi.org/https://doi.org/10.1016/j.astropartphys.2009.09.005} {\bibfield  {journal} {\bibinfo  {journal} {Astroparticle Physics}\ }\textbf {\bibinfo {volume} {32}},\ \bibinfo {pages} {278} (\bibinfo {year} {2009})}\BibitemShut {NoStop}%
\bibitem [{\citenamefont {Tolos}\ and\ \citenamefont {Schaffner-Bielich}(2015)}]{tolos2015dark}%
  \BibitemOpen
  \bibfield  {author} {\bibinfo {author} {\bibfnamefont {L.}~\bibnamefont {Tolos}}\ and\ \bibinfo {author} {\bibfnamefont {J.}~\bibnamefont {Schaffner-Bielich}},\ }\href {https://doi.org/10.1103/PhysRevD.92.123002} {\bibfield  {journal} {\bibinfo  {journal} {Phys. Rev. D}\ }\textbf {\bibinfo {volume} {92}},\ \bibinfo {pages} {123002} (\bibinfo {year} {2015})}\BibitemShut {NoStop}%
\bibitem [{\citenamefont {Dengler}\ \emph {et~al.}(2021)\citenamefont {Dengler}, \citenamefont {Schaffner-Bielich},\ and\ \citenamefont {Tolos}}]{dengler2021erratum}%
  \BibitemOpen
  \bibfield  {author} {\bibinfo {author} {\bibfnamefont {Y.}~\bibnamefont {Dengler}}, \bibinfo {author} {\bibfnamefont {J.}~\bibnamefont {Schaffner-Bielich}},\ and\ \bibinfo {author} {\bibfnamefont {L.}~\bibnamefont {Tolos}},\ }\href {https://doi.org/10.1103/PhysRevD.103.109901} {\bibfield  {journal} {\bibinfo  {journal} {Phys. Rev. D}\ }\textbf {\bibinfo {volume} {103}},\ \bibinfo {pages} {109901} (\bibinfo {year} {2021})}\BibitemShut {NoStop}%
\bibitem [{\citenamefont {Dengler}\ \emph {et~al.}(2022)\citenamefont {Dengler}, \citenamefont {Schaffner-Bielich},\ and\ \citenamefont {Tolos}}]{dengler2022second}%
  \BibitemOpen
  \bibfield  {author} {\bibinfo {author} {\bibfnamefont {Y.}~\bibnamefont {Dengler}}, \bibinfo {author} {\bibfnamefont {J.}~\bibnamefont {Schaffner-Bielich}},\ and\ \bibinfo {author} {\bibfnamefont {L.}~\bibnamefont {Tolos}},\ }\href {https://doi.org/10.1103/PhysRevD.105.043013} {\bibfield  {journal} {\bibinfo  {journal} {Phys. Rev. D}\ }\textbf {\bibinfo {volume} {105}},\ \bibinfo {pages} {043013} (\bibinfo {year} {2022})}\BibitemShut {NoStop}%
\bibitem [{\citenamefont {Biesdorf}\ \emph {et~al.}(2025)\citenamefont {Biesdorf}, \citenamefont {Schaffner-Bielich},\ and\ \citenamefont {Tolos}}]{carline25}%
  \BibitemOpen
  \bibfield  {author} {\bibinfo {author} {\bibfnamefont {C.}~\bibnamefont {Biesdorf}}, \bibinfo {author} {\bibfnamefont {J.}~\bibnamefont {Schaffner-Bielich}},\ and\ \bibinfo {author} {\bibfnamefont {L.}~\bibnamefont {Tolos}},\ }\href {https://doi.org/10.1103/PhysRevD.111.083038} {\bibfield  {journal} {\bibinfo  {journal} {Phys. Rev. D}\ }\textbf {\bibinfo {volume} {111}},\ \bibinfo {pages} {083038} (\bibinfo {year} {2025})}\BibitemShut {NoStop}%
\bibitem [{\citenamefont {Li}\ \emph {et~al.}(2008)\citenamefont {Li}, \citenamefont {Chen},\ and\ \citenamefont {Ko}}]{baoanli08}%
  \BibitemOpen
  \bibfield  {author} {\bibinfo {author} {\bibfnamefont {B.-A.}\ \bibnamefont {Li}}, \bibinfo {author} {\bibfnamefont {L.-W.}\ \bibnamefont {Chen}},\ and\ \bibinfo {author} {\bibfnamefont {C.~M.}\ \bibnamefont {Ko}},\ }\href {https://doi.org/https://doi.org/10.1016/j.physrep.2008.04.005} {\bibfield  {journal} {\bibinfo  {journal} {Phys. Rep.}\ }\textbf {\bibinfo {volume} {464}},\ \bibinfo {pages} {113} (\bibinfo {year} {2008})}\BibitemShut {NoStop}%
\bibitem [{\citenamefont {Dutra}\ \emph {et~al.}(2014)\citenamefont {Dutra} \emph {et~al.}}]{dutra14}%
  \BibitemOpen
  \bibfield  {author} {\bibinfo {author} {\bibfnamefont {M.}~\bibnamefont {Dutra}} \emph {et~al.},\ }\href {https://doi.org/10.1103/PhysRevC.90.055203} {\bibfield  {journal} {\bibinfo  {journal} {Phys. Rev. C}\ }\textbf {\bibinfo {volume} {90}},\ \bibinfo {pages} {055203} (\bibinfo {year} {2014})}\BibitemShut {NoStop}%
\bibitem [{\citenamefont {Sun}\ \emph {et~al.}(2024)\citenamefont {Sun}, \citenamefont {Bhattiprolu},\ and\ \citenamefont {Lattimer}}]{lattimer24}%
  \BibitemOpen
  \bibfield  {author} {\bibinfo {author} {\bibfnamefont {B.}~\bibnamefont {Sun}}, \bibinfo {author} {\bibfnamefont {S.}~\bibnamefont {Bhattiprolu}},\ and\ \bibinfo {author} {\bibfnamefont {J.~M.}\ \bibnamefont {Lattimer}},\ }\href {https://doi.org/10.1103/PhysRevC.109.055801} {\bibfield  {journal} {\bibinfo  {journal} {Phys. Rev. C}\ }\textbf {\bibinfo {volume} {109}},\ \bibinfo {pages} {055801} (\bibinfo {year} {2024})}\BibitemShut {NoStop}%
\bibitem [{\citenamefont {Rios}\ \emph {et~al.}(2009)\citenamefont {Rios}, \citenamefont {Polls},\ and\ \citenamefont {Dickhoff}}]{green}%
  \BibitemOpen
  \bibfield  {author} {\bibinfo {author} {\bibfnamefont {A.}~\bibnamefont {Rios}}, \bibinfo {author} {\bibfnamefont {A.}~\bibnamefont {Polls}},\ and\ \bibinfo {author} {\bibfnamefont {W.~H.}\ \bibnamefont {Dickhoff}},\ }\href {https://doi.org/10.1103/PhysRevC.79.064308} {\bibfield  {journal} {\bibinfo  {journal} {Phys. Rev. C}\ }\textbf {\bibinfo {volume} {79}},\ \bibinfo {pages} {064308} (\bibinfo {year} {2009})}\BibitemShut {NoStop}%
\bibitem [{\citenamefont {Yin}\ \emph {et~al.}(2013)\citenamefont {Yin}, \citenamefont {Li}, \citenamefont {Wang},\ and\ \citenamefont {Zuo}}]{bhf}%
  \BibitemOpen
  \bibfield  {author} {\bibinfo {author} {\bibfnamefont {P.}~\bibnamefont {Yin}}, \bibinfo {author} {\bibfnamefont {J.-Y.}\ \bibnamefont {Li}}, \bibinfo {author} {\bibfnamefont {P.}~\bibnamefont {Wang}},\ and\ \bibinfo {author} {\bibfnamefont {W.}~\bibnamefont {Zuo}},\ }\href {https://doi.org/10.1103/PhysRevC.87.014314} {\bibfield  {journal} {\bibinfo  {journal} {Phys. Rev. C}\ }\textbf {\bibinfo {volume} {87}},\ \bibinfo {pages} {014314} (\bibinfo {year} {2013})}\BibitemShut {NoStop}%
\bibitem [{\citenamefont {Louren\ifmmode~\mbox{\c{c}}\else \c{c}\fi{}o}\ \emph {et~al.}(2024)\citenamefont {Louren\ifmmode~\mbox{\c{c}}\else \c{c}\fi{}o}, \citenamefont {Dutra},\ and\ \citenamefont {Margueron}}]{jerome24}%
  \BibitemOpen
  \bibfield  {author} {\bibinfo {author} {\bibfnamefont {O.}~\bibnamefont {Louren\ifmmode~\mbox{\c{c}}\else \c{c}\fi{}o}}, \bibinfo {author} {\bibfnamefont {M.}~\bibnamefont {Dutra}},\ and\ \bibinfo {author} {\bibfnamefont {J.}~\bibnamefont {Margueron}},\ }\href {https://doi.org/10.1103/PhysRevC.109.055202} {\bibfield  {journal} {\bibinfo  {journal} {Phys. Rev. C}\ }\textbf {\bibinfo {volume} {109}},\ \bibinfo {pages} {055202} (\bibinfo {year} {2024})}\BibitemShut {NoStop}%
\bibitem [{\citenamefont {Souza}\ \emph {et~al.}(2020)\citenamefont {Souza}, \citenamefont {Dutra}, \citenamefont {Lenzi},\ and\ \citenamefont {Louren\ifmmode~\mbox{\c{c}}\else \c{c}\fi{}o}}]{souza2020}%
  \BibitemOpen
  \bibfield  {author} {\bibinfo {author} {\bibfnamefont {L.~A.}\ \bibnamefont {Souza}}, \bibinfo {author} {\bibfnamefont {M.}~\bibnamefont {Dutra}}, \bibinfo {author} {\bibfnamefont {C.~H.}\ \bibnamefont {Lenzi}},\ and\ \bibinfo {author} {\bibfnamefont {O.}~\bibnamefont {Louren\ifmmode~\mbox{\c{c}}\else \c{c}\fi{}o}},\ }\href {https://doi.org/10.1103/PhysRevC.101.065202} {\bibfield  {journal} {\bibinfo  {journal} {Phys. Rev. C}\ }\textbf {\bibinfo {volume} {101}},\ \bibinfo {pages} {065202} (\bibinfo {year} {2020})}\BibitemShut {NoStop}%
\bibitem [{\citenamefont {Dutra}\ \emph {et~al.}(2022)\citenamefont {Dutra}, \citenamefont {Lenzi},\ and\ \citenamefont {Lourenço}}]{dutra_mnras22}%
  \BibitemOpen
  \bibfield  {author} {\bibinfo {author} {\bibfnamefont {M.}~\bibnamefont {Dutra}}, \bibinfo {author} {\bibfnamefont {C.~H.}\ \bibnamefont {Lenzi}},\ and\ \bibinfo {author} {\bibfnamefont {O.}~\bibnamefont {Lourenço}},\ }\href {https://doi.org/10.1093/mnras/stac2986} {\bibfield  {journal} {\bibinfo  {journal} {Monthly Notices of the Royal Astronomical Society}\ }\textbf {\bibinfo {volume} {517}},\ \bibinfo {pages} {4265} (\bibinfo {year} {2022})},\ \Eprint {https://arxiv.org/abs/https://academic.oup.com/mnras/article-pdf/517/3/4265/46711837/stac2986.pdf} {https://academic.oup.com/mnras/article-pdf/517/3/4265/46711837/stac2986.pdf} \BibitemShut {NoStop}%
\bibitem [{\citenamefont {Louren\ifmmode~\mbox{\c{c}}\else \c{c}\fi{}o}\ \emph {et~al.}(2022{\natexlab{a}})\citenamefont {Louren\ifmmode~\mbox{\c{c}}\else \c{c}\fi{}o}, \citenamefont {Frederico},\ and\ \citenamefont {Dutra}}]{lourenco_prd2022_1}%
  \BibitemOpen
  \bibfield  {author} {\bibinfo {author} {\bibfnamefont {O.}~\bibnamefont {Louren\ifmmode~\mbox{\c{c}}\else \c{c}\fi{}o}}, \bibinfo {author} {\bibfnamefont {T.}~\bibnamefont {Frederico}},\ and\ \bibinfo {author} {\bibfnamefont {M.}~\bibnamefont {Dutra}},\ }\href {https://doi.org/10.1103/PhysRevD.105.023008} {\bibfield  {journal} {\bibinfo  {journal} {Phys. Rev. D}\ }\textbf {\bibinfo {volume} {105}},\ \bibinfo {pages} {023008} (\bibinfo {year} {2022}{\natexlab{a}})}\BibitemShut {NoStop}%
\bibitem [{\citenamefont {Louren\ifmmode~\mbox{\c{c}}\else \c{c}\fi{}o}\ \emph {et~al.}(2022{\natexlab{b}})\citenamefont {Louren\ifmmode~\mbox{\c{c}}\else \c{c}\fi{}o}, \citenamefont {Lenzi}, \citenamefont {Frederico},\ and\ \citenamefont {Dutra}}]{lourenco_prd2022_2}%
  \BibitemOpen
  \bibfield  {author} {\bibinfo {author} {\bibfnamefont {O.}~\bibnamefont {Louren\ifmmode~\mbox{\c{c}}\else \c{c}\fi{}o}}, \bibinfo {author} {\bibfnamefont {C.~H.}\ \bibnamefont {Lenzi}}, \bibinfo {author} {\bibfnamefont {T.}~\bibnamefont {Frederico}},\ and\ \bibinfo {author} {\bibfnamefont {M.}~\bibnamefont {Dutra}},\ }\href {https://doi.org/10.1103/PhysRevD.106.043010} {\bibfield  {journal} {\bibinfo  {journal} {Phys. Rev. D}\ }\textbf {\bibinfo {volume} {106}},\ \bibinfo {pages} {043010} (\bibinfo {year} {2022}{\natexlab{b}})}\BibitemShut {NoStop}%
\bibitem [{\citenamefont {Pelicer}\ \emph {et~al.}(2023)\citenamefont {Pelicer}, \citenamefont {Menezes}, \citenamefont {Dutra},\ and\ \citenamefont {Louren\c{c}o}}]{pelicer2023}%
  \BibitemOpen
  \bibfield  {author} {\bibinfo {author} {\bibfnamefont {M.~R.}\ \bibnamefont {Pelicer}}, \bibinfo {author} {\bibfnamefont {D.~P.}\ \bibnamefont {Menezes}}, \bibinfo {author} {\bibfnamefont {M.}~\bibnamefont {Dutra}},\ and\ \bibinfo {author} {\bibfnamefont {O.}~\bibnamefont {Louren\c{c}o}},\ }\bibfield  {journal} {\bibinfo  {journal} {The European Physical Journal A}\ }\textbf {\bibinfo {volume} {59}},\ \href {https://doi.org/10.1140/epja/s10050-023-01122-4} {10.1140/epja/s10050-023-01122-4} (\bibinfo {year} {2023})\BibitemShut {NoStop}%
\bibitem [{\citenamefont {Lalazissis}\ \emph {et~al.}(2009)\citenamefont {Lalazissis}, \citenamefont {Karatzikos}, \citenamefont {Fossion}, \citenamefont {Arteaga}, \citenamefont {Afanasjev},\ and\ \citenamefont {Ring}}]{nl3s}%
  \BibitemOpen
  \bibfield  {author} {\bibinfo {author} {\bibfnamefont {G.}~\bibnamefont {Lalazissis}}, \bibinfo {author} {\bibfnamefont {S.}~\bibnamefont {Karatzikos}}, \bibinfo {author} {\bibfnamefont {R.}~\bibnamefont {Fossion}}, \bibinfo {author} {\bibfnamefont {D.~P.}\ \bibnamefont {Arteaga}}, \bibinfo {author} {\bibfnamefont {A.}~\bibnamefont {Afanasjev}},\ and\ \bibinfo {author} {\bibfnamefont {P.}~\bibnamefont {Ring}},\ }\href {https://doi.org/https://doi.org/10.1016/j.physletb.2008.11.070} {\bibfield  {journal} {\bibinfo  {journal} {Physics Letters B}\ }\textbf {\bibinfo {volume} {671}},\ \bibinfo {pages} {36} (\bibinfo {year} {2009})}\BibitemShut {NoStop}%
\bibitem [{\citenamefont {Carlson}\ \emph {et~al.}(2023)\citenamefont {Carlson}, \citenamefont {Dutra}, \citenamefont {Louren\ifmmode~\mbox{\c{c}}\else \c{c}\fi{}o},\ and\ \citenamefont {Margueron}}]{brett-jerome}%
  \BibitemOpen
  \bibfield  {author} {\bibinfo {author} {\bibfnamefont {B.~V.}\ \bibnamefont {Carlson}}, \bibinfo {author} {\bibfnamefont {M.}~\bibnamefont {Dutra}}, \bibinfo {author} {\bibfnamefont {O.}~\bibnamefont {Louren\ifmmode~\mbox{\c{c}}\else \c{c}\fi{}o}},\ and\ \bibinfo {author} {\bibfnamefont {J.}~\bibnamefont {Margueron}},\ }\href {https://doi.org/10.1103/PhysRevC.107.035805} {\bibfield  {journal} {\bibinfo  {journal} {Phys. Rev. C}\ }\textbf {\bibinfo {volume} {107}},\ \bibinfo {pages} {035805} (\bibinfo {year} {2023})}\BibitemShut {NoStop}%
\bibitem [{\citenamefont {Centelles}\ \emph {et~al.}(1998)\citenamefont {Centelles}, \citenamefont {{Del Estal}},\ and\ \citenamefont {Viñas}}]{nlm}%
  \BibitemOpen
  \bibfield  {author} {\bibinfo {author} {\bibfnamefont {M.}~\bibnamefont {Centelles}}, \bibinfo {author} {\bibfnamefont {M.}~\bibnamefont {{Del Estal}}},\ and\ \bibinfo {author} {\bibfnamefont {X.}~\bibnamefont {Viñas}},\ }\href {https://doi.org/https://doi.org/10.1016/S0375-9474(98)00167-5} {\bibfield  {journal} {\bibinfo  {journal} {Nuclear Physics A}\ }\textbf {\bibinfo {volume} {635}},\ \bibinfo {pages} {193} (\bibinfo {year} {1998})}\BibitemShut {NoStop}%
\bibitem [{\citenamefont {Glendenning}\ and\ \citenamefont {Moszkowski}(1991)}]{gm3}%
  \BibitemOpen
  \bibfield  {author} {\bibinfo {author} {\bibfnamefont {N.~K.}\ \bibnamefont {Glendenning}}\ and\ \bibinfo {author} {\bibfnamefont {S.~A.}\ \bibnamefont {Moszkowski}},\ }\href {https://doi.org/10.1103/PhysRevLett.67.2414} {\bibfield  {journal} {\bibinfo  {journal} {Phys. Rev. Lett.}\ }\textbf {\bibinfo {volume} {67}},\ \bibinfo {pages} {2414} (\bibinfo {year} {1991})}\BibitemShut {NoStop}%
\bibitem [{\citenamefont {Long}\ \emph {et~al.}(2004)\citenamefont {Long}, \citenamefont {Meng}, \citenamefont {Giai},\ and\ \citenamefont {Zhou}}]{pk1}%
  \BibitemOpen
  \bibfield  {author} {\bibinfo {author} {\bibfnamefont {W.}~\bibnamefont {Long}}, \bibinfo {author} {\bibfnamefont {J.}~\bibnamefont {Meng}}, \bibinfo {author} {\bibfnamefont {N.~V.}\ \bibnamefont {Giai}},\ and\ \bibinfo {author} {\bibfnamefont {S.-G.}\ \bibnamefont {Zhou}},\ }\href {https://doi.org/10.1103/PhysRevC.69.034319} {\bibfield  {journal} {\bibinfo  {journal} {Phys. Rev. C}\ }\textbf {\bibinfo {volume} {69}},\ \bibinfo {pages} {034319} (\bibinfo {year} {2004})}\BibitemShut {NoStop}%
\bibitem [{\citenamefont {Toki}\ \emph {et~al.}(1995)\citenamefont {Toki}, \citenamefont {Hirata}, \citenamefont {Sugahara}, \citenamefont {Sumiyoshi},\ and\ \citenamefont {Tanihata}}]{tma}%
  \BibitemOpen
  \bibfield  {author} {\bibinfo {author} {\bibfnamefont {H.}~\bibnamefont {Toki}}, \bibinfo {author} {\bibfnamefont {D.}~\bibnamefont {Hirata}}, \bibinfo {author} {\bibfnamefont {Y.}~\bibnamefont {Sugahara}}, \bibinfo {author} {\bibfnamefont {K.}~\bibnamefont {Sumiyoshi}},\ and\ \bibinfo {author} {\bibfnamefont {I.}~\bibnamefont {Tanihata}},\ }\href {https://doi.org/https://doi.org/10.1016/0375-9474(95)00161-S} {\bibfield  {journal} {\bibinfo  {journal} {Nuclear Physics A}\ }\textbf {\bibinfo {volume} {588}},\ \bibinfo {pages} {c357} (\bibinfo {year} {1995})},\ \bibinfo {note} {proceedings of the Fifth International Symposium on Physics of Unstable Nuclei}\BibitemShut {NoStop}%
\bibitem [{\citenamefont {Bunta}\ and\ \citenamefont {Gmuca}(2003)}]{mb}%
  \BibitemOpen
  \bibfield  {author} {\bibinfo {author} {\bibfnamefont {J.~K. c.~v.}\ \bibnamefont {Bunta}}\ and\ \bibinfo {author} {\bibfnamefont {i.~c.~v.}\ \bibnamefont {Gmuca}},\ }\href {https://doi.org/10.1103/PhysRevC.68.054318} {\bibfield  {journal} {\bibinfo  {journal} {Phys. Rev. C}\ }\textbf {\bibinfo {volume} {68}},\ \bibinfo {pages} {054318} (\bibinfo {year} {2003})}\BibitemShut {NoStop}%
\bibitem [{\citenamefont {Tolman}(1939)}]{tov39}%
  \BibitemOpen
  \bibfield  {author} {\bibinfo {author} {\bibfnamefont {R.~C.}\ \bibnamefont {Tolman}},\ }\href {https://doi.org/10.1103/PhysRev.55.364} {\bibfield  {journal} {\bibinfo  {journal} {Phys. Rev.}\ }\textbf {\bibinfo {volume} {55}},\ \bibinfo {pages} {364} (\bibinfo {year} {1939})}\BibitemShut {NoStop}%
\bibitem [{\citenamefont {Oppenheimer}\ and\ \citenamefont {Volkoff}(1939)}]{tov39a}%
  \BibitemOpen
  \bibfield  {author} {\bibinfo {author} {\bibfnamefont {J.~R.}\ \bibnamefont {Oppenheimer}}\ and\ \bibinfo {author} {\bibfnamefont {G.~M.}\ \bibnamefont {Volkoff}},\ }\href {https://doi.org/10.1103/PhysRev.55.374} {\bibfield  {journal} {\bibinfo  {journal} {Phys. Rev.}\ }\textbf {\bibinfo {volume} {55}},\ \bibinfo {pages} {374} (\bibinfo {year} {1939})}\BibitemShut {NoStop}%
\bibitem [{\citenamefont {Lopez-Honorez}\ \emph {et~al.}(2012)\citenamefont {Lopez-Honorez}, \citenamefont {Schwetz},\ and\ \citenamefont {Zupan}}]{higgsportal1}%
  \BibitemOpen
  \bibfield  {author} {\bibinfo {author} {\bibfnamefont {L.}~\bibnamefont {Lopez-Honorez}}, \bibinfo {author} {\bibfnamefont {T.}~\bibnamefont {Schwetz}},\ and\ \bibinfo {author} {\bibfnamefont {J.}~\bibnamefont {Zupan}},\ }\href {https://doi.org/https://doi.org/10.1016/j.physletb.2012.07.017} {\bibfield  {journal} {\bibinfo  {journal} {Physics Letters B}\ }\textbf {\bibinfo {volume} {716}},\ \bibinfo {pages} {179} (\bibinfo {year} {2012})}\BibitemShut {NoStop}%
\bibitem [{\citenamefont {Arcadi}\ \emph {et~al.}(2020)\citenamefont {Arcadi}, \citenamefont {Djouadi},\ and\ \citenamefont {Raidal}}]{higgsportal2}%
  \BibitemOpen
  \bibfield  {author} {\bibinfo {author} {\bibfnamefont {G.}~\bibnamefont {Arcadi}}, \bibinfo {author} {\bibfnamefont {A.}~\bibnamefont {Djouadi}},\ and\ \bibinfo {author} {\bibfnamefont {M.}~\bibnamefont {Raidal}},\ }\href {https://doi.org/https://doi.org/10.1016/j.physrep.2019.11.003} {\bibfield  {journal} {\bibinfo  {journal} {Physics Reports}\ }\textbf {\bibinfo {volume} {842}},\ \bibinfo {pages} {1} (\bibinfo {year} {2020})},\ \bibinfo {note} {dark Matter through the Higgs portal}\BibitemShut {NoStop}%
\bibitem [{\citenamefont {Das}\ \emph {et~al.}(2022{\natexlab{a}})\citenamefont {Das}, \citenamefont {Kumar}, \citenamefont {Kumar},\ and\ \citenamefont {Patra}}]{rmfdm13}%
  \BibitemOpen
  \bibfield  {author} {\bibinfo {author} {\bibfnamefont {H.~C.}\ \bibnamefont {Das}}, \bibinfo {author} {\bibfnamefont {A.}~\bibnamefont {Kumar}}, \bibinfo {author} {\bibfnamefont {B.}~\bibnamefont {Kumar}},\ and\ \bibinfo {author} {\bibfnamefont {S.~K.}\ \bibnamefont {Patra}},\ }\href {https://doi.org/10.3390/galaxies10010014} {\bibfield  {journal} {\bibinfo  {journal} {Galaxies}\ }\textbf {\bibinfo {volume} {10}},\ \bibinfo {pages} {14} (\bibinfo {year} {2022}{\natexlab{a}})}\BibitemShut {NoStop}%
\bibitem [{\citenamefont {Panotopoulos}\ and\ \citenamefont {Lopes}(2017)}]{rmfdm2}%
  \BibitemOpen
  \bibfield  {author} {\bibinfo {author} {\bibfnamefont {G.}~\bibnamefont {Panotopoulos}}\ and\ \bibinfo {author} {\bibfnamefont {I.}~\bibnamefont {Lopes}},\ }\href {https://doi.org/10.1103/PhysRevD.96.083004} {\bibfield  {journal} {\bibinfo  {journal} {Phys. Rev. D}\ }\textbf {\bibinfo {volume} {96}},\ \bibinfo {pages} {083004} (\bibinfo {year} {2017})}\BibitemShut {NoStop}%
\bibitem [{\citenamefont {Das}\ \emph {et~al.}(2019)\citenamefont {Das}, \citenamefont {Malik},\ and\ \citenamefont {Nayak}}]{rmfdm3}%
  \BibitemOpen
  \bibfield  {author} {\bibinfo {author} {\bibfnamefont {A.}~\bibnamefont {Das}}, \bibinfo {author} {\bibfnamefont {T.}~\bibnamefont {Malik}},\ and\ \bibinfo {author} {\bibfnamefont {A.~C.}\ \bibnamefont {Nayak}},\ }\href {https://doi.org/10.1103/PhysRevD.99.043016} {\bibfield  {journal} {\bibinfo  {journal} {Phys. Rev. D}\ }\textbf {\bibinfo {volume} {99}},\ \bibinfo {pages} {043016} (\bibinfo {year} {2019})}\BibitemShut {NoStop}%
\bibitem [{\citenamefont {Quddus}\ \emph {et~al.}(2020)\citenamefont {Quddus}, \citenamefont {Panotopoulos}, \citenamefont {Kumar}, \citenamefont {Ahmad},\ and\ \citenamefont {Patra}}]{abdul}%
  \BibitemOpen
  \bibfield  {author} {\bibinfo {author} {\bibfnamefont {A.}~\bibnamefont {Quddus}}, \bibinfo {author} {\bibfnamefont {G.}~\bibnamefont {Panotopoulos}}, \bibinfo {author} {\bibfnamefont {B.}~\bibnamefont {Kumar}}, \bibinfo {author} {\bibfnamefont {S.}~\bibnamefont {Ahmad}},\ and\ \bibinfo {author} {\bibfnamefont {S.~K.}\ \bibnamefont {Patra}},\ }\href {https://doi.org/10.1088/1361-6471/ab9d36} {\bibfield  {journal} {\bibinfo  {journal} {Journal of Physics G: Nuclear and Particle Physics}\ }\textbf {\bibinfo {volume} {47}},\ \bibinfo {pages} {095202} (\bibinfo {year} {2020})}\BibitemShut {NoStop}%
\bibitem [{\citenamefont {Das}\ \emph {et~al.}(2020)\citenamefont {Das}, \citenamefont {Kumar}, \citenamefont {Kumar}, \citenamefont {Biswal}, \citenamefont {Nakatsukasa}, \citenamefont {Li},\ and\ \citenamefont {Patra}}]{rmfdm6}%
  \BibitemOpen
  \bibfield  {author} {\bibinfo {author} {\bibfnamefont {H.~C.}\ \bibnamefont {Das}}, \bibinfo {author} {\bibfnamefont {A.}~\bibnamefont {Kumar}}, \bibinfo {author} {\bibfnamefont {B.}~\bibnamefont {Kumar}}, \bibinfo {author} {\bibfnamefont {S.~K.}\ \bibnamefont {Biswal}}, \bibinfo {author} {\bibfnamefont {T.}~\bibnamefont {Nakatsukasa}}, \bibinfo {author} {\bibfnamefont {A.}~\bibnamefont {Li}},\ and\ \bibinfo {author} {\bibfnamefont {S.~K.}\ \bibnamefont {Patra}},\ }\href {https://doi.org/10.1093/mnras/staa1435} {\bibfield  {journal} {\bibinfo  {journal} {Monthly Notices of the Royal Astronomical Society}\ }\textbf {\bibinfo {volume} {495}},\ \bibinfo {pages} {4893} (\bibinfo {year} {2020})}\BibitemShut {NoStop}%
\bibitem [{\citenamefont {Das}\ \emph {et~al.}(2021{\natexlab{a}})\citenamefont {Das}, \citenamefont {Kumar},\ and\ \citenamefont {Patra}}]{rmfdm11}%
  \BibitemOpen
  \bibfield  {author} {\bibinfo {author} {\bibfnamefont {H.~C.}\ \bibnamefont {Das}}, \bibinfo {author} {\bibfnamefont {A.}~\bibnamefont {Kumar}},\ and\ \bibinfo {author} {\bibfnamefont {S.~K.}\ \bibnamefont {Patra}},\ }\href {https://doi.org/10.1103/PhysRevD.104.063028} {\bibfield  {journal} {\bibinfo  {journal} {Phys. Rev. D}\ }\textbf {\bibinfo {volume} {104}},\ \bibinfo {pages} {063028} (\bibinfo {year} {2021}{\natexlab{a}})}\BibitemShut {NoStop}%
\bibitem [{\citenamefont {Das}\ \emph {et~al.}(2021{\natexlab{b}})\citenamefont {Das}, \citenamefont {Kumar}, \citenamefont {Biswal},\ and\ \citenamefont {Patra}}]{rmfdm10}%
  \BibitemOpen
  \bibfield  {author} {\bibinfo {author} {\bibfnamefont {H.~C.}\ \bibnamefont {Das}}, \bibinfo {author} {\bibfnamefont {A.}~\bibnamefont {Kumar}}, \bibinfo {author} {\bibfnamefont {S.~K.}\ \bibnamefont {Biswal}},\ and\ \bibinfo {author} {\bibfnamefont {S.~K.}\ \bibnamefont {Patra}},\ }\href {https://doi.org/10.1103/PhysRevD.104.123006} {\bibfield  {journal} {\bibinfo  {journal} {Phys. Rev. D}\ }\textbf {\bibinfo {volume} {104}},\ \bibinfo {pages} {123006} (\bibinfo {year} {2021}{\natexlab{b}})}\BibitemShut {NoStop}%
\bibitem [{\citenamefont {Das}\ \emph {et~al.}(2021{\natexlab{c}})\citenamefont {Das}, \citenamefont {Kumar},\ and\ \citenamefont {Patra}}]{rmfdm8}%
  \BibitemOpen
  \bibfield  {author} {\bibinfo {author} {\bibfnamefont {H.~C.}\ \bibnamefont {Das}}, \bibinfo {author} {\bibfnamefont {A.}~\bibnamefont {Kumar}},\ and\ \bibinfo {author} {\bibfnamefont {S.~K.}\ \bibnamefont {Patra}},\ }\href {https://doi.org/10.1093/mnras/stab2387} {\bibfield  {journal} {\bibinfo  {journal} {Monthly Notices of the Royal Astronomical Society}\ }\textbf {\bibinfo {volume} {507}},\ \bibinfo {pages} {4053} (\bibinfo {year} {2021}{\natexlab{c}})}\BibitemShut {NoStop}%
\bibitem [{\citenamefont {Das}\ \emph {et~al.}(2021{\natexlab{d}})\citenamefont {Das}, \citenamefont {Kumar}, \citenamefont {Kumar}, \citenamefont {Biswal},\ and\ \citenamefont {Patra}}]{rmfdm7}%
  \BibitemOpen
  \bibfield  {author} {\bibinfo {author} {\bibfnamefont {H.}~\bibnamefont {Das}}, \bibinfo {author} {\bibfnamefont {A.}~\bibnamefont {Kumar}}, \bibinfo {author} {\bibfnamefont {B.}~\bibnamefont {Kumar}}, \bibinfo {author} {\bibfnamefont {S.}~\bibnamefont {Biswal}},\ and\ \bibinfo {author} {\bibfnamefont {S.}~\bibnamefont {Patra}},\ }\href {https://doi.org/10.1088/1475-7516/2021/01/007} {\bibfield  {journal} {\bibinfo  {journal} {Journal of Cosmology and Astroparticle Physics}\ }\textbf {\bibinfo {volume} {2021}}\bibinfo  {number} { (01)},\ \bibinfo {pages} {007}}\BibitemShut {NoStop}%
\bibitem [{\citenamefont {Kumar}\ \emph {et~al.}(2022)\citenamefont {Kumar}, \citenamefont {Das},\ and\ \citenamefont {Patra}}]{rmfdm12}%
  \BibitemOpen
\bibfield  {number} {  }\bibfield  {author} {\bibinfo {author} {\bibfnamefont {A.}~\bibnamefont {Kumar}}, \bibinfo {author} {\bibfnamefont {H.~C.}\ \bibnamefont {Das}},\ and\ \bibinfo {author} {\bibfnamefont {S.~K.}\ \bibnamefont {Patra}},\ }\href {https://doi.org/10.1093/mnras/stac1013} {\bibfield  {journal} {\bibinfo  {journal} {Monthly Notices of the Royal Astronomical Society}\ }\textbf {\bibinfo {volume} {513}},\ \bibinfo {pages} {1820} (\bibinfo {year} {2022})}\BibitemShut {NoStop}%
\bibitem [{\citenamefont {Routaray}\ \emph {et~al.}(2023)\citenamefont {Routaray}, \citenamefont {Mohanty}, \citenamefont {Das}, \citenamefont {Ghosh}, \citenamefont {Kalita}, \citenamefont {Parmar},\ and\ \citenamefont {Kumar}}]{nitr}%
  \BibitemOpen
  \bibfield  {author} {\bibinfo {author} {\bibfnamefont {P.}~\bibnamefont {Routaray}}, \bibinfo {author} {\bibfnamefont {S.~R.}\ \bibnamefont {Mohanty}}, \bibinfo {author} {\bibfnamefont {H.}~\bibnamefont {Das}}, \bibinfo {author} {\bibfnamefont {S.}~\bibnamefont {Ghosh}}, \bibinfo {author} {\bibfnamefont {P.}~\bibnamefont {Kalita}}, \bibinfo {author} {\bibfnamefont {V.}~\bibnamefont {Parmar}},\ and\ \bibinfo {author} {\bibfnamefont {B.}~\bibnamefont {Kumar}},\ }\href {https://doi.org/10.1088/1475-7516/2023/10/073} {\bibfield  {journal} {\bibinfo  {journal} {Journal of Cosmology and Astroparticle Physics}\ }\textbf {\bibinfo {volume} {2023}}\bibinfo  {number} { (10)},\ \bibinfo {pages} {073}}\BibitemShut {NoStop}%
\bibitem [{\citenamefont {Akerib}\ \emph {et~al.}(2017)\citenamefont {Akerib} \emph {et~al.}}]{LUX:2017ree}%
  \BibitemOpen
\bibfield  {number} {  }\bibfield  {author} {\bibinfo {author} {\bibfnamefont {D.~S.}\ \bibnamefont {Akerib}} \emph {et~al.} (\bibinfo {collaboration} {LUX}),\ }\href {https://doi.org/10.1103/PhysRevLett.118.251302} {\bibfield  {journal} {\bibinfo  {journal} {Phys. Rev. Lett.}\ }\textbf {\bibinfo {volume} {118}},\ \bibinfo {pages} {251302} (\bibinfo {year} {2017})}\BibitemShut {NoStop}%
\bibitem [{\citenamefont {Aprile}\ \emph {et~al.}(2018)\citenamefont {Aprile} \emph {et~al.}}]{XENON:2018voc}%
  \BibitemOpen
  \bibfield  {author} {\bibinfo {author} {\bibfnamefont {E.}~\bibnamefont {Aprile}} \emph {et~al.} (\bibinfo {collaboration} {XENON}),\ }\href {https://doi.org/10.1103/PhysRevLett.121.111302} {\bibfield  {journal} {\bibinfo  {journal} {Phys. Rev. Lett.}\ }\textbf {\bibinfo {volume} {121}},\ \bibinfo {pages} {111302} (\bibinfo {year} {2018})}\BibitemShut {NoStop}%
\bibitem [{\citenamefont {Cui}\ \emph {et~al.}(2017)\citenamefont {Cui} \emph {et~al.}}]{PandaX-II:2017hlx}%
  \BibitemOpen
  \bibfield  {author} {\bibinfo {author} {\bibfnamefont {X.}~\bibnamefont {Cui}} \emph {et~al.} (\bibinfo {collaboration} {PandaX-II}),\ }\href {https://doi.org/10.1103/PhysRevLett.119.181302} {\bibfield  {journal} {\bibinfo  {journal} {Phys. Rev. Lett.}\ }\textbf {\bibinfo {volume} {119}},\ \bibinfo {pages} {181302} (\bibinfo {year} {2017})}\BibitemShut {NoStop}%
\bibitem [{\citenamefont {Billard}\ \emph {et~al.}(2022)\citenamefont {Billard} \emph {et~al.}}]{Billard:2021uyg}%
  \BibitemOpen
  \bibfield  {author} {\bibinfo {author} {\bibfnamefont {J.}~\bibnamefont {Billard}} \emph {et~al.},\ }\href {https://doi.org/10.1088/1361-6633/ac5754} {\bibfield  {journal} {\bibinfo  {journal} {Rept. Prog. Phys.}\ }\textbf {\bibinfo {volume} {85}},\ \bibinfo {pages} {056201} (\bibinfo {year} {2022})}\BibitemShut {NoStop}%
\bibitem [{\citenamefont {Schumann}(2019)}]{Schumann:2019eaa}%
  \BibitemOpen
  \bibfield  {author} {\bibinfo {author} {\bibfnamefont {M.}~\bibnamefont {Schumann}},\ }\href {https://doi.org/10.1088/1361-6471/ab2ea5} {\bibfield  {journal} {\bibinfo  {journal} {J. Phys. G}\ }\textbf {\bibinfo {volume} {46}},\ \bibinfo {pages} {103003} (\bibinfo {year} {2019})}\BibitemShut {NoStop}%
\bibitem [{\citenamefont {Xiang}\ \emph {et~al.}(2014)\citenamefont {Xiang}, \citenamefont {Jiang}, \citenamefont {Zhang},\ and\ \citenamefont {Yang}}]{qian14}%
  \BibitemOpen
  \bibfield  {author} {\bibinfo {author} {\bibfnamefont {Q.-F.}\ \bibnamefont {Xiang}}, \bibinfo {author} {\bibfnamefont {W.-Z.}\ \bibnamefont {Jiang}}, \bibinfo {author} {\bibfnamefont {D.-R.}\ \bibnamefont {Zhang}},\ and\ \bibinfo {author} {\bibfnamefont {R.-Y.}\ \bibnamefont {Yang}},\ }\href {https://doi.org/10.1103/PhysRevC.89.025803} {\bibfield  {journal} {\bibinfo  {journal} {Phys. Rev. C}\ }\textbf {\bibinfo {volume} {89}},\ \bibinfo {pages} {025803} (\bibinfo {year} {2014})}\BibitemShut {NoStop}%
\bibitem [{\citenamefont {Collier}\ \emph {et~al.}(2022)\citenamefont {Collier}, \citenamefont {Croon},\ and\ \citenamefont {Leane}}]{Collier:2022cpr}%
  \BibitemOpen
  \bibfield  {author} {\bibinfo {author} {\bibfnamefont {M.}~\bibnamefont {Collier}}, \bibinfo {author} {\bibfnamefont {D.}~\bibnamefont {Croon}},\ and\ \bibinfo {author} {\bibfnamefont {R.~K.}\ \bibnamefont {Leane}},\ }\href {https://doi.org/10.1103/PhysRevD.106.123027} {\bibfield  {journal} {\bibinfo  {journal} {Phys. Rev. D}\ }\textbf {\bibinfo {volume} {106}},\ \bibinfo {pages} {123027} (\bibinfo {year} {2022})}\BibitemShut {NoStop}%
\bibitem [{\citenamefont {Shakeri}\ and\ \citenamefont {Karkevandi}(2024)}]{Shakeri:2022dwg}%
  \BibitemOpen
  \bibfield  {author} {\bibinfo {author} {\bibfnamefont {S.}~\bibnamefont {Shakeri}}\ and\ \bibinfo {author} {\bibfnamefont {D.~R.}\ \bibnamefont {Karkevandi}},\ }\href {https://doi.org/10.1103/PhysRevD.109.043029} {\bibfield  {journal} {\bibinfo  {journal} {Phys. Rev. D}\ }\textbf {\bibinfo {volume} {109}},\ \bibinfo {pages} {043029} (\bibinfo {year} {2024})}\BibitemShut {NoStop}%
\bibitem [{\citenamefont {Miao}\ \emph {et~al.}(2022)\citenamefont {Miao}, \citenamefont {Zhu}, \citenamefont {Li},\ and\ \citenamefont {Huang}}]{Miao:2022rqj}%
  \BibitemOpen
  \bibfield  {author} {\bibinfo {author} {\bibfnamefont {Z.}~\bibnamefont {Miao}}, \bibinfo {author} {\bibfnamefont {Y.}~\bibnamefont {Zhu}}, \bibinfo {author} {\bibfnamefont {A.}~\bibnamefont {Li}},\ and\ \bibinfo {author} {\bibfnamefont {F.}~\bibnamefont {Huang}},\ }\href {https://doi.org/10.3847/1538-4357/ac8544} {\bibfield  {journal} {\bibinfo  {journal} {Astrophys. J.}\ }\textbf {\bibinfo {volume} {936}},\ \bibinfo {pages} {69} (\bibinfo {year} {2022})}\BibitemShut {NoStop}%
\bibitem [{\citenamefont {Emma}\ \emph {et~al.}(2022)\citenamefont {Emma}, \citenamefont {Schianchi}, \citenamefont {Pannarale}, \citenamefont {Sagun},\ and\ \citenamefont {Dietrich}}]{Emma:2022xjs}%
  \BibitemOpen
  \bibfield  {author} {\bibinfo {author} {\bibfnamefont {M.}~\bibnamefont {Emma}}, \bibinfo {author} {\bibfnamefont {F.}~\bibnamefont {Schianchi}}, \bibinfo {author} {\bibfnamefont {F.}~\bibnamefont {Pannarale}}, \bibinfo {author} {\bibfnamefont {V.}~\bibnamefont {Sagun}},\ and\ \bibinfo {author} {\bibfnamefont {T.}~\bibnamefont {Dietrich}},\ }\href {https://doi.org/10.3390/particles5030024} {\bibfield  {journal} {\bibinfo  {journal} {Particles}\ }\textbf {\bibinfo {volume} {5}},\ \bibinfo {pages} {273} (\bibinfo {year} {2022})}\BibitemShut {NoStop}%
\bibitem [{\citenamefont {Hong}\ and\ \citenamefont {Ren}(2024)}]{Hong:2024sey}%
  \BibitemOpen
  \bibfield  {author} {\bibinfo {author} {\bibfnamefont {B.}~\bibnamefont {Hong}}\ and\ \bibinfo {author} {\bibfnamefont {Z.}~\bibnamefont {Ren}},\ }\href {https://doi.org/10.1103/PhysRevD.109.023002} {\bibfield  {journal} {\bibinfo  {journal} {Phys. Rev. D}\ }\textbf {\bibinfo {volume} {109}},\ \bibinfo {pages} {023002} (\bibinfo {year} {2024})}\BibitemShut {NoStop}%
\bibitem [{\citenamefont {Karkevandi}\ \emph {et~al.}(2022)\citenamefont {Karkevandi}, \citenamefont {Shakeri}, \citenamefont {Sagun},\ and\ \citenamefont {Ivanytskyi}}]{Karkevandi:2021ygv}%
  \BibitemOpen
  \bibfield  {author} {\bibinfo {author} {\bibfnamefont {D.~R.}\ \bibnamefont {Karkevandi}}, \bibinfo {author} {\bibfnamefont {S.}~\bibnamefont {Shakeri}}, \bibinfo {author} {\bibfnamefont {V.}~\bibnamefont {Sagun}},\ and\ \bibinfo {author} {\bibfnamefont {O.}~\bibnamefont {Ivanytskyi}},\ }\href {https://doi.org/10.1103/PhysRevD.105.023001} {\bibfield  {journal} {\bibinfo  {journal} {Phys. Rev. D}\ }\textbf {\bibinfo {volume} {105}},\ \bibinfo {pages} {023001} (\bibinfo {year} {2022})}\BibitemShut {NoStop}%
\bibitem [{\citenamefont {Liu}\ \emph {et~al.}(2023)\citenamefont {Liu}, \citenamefont {Wei}, \citenamefont {Li}, \citenamefont {Burgio},\ and\ \citenamefont {Schulze}}]{Liu:2023ecz}%
  \BibitemOpen
  \bibfield  {author} {\bibinfo {author} {\bibfnamefont {H.-M.}\ \bibnamefont {Liu}}, \bibinfo {author} {\bibfnamefont {J.-B.}\ \bibnamefont {Wei}}, \bibinfo {author} {\bibfnamefont {Z.-H.}\ \bibnamefont {Li}}, \bibinfo {author} {\bibfnamefont {G.~F.}\ \bibnamefont {Burgio}},\ and\ \bibinfo {author} {\bibfnamefont {H.~J.}\ \bibnamefont {Schulze}},\ }\href {https://doi.org/10.1016/j.dark.2023.101338} {\bibfield  {journal} {\bibinfo  {journal} {Phys. Dark Univ.}\ }\textbf {\bibinfo {volume} {42}},\ \bibinfo {pages} {101338} (\bibinfo {year} {2023})}\BibitemShut {NoStop}%
\bibitem [{\citenamefont {Ivanytskyi}\ \emph {et~al.}(2020)\citenamefont {Ivanytskyi}, \citenamefont {Sagun},\ and\ \citenamefont {Lopes}}]{Ivanytskyi:2019wxd}%
  \BibitemOpen
  \bibfield  {author} {\bibinfo {author} {\bibfnamefont {O.}~\bibnamefont {Ivanytskyi}}, \bibinfo {author} {\bibfnamefont {V.}~\bibnamefont {Sagun}},\ and\ \bibinfo {author} {\bibfnamefont {I.}~\bibnamefont {Lopes}},\ }\href {https://doi.org/10.1103/PhysRevD.102.063028} {\bibfield  {journal} {\bibinfo  {journal} {Phys. Rev. D}\ }\textbf {\bibinfo {volume} {102}},\ \bibinfo {pages} {063028} (\bibinfo {year} {2020})}\BibitemShut {NoStop}%
\bibitem [{\citenamefont {Buras-Stubbs}\ and\ \citenamefont {Lopes}(2024)}]{Buras-Stubbs:2024don}%
  \BibitemOpen
  \bibfield  {author} {\bibinfo {author} {\bibfnamefont {Z.}~\bibnamefont {Buras-Stubbs}}\ and\ \bibinfo {author} {\bibfnamefont {I.}~\bibnamefont {Lopes}},\ }\href {https://doi.org/10.1103/PhysRevD.109.043043} {\bibfield  {journal} {\bibinfo  {journal} {Phys. Rev. D}\ }\textbf {\bibinfo {volume} {109}},\ \bibinfo {pages} {043043} (\bibinfo {year} {2024})}\BibitemShut {NoStop}%
\bibitem [{\citenamefont {Rutherford}\ \emph {et~al.}(2023)\citenamefont {Rutherford}, \citenamefont {Raaijmakers}, \citenamefont {Prescod-Weinstein},\ and\ \citenamefont {Watts}}]{Rutherford:2022xeb}%
  \BibitemOpen
  \bibfield  {author} {\bibinfo {author} {\bibfnamefont {N.}~\bibnamefont {Rutherford}}, \bibinfo {author} {\bibfnamefont {G.}~\bibnamefont {Raaijmakers}}, \bibinfo {author} {\bibfnamefont {C.}~\bibnamefont {Prescod-Weinstein}},\ and\ \bibinfo {author} {\bibfnamefont {A.}~\bibnamefont {Watts}},\ }\href {https://doi.org/10.1103/PhysRevD.107.103051} {\bibfield  {journal} {\bibinfo  {journal} {Phys. Rev. D}\ }\textbf {\bibinfo {volume} {107}},\ \bibinfo {pages} {103051} (\bibinfo {year} {2023})}\BibitemShut {NoStop}%
\bibitem [{\citenamefont {Thakur}\ \emph {et~al.}(2024{\natexlab{a}})\citenamefont {Thakur}, \citenamefont {Malik}, \citenamefont {Das}, \citenamefont {Jha},\ and\ \citenamefont {Provid\^encia}}]{Thakur:2023aqm}%
  \BibitemOpen
  \bibfield  {author} {\bibinfo {author} {\bibfnamefont {P.}~\bibnamefont {Thakur}}, \bibinfo {author} {\bibfnamefont {T.}~\bibnamefont {Malik}}, \bibinfo {author} {\bibfnamefont {A.}~\bibnamefont {Das}}, \bibinfo {author} {\bibfnamefont {T.~K.}\ \bibnamefont {Jha}},\ and\ \bibinfo {author} {\bibfnamefont {C.}~\bibnamefont {Provid\^encia}},\ }\href {https://doi.org/10.1103/PhysRevD.109.043030} {\bibfield  {journal} {\bibinfo  {journal} {Phys. Rev. D}\ }\textbf {\bibinfo {volume} {109}},\ \bibinfo {pages} {043030} (\bibinfo {year} {2024}{\natexlab{a}})}\BibitemShut {NoStop}%
\bibitem [{\citenamefont {Mahapatra}\ \emph {et~al.}(2024)\citenamefont {Mahapatra}, \citenamefont {Singha}, \citenamefont {Hazarika},\ and\ \citenamefont {Das}}]{Mahapatra:2024ywx}%
  \BibitemOpen
  \bibfield  {author} {\bibinfo {author} {\bibfnamefont {P.}~\bibnamefont {Mahapatra}}, \bibinfo {author} {\bibfnamefont {C.}~\bibnamefont {Singha}}, \bibinfo {author} {\bibfnamefont {A.}~\bibnamefont {Hazarika}},\ and\ \bibinfo {author} {\bibfnamefont {P.~K.}\ \bibnamefont {Das}},\ }\href@noop {} {\bibinfo {title} {{Implications of Fermionic Dark Matter Interactions on Anisotropic Neutron Stars}}} (\bibinfo {year} {2024})\BibitemShut {NoStop}%
\bibitem [{\citenamefont {Thakur}\ \emph {et~al.}(2024{\natexlab{b}})\citenamefont {Thakur}, \citenamefont {Malik},\ and\ \citenamefont {Jha}}]{Thakur:2024mxs}%
  \BibitemOpen
  \bibfield  {author} {\bibinfo {author} {\bibfnamefont {P.}~\bibnamefont {Thakur}}, \bibinfo {author} {\bibfnamefont {T.}~\bibnamefont {Malik}},\ and\ \bibinfo {author} {\bibfnamefont {T.~K.}\ \bibnamefont {Jha}},\ }\href {https://doi.org/10.3390/particles7010005} {\bibfield  {journal} {\bibinfo  {journal} {Particles}\ }\textbf {\bibinfo {volume} {7}},\ \bibinfo {pages} {80} (\bibinfo {year} {2024}{\natexlab{b}})}\BibitemShut {NoStop}%
\bibitem [{\citenamefont {Liu}\ \emph {et~al.}(2025)\citenamefont {Liu}, \citenamefont {Mahapatra}, \citenamefont {Huang}, \citenamefont {Hazarika}, \citenamefont {Singha},\ and\ \citenamefont {Das}}]{liu25}%
  \BibitemOpen
  \bibfield  {author} {\bibinfo {author} {\bibfnamefont {X.-Z.}\ \bibnamefont {Liu}}, \bibinfo {author} {\bibfnamefont {P.}~\bibnamefont {Mahapatra}}, \bibinfo {author} {\bibfnamefont {C.}~\bibnamefont {Huang}}, \bibinfo {author} {\bibfnamefont {A.}~\bibnamefont {Hazarika}}, \bibinfo {author} {\bibfnamefont {C.}~\bibnamefont {Singha}},\ and\ \bibinfo {author} {\bibfnamefont {P.~K.}\ \bibnamefont {Das}},\ }\href {https://doi.org/10.1103/zhs6-487x} {\bibfield  {journal} {\bibinfo  {journal} {Phys. Rev. D}\ }\textbf {\bibinfo {volume} {112}},\ \bibinfo {pages} {083032} (\bibinfo {year} {2025})}\BibitemShut {NoStop}%
\bibitem [{\citenamefont {Barbat}\ \emph {et~al.}(2024)\citenamefont {Barbat}, \citenamefont {Schaffner-Bielich},\ and\ \citenamefont {Tolos}}]{laura24}%
  \BibitemOpen
  \bibfield  {author} {\bibinfo {author} {\bibfnamefont {M.~F.}\ \bibnamefont {Barbat}}, \bibinfo {author} {\bibfnamefont {J.}~\bibnamefont {Schaffner-Bielich}},\ and\ \bibinfo {author} {\bibfnamefont {L.}~\bibnamefont {Tolos}},\ }\href {https://doi.org/10.1103/PhysRevD.110.023013} {\bibfield  {journal} {\bibinfo  {journal} {Phys. Rev. D}\ }\textbf {\bibinfo {volume} {110}},\ \bibinfo {pages} {023013} (\bibinfo {year} {2024})}\BibitemShut {NoStop}%
\bibitem [{\citenamefont {Narain}\ \emph {et~al.}(2006)\citenamefont {Narain}, \citenamefont {Schaffner-Bielich},\ and\ \citenamefont {Mishustin}}]{jurgen06}%
  \BibitemOpen
  \bibfield  {author} {\bibinfo {author} {\bibfnamefont {G.}~\bibnamefont {Narain}}, \bibinfo {author} {\bibfnamefont {J.}~\bibnamefont {Schaffner-Bielich}},\ and\ \bibinfo {author} {\bibfnamefont {I.~N.}\ \bibnamefont {Mishustin}},\ }\href {https://doi.org/10.1103/PhysRevD.74.063003} {\bibfield  {journal} {\bibinfo  {journal} {Phys. Rev. D}\ }\textbf {\bibinfo {volume} {74}},\ \bibinfo {pages} {063003} (\bibinfo {year} {2006})}\BibitemShut {NoStop}%
\bibitem [{\citenamefont {Navas}\ \emph {et~al.}(2024)\citenamefont {Navas}, \citenamefont {Amsler}, \citenamefont {Gutsche}, \citenamefont {Hanhart}, \citenamefont {Hern\'andez-Rey}, \citenamefont {Louren\ifmmode~\mbox{\c{c}}\else \c{c}\fi{}o}, \citenamefont {Masoni}, \citenamefont {Mikhasenko}, \citenamefont {Mitchell}, \citenamefont {Patrignani} \emph {et~al.}}]{pdg24}%
  \BibitemOpen
  \bibfield  {author} {\bibinfo {author} {\bibfnamefont {S.}~\bibnamefont {Navas}}, \bibinfo {author} {\bibfnamefont {C.}~\bibnamefont {Amsler}}, \bibinfo {author} {\bibfnamefont {T.}~\bibnamefont {Gutsche}}, \bibinfo {author} {\bibfnamefont {C.}~\bibnamefont {Hanhart}}, \bibinfo {author} {\bibfnamefont {J.~J.}\ \bibnamefont {Hern\'andez-Rey}}, \bibinfo {author} {\bibfnamefont {C.}~\bibnamefont {Louren\ifmmode~\mbox{\c{c}}\else \c{c}\fi{}o}}, \bibinfo {author} {\bibfnamefont {A.}~\bibnamefont {Masoni}}, \bibinfo {author} {\bibfnamefont {M.}~\bibnamefont {Mikhasenko}}, \bibinfo {author} {\bibfnamefont {R.~E.}\ \bibnamefont {Mitchell}}, \bibinfo {author} {\bibfnamefont {C.}~\bibnamefont {Patrignani}}, \emph {et~al.} (\bibinfo {collaboration} {Particle Data Group Collaboration}),\ }\href {https://doi.org/10.1103/PhysRevD.110.030001} {\bibfield  {journal} {\bibinfo  {journal} {Phys. Rev. D}\ }\textbf {\bibinfo {volume} {110}},\ \bibinfo {pages} {030001} (\bibinfo {year} {2024})}\BibitemShut {NoStop}%
\bibitem [{\citenamefont {Das}\ \emph {et~al.}(2022{\natexlab{b}})\citenamefont {Das}, \citenamefont {Malik},\ and\ \citenamefont {Nayak}}]{rmfdm1}%
  \BibitemOpen
  \bibfield  {author} {\bibinfo {author} {\bibfnamefont {A.}~\bibnamefont {Das}}, \bibinfo {author} {\bibfnamefont {T.}~\bibnamefont {Malik}},\ and\ \bibinfo {author} {\bibfnamefont {A.~C.}\ \bibnamefont {Nayak}},\ }\href {https://doi.org/10.1103/PhysRevD.105.123034} {\bibfield  {journal} {\bibinfo  {journal} {Phys. Rev. D}\ }\textbf {\bibinfo {volume} {105}},\ \bibinfo {pages} {123034} (\bibinfo {year} {2022}{\natexlab{b}})}\BibitemShut {NoStop}%
\bibitem [{\citenamefont {Calmet}\ and\ \citenamefont {Kuipers}(2021)}]{calmet21}%
  \BibitemOpen
  \bibfield  {author} {\bibinfo {author} {\bibfnamefont {X.}~\bibnamefont {Calmet}}\ and\ \bibinfo {author} {\bibfnamefont {F.}~\bibnamefont {Kuipers}},\ }\href {https://doi.org/https://doi.org/10.1016/j.physletb.2021.136068} {\bibfield  {journal} {\bibinfo  {journal} {Physics Letters B}\ }\textbf {\bibinfo {volume} {814}},\ \bibinfo {pages} {136068} (\bibinfo {year} {2021})}\BibitemShut {NoStop}%
\bibitem [{\citenamefont {{Baym}}\ \emph {et~al.}(1971)\citenamefont {{Baym}}, \citenamefont {{Pethick}},\ and\ \citenamefont {{Sutherland}}}]{bps}%
  \BibitemOpen
  \bibfield  {author} {\bibinfo {author} {\bibfnamefont {G.}~\bibnamefont {{Baym}}}, \bibinfo {author} {\bibfnamefont {C.}~\bibnamefont {{Pethick}}},\ and\ \bibinfo {author} {\bibfnamefont {P.}~\bibnamefont {{Sutherland}}},\ }\href {https://doi.org/10.1086/151216} {\bibfield  {journal} {\bibinfo  {journal} {Astrophys. J.}\ }\textbf {\bibinfo {volume} {170}},\ \bibinfo {pages} {299} (\bibinfo {year} {1971})}\BibitemShut {NoStop}%
\bibitem [{\citenamefont {Chabanat}\ \emph {et~al.}(1998)\citenamefont {Chabanat}, \citenamefont {Bonche}, \citenamefont {Haensel}, \citenamefont {Meyer},\ and\ \citenamefont {Schaeffer}}]{sly4}%
  \BibitemOpen
  \bibfield  {author} {\bibinfo {author} {\bibfnamefont {E.}~\bibnamefont {Chabanat}}, \bibinfo {author} {\bibfnamefont {P.}~\bibnamefont {Bonche}}, \bibinfo {author} {\bibfnamefont {P.}~\bibnamefont {Haensel}}, \bibinfo {author} {\bibfnamefont {J.}~\bibnamefont {Meyer}},\ and\ \bibinfo {author} {\bibfnamefont {R.}~\bibnamefont {Schaeffer}},\ }\href {https://doi.org/https://doi.org/10.1016/S0375-9474(98)00180-8} {\bibfield  {journal} {\bibinfo  {journal} {Nuclear Physics A}\ }\textbf {\bibinfo {volume} {635}},\ \bibinfo {pages} {231} (\bibinfo {year} {1998})}\BibitemShut {NoStop}%
\bibitem [{\citenamefont {{Douchin, F.}}\ and\ \citenamefont {{Haensel, P.}}(2001)}]{douchin}%
  \BibitemOpen
  \bibfield  {author} {\bibinfo {author} {\bibnamefont {{Douchin, F.}}}\ and\ \bibinfo {author} {\bibnamefont {{Haensel, P.}}},\ }\href {https://doi.org/10.1051/0004-6361:20011402} {\bibfield  {journal} {\bibinfo  {journal} {A\&A}\ }\textbf {\bibinfo {volume} {380}},\ \bibinfo {pages} {151} (\bibinfo {year} {2001})}\BibitemShut {NoStop}%
\bibitem [{\citenamefont {Hippert}\ \emph {et~al.}(2023)\citenamefont {Hippert}, \citenamefont {Dillingham}, \citenamefont {Tan}, \citenamefont {Curtin}, \citenamefont {Noronha-Hostler},\ and\ \citenamefont {Yunes}}]{mauricio23}%
  \BibitemOpen
  \bibfield  {author} {\bibinfo {author} {\bibfnamefont {M.}~\bibnamefont {Hippert}}, \bibinfo {author} {\bibfnamefont {E.}~\bibnamefont {Dillingham}}, \bibinfo {author} {\bibfnamefont {H.}~\bibnamefont {Tan}}, \bibinfo {author} {\bibfnamefont {D.}~\bibnamefont {Curtin}}, \bibinfo {author} {\bibfnamefont {J.}~\bibnamefont {Noronha-Hostler}},\ and\ \bibinfo {author} {\bibfnamefont {N.}~\bibnamefont {Yunes}},\ }\href {https://doi.org/10.1103/PhysRevD.107.115028} {\bibfield  {journal} {\bibinfo  {journal} {Phys. Rev. D}\ }\textbf {\bibinfo {volume} {107}},\ \bibinfo {pages} {115028} (\bibinfo {year} {2023})}\BibitemShut {NoStop}%
\bibitem [{\citenamefont {Pitz}\ and\ \citenamefont {Schaffner-Bielich}(2025)}]{pitz25}%
  \BibitemOpen
  \bibfield  {author} {\bibinfo {author} {\bibfnamefont {S.~L.}\ \bibnamefont {Pitz}}\ and\ \bibinfo {author} {\bibfnamefont {J.}~\bibnamefont {Schaffner-Bielich}},\ }\href {https://doi.org/10.1103/PhysRevD.111.043050} {\bibfield  {journal} {\bibinfo  {journal} {Phys. Rev. D}\ }\textbf {\bibinfo {volume} {111}},\ \bibinfo {pages} {043050} (\bibinfo {year} {2025})}\BibitemShut {NoStop}%
\bibitem [{\citenamefont {Salmi}\ \emph {et~al.}(2024)\citenamefont {Salmi}, \citenamefont {Choudhury}, \citenamefont {Kini}, \citenamefont {Riley}, \citenamefont {Vinciguerra}, \citenamefont {Watts}, \citenamefont {Wolff}, \citenamefont {Arzoumanian}, \citenamefont {Bogdanov}, \citenamefont {Chakrabarty}, \citenamefont {Gendreau}, \citenamefont {Guillot}, \citenamefont {Ho}, \citenamefont {Huppenkothen}, \citenamefont {Ludlam}, \citenamefont {Morsink},\ and\ \citenamefont {Ray}}]{Salmi_2024}%
  \BibitemOpen
  \bibfield  {author} {\bibinfo {author} {\bibfnamefont {T.}~\bibnamefont {Salmi}}, \bibinfo {author} {\bibfnamefont {D.}~\bibnamefont {Choudhury}}, \bibinfo {author} {\bibfnamefont {Y.}~\bibnamefont {Kini}}, \bibinfo {author} {\bibfnamefont {T.~E.}\ \bibnamefont {Riley}}, \bibinfo {author} {\bibfnamefont {S.}~\bibnamefont {Vinciguerra}}, \bibinfo {author} {\bibfnamefont {A.~L.}\ \bibnamefont {Watts}}, \bibinfo {author} {\bibfnamefont {M.~T.}\ \bibnamefont {Wolff}}, \bibinfo {author} {\bibfnamefont {Z.}~\bibnamefont {Arzoumanian}}, \bibinfo {author} {\bibfnamefont {S.}~\bibnamefont {Bogdanov}}, \bibinfo {author} {\bibfnamefont {D.}~\bibnamefont {Chakrabarty}}, \bibinfo {author} {\bibfnamefont {K.}~\bibnamefont {Gendreau}}, \bibinfo {author} {\bibfnamefont {S.}~\bibnamefont {Guillot}}, \bibinfo {author} {\bibfnamefont {W.~C.~G.}\ \bibnamefont {Ho}}, \bibinfo {author} {\bibfnamefont {D.}~\bibnamefont {Huppenkothen}}, \bibinfo {author} {\bibfnamefont {R.~M.}\ \bibnamefont {Ludlam}}, \bibinfo {author}
  {\bibfnamefont {S.~M.}\ \bibnamefont {Morsink}},\ and\ \bibinfo {author} {\bibfnamefont {P.~S.}\ \bibnamefont {Ray}},\ }\href {https://doi.org/10.3847/1538-4357/ad5f1f} {\bibfield  {journal} {\bibinfo  {journal} {The Astrophysical Journal}\ }\textbf {\bibinfo {volume} {974}},\ \bibinfo {pages} {294} (\bibinfo {year} {2024})}\BibitemShut {NoStop}%
\bibitem [{\citenamefont {Dittmann}\ \emph {et~al.}(2024)\citenamefont {Dittmann}, \citenamefont {Miller}, \citenamefont {Lamb}, \citenamefont {Holt}, \citenamefont {Chirenti}, \citenamefont {Wolff}, \citenamefont {Bogdanov}, \citenamefont {Guillot}, \citenamefont {Ho}, \citenamefont {Morsink}, \citenamefont {Arzoumanian},\ and\ \citenamefont {Gendreau}}]{Dittmann_2024}%
  \BibitemOpen
  \bibfield  {author} {\bibinfo {author} {\bibfnamefont {A.~J.}\ \bibnamefont {Dittmann}}, \bibinfo {author} {\bibfnamefont {M.~C.}\ \bibnamefont {Miller}}, \bibinfo {author} {\bibfnamefont {F.~K.}\ \bibnamefont {Lamb}}, \bibinfo {author} {\bibfnamefont {I.~M.}\ \bibnamefont {Holt}}, \bibinfo {author} {\bibfnamefont {C.}~\bibnamefont {Chirenti}}, \bibinfo {author} {\bibfnamefont {M.~T.}\ \bibnamefont {Wolff}}, \bibinfo {author} {\bibfnamefont {S.}~\bibnamefont {Bogdanov}}, \bibinfo {author} {\bibfnamefont {S.}~\bibnamefont {Guillot}}, \bibinfo {author} {\bibfnamefont {W.~C.~G.}\ \bibnamefont {Ho}}, \bibinfo {author} {\bibfnamefont {S.~M.}\ \bibnamefont {Morsink}}, \bibinfo {author} {\bibfnamefont {Z.}~\bibnamefont {Arzoumanian}},\ and\ \bibinfo {author} {\bibfnamefont {K.~C.}\ \bibnamefont {Gendreau}},\ }\href {https://doi.org/10.3847/1538-4357/ad5f1e} {\bibfield  {journal} {\bibinfo  {journal} {The Astrophysical Journal}\ }\textbf {\bibinfo {volume} {974}},\ \bibinfo {pages} {295} (\bibinfo {year}
  {2024})}\BibitemShut {NoStop}%
\bibitem [{\citenamefont {Mukhopadhyay}\ and\ \citenamefont {Schaffner-Bielich}(2016)}]{jurgen16}%
  \BibitemOpen
  \bibfield  {author} {\bibinfo {author} {\bibfnamefont {P.}~\bibnamefont {Mukhopadhyay}}\ and\ \bibinfo {author} {\bibfnamefont {J.}~\bibnamefont {Schaffner-Bielich}},\ }\href {https://doi.org/10.1103/PhysRevD.93.083009} {\bibfield  {journal} {\bibinfo  {journal} {Phys. Rev. D}\ }\textbf {\bibinfo {volume} {93}},\ \bibinfo {pages} {083009} (\bibinfo {year} {2016})}\BibitemShut {NoStop}%
\bibitem [{\citenamefont {Ellis}\ \emph {et~al.}(2018)\citenamefont {Ellis}, \citenamefont {H\"utsi}, \citenamefont {Kannike}, \citenamefont {Marzola}, \citenamefont {Raidal},\ and\ \citenamefont {Vaskonen}}]{ellis18}%
  \BibitemOpen
  \bibfield  {author} {\bibinfo {author} {\bibfnamefont {J.}~\bibnamefont {Ellis}}, \bibinfo {author} {\bibfnamefont {G.}~\bibnamefont {H\"utsi}}, \bibinfo {author} {\bibfnamefont {K.}~\bibnamefont {Kannike}}, \bibinfo {author} {\bibfnamefont {L.}~\bibnamefont {Marzola}}, \bibinfo {author} {\bibfnamefont {M.}~\bibnamefont {Raidal}},\ and\ \bibinfo {author} {\bibfnamefont {V.}~\bibnamefont {Vaskonen}},\ }\href {https://doi.org/10.1103/PhysRevD.97.123007} {\bibfield  {journal} {\bibinfo  {journal} {Phys. Rev. D}\ }\textbf {\bibinfo {volume} {97}},\ \bibinfo {pages} {123007} (\bibinfo {year} {2018})}\BibitemShut {NoStop}%
\bibitem [{\citenamefont {Ray}\ \emph {et~al.}(2019)\citenamefont {Ray}, \citenamefont {Arzoumanian}, \citenamefont {Ballantyne}, \citenamefont {Bozzo}, \citenamefont {Brandt}, \citenamefont {Brenneman}, \citenamefont {Chakrabarty}, \citenamefont {Christophersen}, \citenamefont {DeRosa}, \citenamefont {Feroci}, \citenamefont {Gendreau}, \citenamefont {Goldstein}, \citenamefont {Hartmann} \emph {et~al.}}]{strobex}%
  \BibitemOpen
  \bibfield  {author} {\bibinfo {author} {\bibfnamefont {P.~S.}\ \bibnamefont {Ray}}, \bibinfo {author} {\bibfnamefont {Z.}~\bibnamefont {Arzoumanian}}, \bibinfo {author} {\bibfnamefont {D.}~\bibnamefont {Ballantyne}}, \bibinfo {author} {\bibfnamefont {E.}~\bibnamefont {Bozzo}}, \bibinfo {author} {\bibfnamefont {S.}~\bibnamefont {Brandt}}, \bibinfo {author} {\bibfnamefont {L.}~\bibnamefont {Brenneman}}, \bibinfo {author} {\bibfnamefont {D.}~\bibnamefont {Chakrabarty}}, \bibinfo {author} {\bibfnamefont {M.}~\bibnamefont {Christophersen}}, \bibinfo {author} {\bibfnamefont {A.}~\bibnamefont {DeRosa}}, \bibinfo {author} {\bibfnamefont {M.}~\bibnamefont {Feroci}}, \bibinfo {author} {\bibfnamefont {K.}~\bibnamefont {Gendreau}}, \bibinfo {author} {\bibfnamefont {A.}~\bibnamefont {Goldstein}}, \bibinfo {author} {\bibfnamefont {D.}~\bibnamefont {Hartmann}}, \emph {et~al.},\ }\href {https://arxiv.org/abs/1903.03035} {\bibinfo {title} {Strobe-x: X-ray timing and spectroscopy on dynamical timescales from microseconds to
  years}} (\bibinfo {year} {2019}),\ \Eprint {https://arxiv.org/abs/1903.03035} {arXiv:1903.03035 [astro-ph.IM]} \BibitemShut {NoStop}%
\bibitem [{\citenamefont {in't Zand}\ \emph {et~al.}(2019)\citenamefont {in't Zand}, \citenamefont {{Bozzo}}, \citenamefont {{Qu}}, \citenamefont {{Li}}, \citenamefont {{Amati}}, \citenamefont {{Chen}}, \citenamefont {{Donnarumma}}, \citenamefont {{Doroshenko}}, \citenamefont {{Drake}}, \citenamefont {{Hernanz}} \emph {et~al.}}]{zand2019}%
  \BibitemOpen
  \bibfield  {author} {\bibinfo {author} {\bibfnamefont {J.~J.~M.}\ \bibnamefont {in't Zand}}, \bibinfo {author} {\bibfnamefont {E.}~\bibnamefont {{Bozzo}}}, \bibinfo {author} {\bibfnamefont {J.}~\bibnamefont {{Qu}}}, \bibinfo {author} {\bibfnamefont {X.-D.}\ \bibnamefont {{Li}}}, \bibinfo {author} {\bibfnamefont {L.}~\bibnamefont {{Amati}}}, \bibinfo {author} {\bibfnamefont {Y.}~\bibnamefont {{Chen}}}, \bibinfo {author} {\bibfnamefont {I.}~\bibnamefont {{Donnarumma}}}, \bibinfo {author} {\bibfnamefont {V.}~\bibnamefont {{Doroshenko}}}, \bibinfo {author} {\bibfnamefont {S.~A.}\ \bibnamefont {{Drake}}}, \bibinfo {author} {\bibfnamefont {M.}~\bibnamefont {{Hernanz}}}, \emph {et~al.},\ }\href {https://doi.org/10.1007/s11433-017-9186-1} {\bibfield  {journal} {\bibinfo  {journal} {Science China Physics, Mechanics, and Astronomy}\ }\textbf {\bibinfo {volume} {62}},\ \bibinfo {eid} {29506} (\bibinfo {year} {2019})}\BibitemShut {NoStop}%
\bibitem [{\citenamefont {Cassano}\ \emph {et~al.}(2018)\citenamefont {Cassano}, \citenamefont {Fender}, \citenamefont {Ferrari}, \citenamefont {Merloni}, \citenamefont {Akahori}, \citenamefont {Akamatsu}, \citenamefont {Ascasibar}, \citenamefont {Ballantyne}, \citenamefont {Brunetti}, \citenamefont {Corbelli}, \citenamefont {Croston}, \citenamefont {Donnarumma}, \citenamefont {Ettori}, \citenamefont {Ferdman} \emph {et~al.}}]{athena}%
  \BibitemOpen
  \bibfield  {author} {\bibinfo {author} {\bibfnamefont {R.}~\bibnamefont {Cassano}}, \bibinfo {author} {\bibfnamefont {R.}~\bibnamefont {Fender}}, \bibinfo {author} {\bibfnamefont {C.}~\bibnamefont {Ferrari}}, \bibinfo {author} {\bibfnamefont {A.}~\bibnamefont {Merloni}}, \bibinfo {author} {\bibfnamefont {T.}~\bibnamefont {Akahori}}, \bibinfo {author} {\bibfnamefont {H.}~\bibnamefont {Akamatsu}}, \bibinfo {author} {\bibfnamefont {Y.}~\bibnamefont {Ascasibar}}, \bibinfo {author} {\bibfnamefont {D.}~\bibnamefont {Ballantyne}}, \bibinfo {author} {\bibfnamefont {G.}~\bibnamefont {Brunetti}}, \bibinfo {author} {\bibfnamefont {E.}~\bibnamefont {Corbelli}}, \bibinfo {author} {\bibfnamefont {J.}~\bibnamefont {Croston}}, \bibinfo {author} {\bibfnamefont {I.}~\bibnamefont {Donnarumma}}, \bibinfo {author} {\bibfnamefont {S.}~\bibnamefont {Ettori}}, \bibinfo {author} {\bibfnamefont {R.}~\bibnamefont {Ferdman}}, \emph {et~al.},\ }\href {https://arxiv.org/abs/1807.09080} {\bibinfo {title} {Ska-athena synergy white paper}}
  (\bibinfo {year} {2018}),\ \Eprint {https://arxiv.org/abs/1807.09080} {arXiv:1807.09080 [astro-ph.HE]} \BibitemShut {NoStop}%
\bibitem [{\citenamefont {Barcons}\ \emph {et~al.}(2024)\citenamefont {Barcons}, \citenamefont {Nandra}, \citenamefont {Pointecouteau}, \citenamefont {den Herder}, \citenamefont {Barret},\ and\ \citenamefont {et~al.}}]{newathena2024}%
  \BibitemOpen
  \bibfield  {author} {\bibinfo {author} {\bibfnamefont {X.}~\bibnamefont {Barcons}}, \bibinfo {author} {\bibfnamefont {K.}~\bibnamefont {Nandra}}, \bibinfo {author} {\bibfnamefont {E.}~\bibnamefont {Pointecouteau}}, \bibinfo {author} {\bibfnamefont {J.-W.}\ \bibnamefont {den Herder}}, \bibinfo {author} {\bibfnamefont {D.}~\bibnamefont {Barret}},\ and\ \bibinfo {author} {\bibnamefont {et~al.}},\ }\href {https://doi.org/10.1038/s41550-024-02416-3} {\bibfield  {journal} {\bibinfo  {journal} {Nature Astronomy}\ }\textbf {\bibinfo {volume} {8}},\ \bibinfo {pages} {856} (\bibinfo {year} {2024})}\BibitemShut {NoStop}%
\bibitem [{\citenamefont {Santangelo}\ \emph {et~al.}(2024)\citenamefont {Santangelo}, \citenamefont {Zhang}, \citenamefont {Feroci}, \citenamefont {Hernanz}, \citenamefont {Lu},\ and\ \citenamefont {Xu}}]{Santangelo2024}%
  \BibitemOpen
  \bibfield  {author} {\bibinfo {author} {\bibfnamefont {A.}~\bibnamefont {Santangelo}}, \bibinfo {author} {\bibfnamefont {S.-N.}\ \bibnamefont {Zhang}}, \bibinfo {author} {\bibfnamefont {M.}~\bibnamefont {Feroci}}, \bibinfo {author} {\bibfnamefont {M.}~\bibnamefont {Hernanz}}, \bibinfo {author} {\bibfnamefont {F.}~\bibnamefont {Lu}},\ and\ \bibinfo {author} {\bibfnamefont {Y.}~\bibnamefont {Xu}},\ }\bibinfo {title} {The enhanced x-ray timing and polarimetry mission: extp},\ in\ \href {https://doi.org/10.1007/978-981-19-6960-7_34} {\emph {\bibinfo {booktitle} {Handbook of X-ray and Gamma-ray Astrophysics}}},\ \bibinfo {editor} {edited by\ \bibinfo {editor} {\bibfnamefont {C.}~\bibnamefont {Bambi}}\ and\ \bibinfo {editor} {\bibfnamefont {A.}~\bibnamefont {Santangelo}}}\ (\bibinfo  {publisher} {Springer Nature Singapore},\ \bibinfo {address} {Singapore},\ \bibinfo {year} {2024})\ pp.\ \bibinfo {pages} {1201--1229}\BibitemShut {NoStop}%
\bibitem [{\citenamefont {Koehn}\ \emph {et~al.}(2024)\citenamefont {Koehn}, \citenamefont {Giangrandi}, \citenamefont {Kunert}, \citenamefont {Somasundaram}, \citenamefont {Sagun},\ and\ \citenamefont {Dietrich}}]{Koehn2024}%
  \BibitemOpen
  \bibfield  {author} {\bibinfo {author} {\bibfnamefont {H.}~\bibnamefont {Koehn}}, \bibinfo {author} {\bibfnamefont {E.}~\bibnamefont {Giangrandi}}, \bibinfo {author} {\bibfnamefont {N.}~\bibnamefont {Kunert}}, \bibinfo {author} {\bibfnamefont {R.}~\bibnamefont {Somasundaram}}, \bibinfo {author} {\bibfnamefont {V.}~\bibnamefont {Sagun}},\ and\ \bibinfo {author} {\bibfnamefont {T.}~\bibnamefont {Dietrich}},\ }\href {https://doi.org/10.1103/PhysRevD.110.103033} {\bibfield  {journal} {\bibinfo  {journal} {Phys. Rev. D}\ }\textbf {\bibinfo {volume} {110}},\ \bibinfo {pages} {103033} (\bibinfo {year} {2024})}\BibitemShut {NoStop}%
\end{thebibliography}%

\end{document}